\documentclass[aps,prx,twocolumn,showpacs,superscriptaddress,longbibliography]{revtex4-2}
\usepackage{amsmath,amssymb,graphicx}
\usepackage{graphicx,epstopdf}
\usepackage{gensymb}
\usepackage[dvipsnames]{xcolor}
\usepackage[T1]{fontenc}
\usepackage[utf8]{inputenc}
\newcommand{\be}{\begin{equation}}
\newcommand{\ee}{\end{equation}}
\newcommand{\bea}{\begin{eqnarray}}
\newcommand{\eea}{\end{eqnarray}}
\newcommand{\bse}{\begin{subequations}}
\newcommand{\ese}{\end{subequations}}

\usepackage{multibib}
\usepackage{color}
\usepackage[colorlinks,bookmarks=false,citecolor=darkblue,linkcolor=red,urlcolor=blue]{hyperref}

\definecolor{darkred}{rgb}{0.7,0.0,0.0}

\definecolor{darkblue}{rgb}{0,0.02,0.45}

\definecolor{darkgreen}{rgb}{0.02,0.45,0.0}

\definecolor{violet}{rgb}{0.8,0.2,0.6}

\begin{document}

\title{Crystal electric field excitations and spin dynamics in a spin-orbit coupled distorted honeycomb magnet BiErGeO$_{5}$}
\author{S. Mohanty}
\author{S. Guchhait}
\affiliation{School of Physics, Indian Institute of Science Education and Research Thiruvananthapuram, Thiruvananthapuram-695551, India}
\author{S. S. Islam}
\email{shams.islam@psi.ch}
\affiliation{PSI Center for Neutron and Muon Sciences CNM, 5232 Villigen PSI, Switzerland}
\author{Surya P. Patra}
\affiliation{School of Physics, Indian Institute of Science Education and Research Thiruvananthapuram, Thiruvananthapuram-695551, India}
\author{M. P. Saravanan}
\affiliation{UGC-DAE Consortium for Scientific Research, University Campus, Khandwa Road, Indore 452001, India}
\author{J. A. Krieger}
\affiliation{PSI Center for Neutron and Muon Sciences CNM, 5232 Villigen PSI, Switzerland}
\author{T. J. Hicken}
\affiliation{PSI Center for Neutron and Muon Sciences CNM, 5232 Villigen PSI, Switzerland}
\author{H. Luetkens}
\affiliation{PSI Center for Neutron and Muon Sciences CNM, 5232 Villigen PSI, Switzerland}
\author{D. T. Adroja}
\affiliation{ISIS Neutron and Muon Source, Science and Technology Facilities Council, Rutherford Appleton Laboratory, Didcot OX11 0QX, United Kingdom}
\author{Gøran J. Nilsen}
\affiliation{ISIS Neutron and Muon Source, Science and Technology Facilities Council, Rutherford Appleton Laboratory, Didcot OX11 0QX, United Kingdom}
\author{M. D. Le}
\affiliation{ISIS Neutron and Muon Source, Science and Technology Facilities Council, Rutherford Appleton Laboratory, Didcot OX11 0QX, United Kingdom}
\author{R. Nath}
\email{rameshchandra.nath@gmail.com}
\affiliation{School of Physics, Indian Institute of Science Education and Research Thiruvananthapuram, Thiruvananthapuram-695551, India}
\date{\today}

\begin{abstract}
The magnetic properties and crystal electric field (CEF) scheme of BiErGeO$_{5}$ are investigated via magnetization, heat capacity, muon spin relaxation ($\mu$SR), and inelastic neutron scattering (INS) experiments on a polycrystalline sample. The Er$^{3+}$ ions form a quasi-two-dimensional distorted honeycomb network with a Kramers doublet ground state. Magnetic susceptibility and heat capacity reveal short-range antiferromagnetic correlations, manifested as a broad maximum around 1.4~K. Heat-capacity data further confirm the onset of a magnetic long-range order at $T_{\rm N} \simeq 0.4$~K. The INS spectra exhibit eight CEF excitations and the CEF analysis yields the $g$-factor anisotropy with $g_{xy}/g_{z} \simeq 1.38$ and exchange anisotropy with $\mathcal{J}_{xy} \simeq 2.96$~K and $\mathcal{J}_{z} \simeq 1.56$~K. The experimental temperature and field dependent magnetization and heat capacity are also reproduced by the simulation using CEF energy scheme. Zero-field $\mu$SR measurements down to 30~mK, do not exhibit coherent oscillations or a static $1/3$ tail. The spectra are well described by two exponential relaxation components, indicating two magnetically inequivalent muon environments. The relaxation rates display a nearly temperature-independent plateau below $T_{\rm N}$ and follow an Orbach-type activated behavior at higher temperatures involving excited CEF levels, consistent with the INS results. Longitudinal-field $\mu$SR measurements reveal only weak decoupling up to 1.5~T, indicating the presence of slow spin fluctuations that persist well below $T_{\rm N}$. Overall, BiErGeO$_5$ emerges as an anisotropic spin–orbit-coupled honeycomb magnet in which strong CEF-driven anisotropy and reduced dimensionality give rise to unconventional low-temperature magnetic behavior distinct from its Yb counterpart BiYbGeO$_5$.
\end{abstract}

\maketitle

\section{Introduction}
Rare-earth (4$f$) magnets provide a powerful platform for exploring unconventional quantum phenomena arising from spin-orbit coupling (SOC) and crystal electric field (CEF). These effects host highly anisotropic super-exchange pathways and offer a tunable route to stabilize nontrivial magnetic ground states~\cite{Elliott167,Fulde589,Li097201}. The CEF controls the size and anisotropy of the magnetic moment and determines the nature of the low-lying excitations. The composition of the CEF wave functions can be expressed using $\lvert J, J_{z} \rangle$ basis ($J$ and $J_z$ indicates the total angular momentum and $z$ component of the angular momentum operator, respectively) which provides direct information about the quantum fluctuations~\cite{Rau144417,Tomasello155120,Petit060410}. Here, the $\lvert J, J_{z} \rangle$ basis is valid because the spin-orbit coupling is much stronger than the CEF. When the CEF has a small $\lvert J_z \rvert$ component, enhanced fluctuations are supported, which gives rise to exotic states like a quantum spin liquid (QSL)~\cite{Gao024424,Sibille711}. Therefore, it is necessary to analyze the CEF scheme, especially in rare-earth based systems to understand the low-temperature magnetic behavior. 

Among the $4f$ ions, Yb$^{3+}$ based magnets have been extensively studied with a rich variety of quantum phases~\cite{Li107202,Somesh064421,Lhotel024427,Guchhait214437,Sebastian104428}. Similarly, Er$^{3+}$ based compounds have also gained considerable attention to understand anisotropic driven quantum effects at low temperatures. For example, depending on the CEF environment, XY and Ising like anisotropies are realized in different lattice geometries with Er$^{3+}$ ions~\cite{Yahne104423,Scheie144432,Gaudet024415,Champion020401,Cai094432,Cai184415}. ErMgGaO$_4$ is one of the most celebrated compounds without any magnetic long-range order (LRO) down to 25~mK~\cite{Cai094432}. Similarly, the triangular delafossite series with general formula $A$Er$X_2$ ($A$ = Na, K, and Cs, and $X$ = S, Se, and Te) is being pursued vigorously~\cite{Gao024424,Xing114413,Scheie144432,Xing144413,DingL100411,Xie117505,Liu10,Liu2021,Liu107504,whitelock2025}. Most of these compounds exhibit low-temperature magnetic ordering despite strong geometric frustration. Recently, two polymorphs of a distorted honeycomb compound Er$_2$Si$_2$O$_7$ have attracted increasing attention. The C-polymorph hosts a simple collinear N\'eel antiferromagnetic (AFM) order with moderate anisotropy and conventional excitations~\cite{Hester125804}. In contrast, the D-polymorph exhibits a four-sublattice noncollinear Ising-like structure accompanied by gapped excitations~\cite{Islam094420,Hester405801}.

An additional motivation for studying 4$f$ honeycomb magnets lies in the prospect of realizing bond-dependent Kitaev interactions~\cite{Kitaev2,Trebst1}. Numerous honeycomb systems have been proposed but a QSL with pure Kitaev Hamiltonian is yet to be realized experimentally in real materials~\cite{Hermanns17}. This encourages the discovery of new honeycomb lattices with strong SOC. Our recent work on BiYbGeO$_5$ (a structural analogue of the compound under investigation) revealed a disordered ground state down to 100~mK with anisotropic exchange interactions~\cite{Mohanty134408}. Here, Yb$^{3+}$ ions become dimerized at low temperatures yielding a tiny spin gap. Our present work examines how substituting Yb$^{3+}$ with Er$^{3+}$ influences the magnetic ground state in this structural family and it also provides an opportunity to investigate the role of the rare-earth ion in shaping the correlated behavior of honeycomb quantum magnets.

\begin{figure}
	\includegraphics[width=\linewidth]{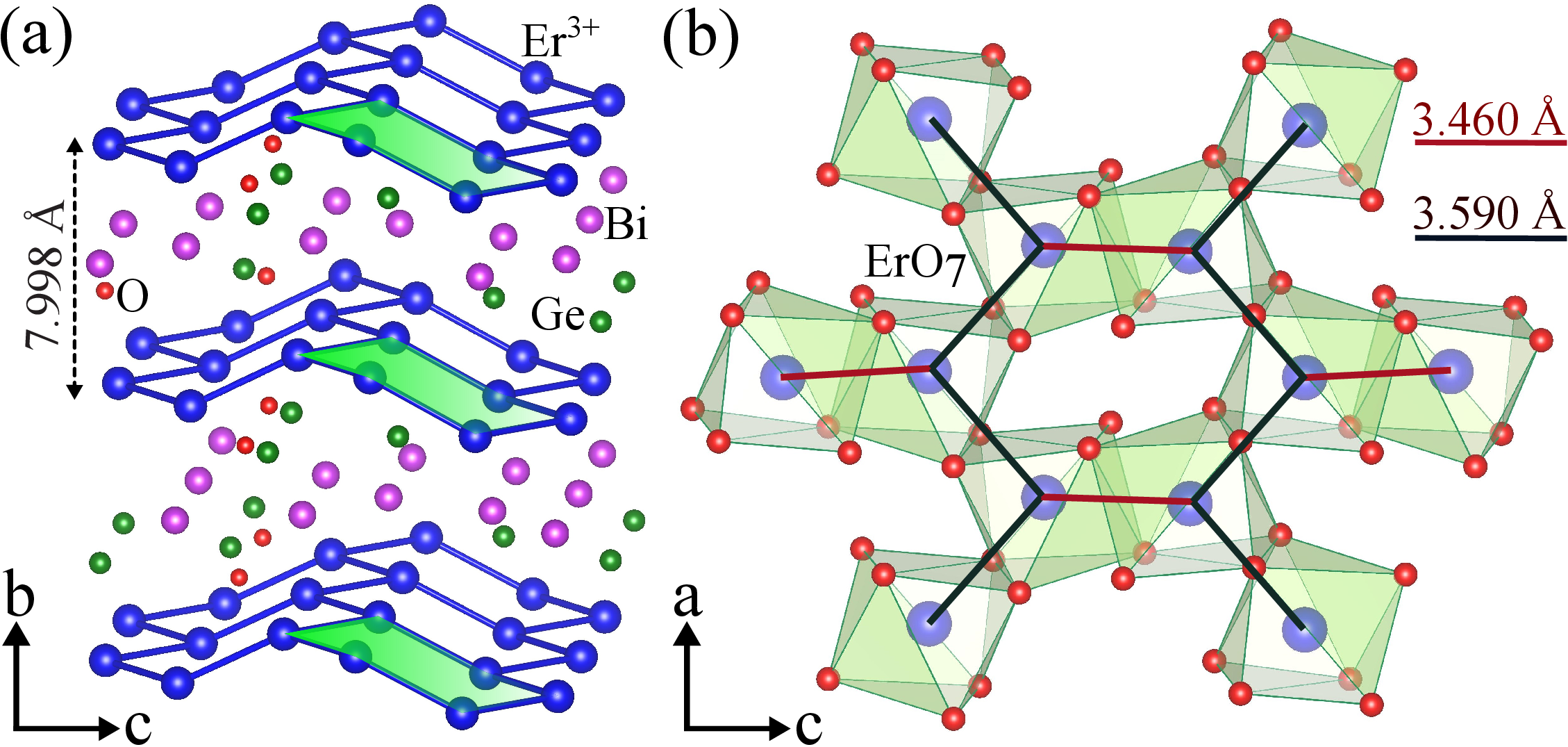}
	\caption{\label{Fig1} (a) Crystal structure of BiErGeO$_5$ viewed along the $a$-axis, showing the separation of adjacent honeycomb layers. (b) A section of the honeycomb layer in the $ac$-plane formed by edge-shared ErO$_7$ polyhedra and highlights the anisotropic bond lengths.}
\end{figure}
Herein, we investigate the low temperature magnetic properties and CEF excitations of a spin-orbit coupled quantum magnet BiErGeO$_5$. Figure~\ref{Fig1} indicates the crystal structure of BiErGeO$_5$ that features a quasi two-dimensional (2D) distorted honeycomb lattice of Er$^{3+}$ ions in the $ac$-plane. The edge-shared distorted ErO$_7$ polyhedra constituting the honeycomb layers are strongly buckled. These layers are well isolated from each other along the $b$-direction by the interstitial non-magnetic Bi$^{3+}$ and Ge$^{4+}$ ions. In each honeycomb unit, the Er$^{3+}$~-~Er$^{3+}$ distances are found to be anisotropic with bond lengths of $\sim3.460$~\AA~and $\sim3.590$~\AA. The thermodynamic data shows a magnetic LRO around 0.4~K followed by a short-range broad maxima. We have modeled the INS spectra using the CEF Hamiltonian and extracted the information about the CEF energy levels. Finally, all the thermodynamic data are reproduced using the CEF parameters. Interestingly, the $\mu$SR relaxation rates suggest the presence of slow spin fluctuations well below the magnetic LRO.

\section{Experimental Details}
\begin{figure}
	\includegraphics[width=\columnwidth]{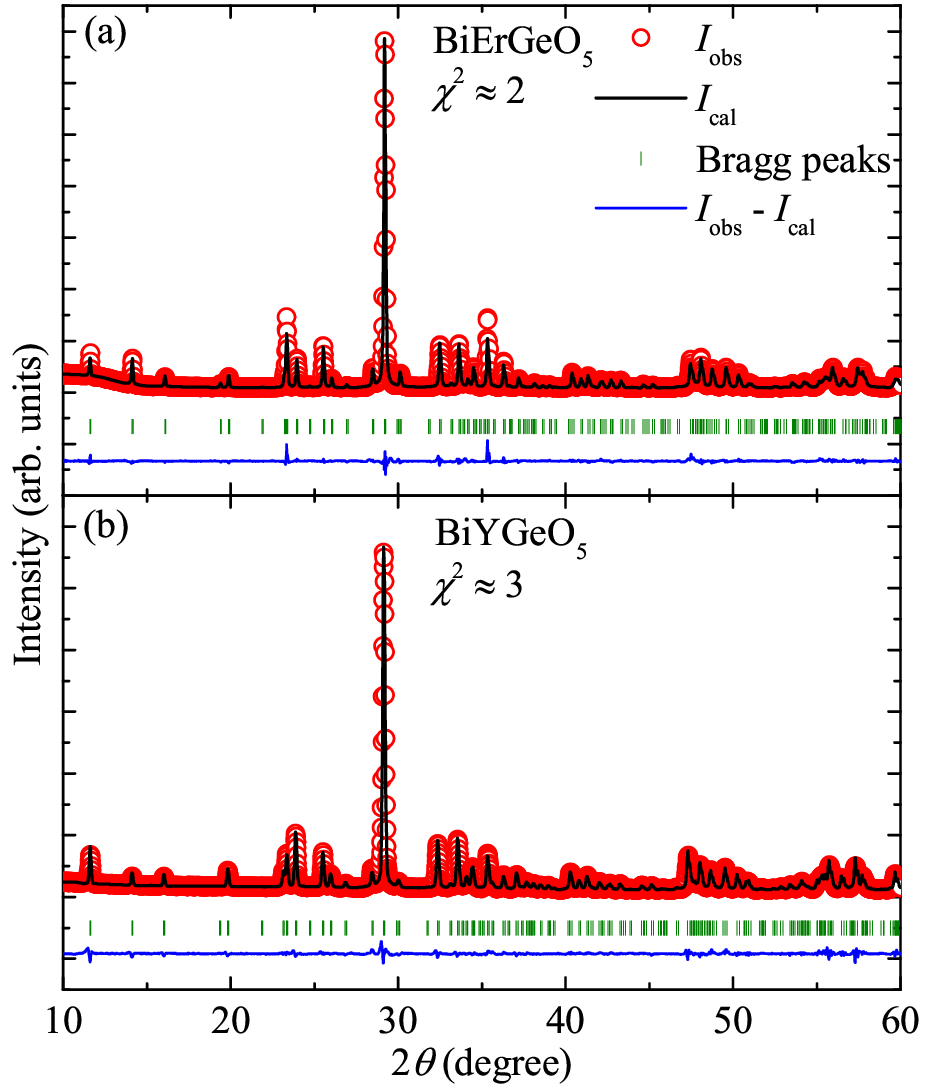}
	\caption{\label{Fig2} Powder XRD pattern (open circles) at $T = 300$~K for (a) BiErGeO$_5$ and (b) BiYGeO$_5$. The black solid line represents the Rietveld fit of the data. Expected Bragg positions are indicated by green vertical bars and the solid blue line at the bottom denotes the difference between experimental and calculated intensities. The goodness-of-fit is achieved to be $\chi^{2} \simeq 2$ and 3, respectively.}
\end{figure}

Polycrystalline samples of BiErGeO$_5$ and its iso-structural nonmagnetic analog BiYGeO$_5$ were synthesized via the conventional solid state synthesis technique. Bi$_{2}$O$_{3}$ (99.99~\%, Sigma Aldrich), GeO$_{2}$(99.99~\%, Sigma Aldrich), Er$_2$O$_{3}$ (99.99~\%, Sigma Aldrich), and Y$_2$O$_{3}$ (99.99~\%, Sigma Aldrich) were used as starting precursors. The stoichiometric mixtures of the starting materials were thoroughly ground, pressed into pellets, heated in a platinum crucible at 950$\degree$C for two days in air, and then quenched. Excess 4~\% Bi$_{2}$O$_{3}$ was used as the starating material to get the pure phase of Bi$Re$GeO$_{5}$ ($Re$ = Er and Y). In order to confirm the phase purity and crystal structure of the samples, powder x-ray diffraction (XRD) measurement was performed at room temperature using the PANalytical x-ray diffractometer (Cu~\textit{K$_{\alpha}$} radiation, $\lambda_{\rm av} \simeq 1.5418$~\AA). Rietveld refinement of the XRD data was carried out using \verb"FullProf" software package~\cite{Carvajal55}, that confirms the formation of pure phase as shown in Fig.~\ref{Fig2}. All the peaks in the XRD data could be appropriately indexed by the orthorhombic space group $Pbca$ (No.~61). The obtained lattice parameters at room temperature are [$a = 5.3249(1)$~\AA, $b = 15.2256(2)$~\AA, $c = 11.0167(2)$~\AA, and $V_{\rm cell} = 893.17(3)$] and [$a = 5.3380(3)$~\AA, $b = 15.2266(3)$~\AA, $c = 11.0594(3)$~\AA, and $V_{\rm cell} = 898.90(3)$] for BiErGeO$_{5}$ and BiYGeO$_{5}$, respectively. These lattice constants are comparable with the previous report~\cite{Cascales262}. The atomic coordinates of different atoms for both the compounds after the refinement are tabulated in Table~\ref{Table1}.

\begin{table}[h]
	\caption{Listed are the Wyckoff positions and the refined atomic coordinates ($x$, $y$, and $z$) for each atom at room temperature for BiErGeO$_5$ and BiYGeO$_5$.}
	\label{Table1}
	\begin{ruledtabular}
		\begin{tabular}{cccccc}
			Atom & Site & $x$ & $y$ & $z$ & Occupancy \\\hline
			Bi1 & 8c & $0.9549(3)$ & $0.2393(1)$ & $0.1473(1)$ & 1\\
			&    & $0.9552(3)$ & $0.2383(3)$ & $0.1470(3)$ & 1\\
			Er1 & 8c & $0.0077(5)$ & $0.0525(1)$ & $0.3610(2)$ & 1\\
			Y1   &    & $0.0043(2$) & $0.0528(3)$ & $0.3611(3)$ & 1\\
			Ge1 & 8c & $0.0109(1)$ & $0.4024(2)$ & $0.4009(3)$ & 1\\
			&    & $0.0098(2)$ & $0.4055(3)$ & $0.4055(3)$ & 1\\
			O1 & 8c & $0.0305(4)$ & $0.2825(1)$ & $0.3584(1)$ & 1\\
			&    & $0.0660(2)$ & $0.3193(2)$ & $0.3398(2)$ & 1\\
			O2 & 8c & $0.3569(4)$ & $0.4233(1)$ & $0.4025(2)$ & 1\\
			&    & $0.3203(2)$ & $0.4402(2)$ & $0.4496(2)$ & 1\\
			O3 & 8c & $0.3261(3)$ & $0.0862(2)$ & $0.4911(1)$ & 1\\
			&    & $0.2991(2)$ & $0.1085(2)$ & $0.4920(2)$ & 1\\
			O4 & 8c & $0.2595(6)$ & $0.1593(1)$ & $0.2483(4)$ & 1\\
			&    & $0.2702(2)$ & $0.1447(2)$ & $0.2442(2)$ & 1\\
			O5 & 8c & $0.3756(3)$ & $0.4847(1)$ & $0.1850(1)$ & 1\\
			&    & $0.3599(2)$ & $0.4797(2)$ & $0.1969(2)$ & 1\\
		\end{tabular}
	\end{ruledtabular}
\end{table}
Magnetization ($M$) was measured using the SQUID magnetometer (MPMS-3, Quantum Design) down to 0.4~K using an additional $^3$He (iHelium3, Quantum Design, Japan) attachment. Heat capacity ($C_{\rm p}$) for $T > 1.8$~K was measured on a small piece of pellet using the relaxation technique in the physical property measurement system (PPMS, Quantum Design). A dilution refrigerator insert to the PPMS was used to measure below 1.8~K and down to 100~mK.

The direct geometry time-of-flight (TOF) spectrometer MARI~\cite{Le168646} at the ISIS Facility, Rutherford Appleton Laboratory, United Kingdom was used to measure the zero-field inelastic neutron scattering (INS) experiment. Powder samples with total mass of 4.5~g of BiErGeO$_5$ and BiYGeO$_5$ were packed in an annular geometry inside Al cans, which were cooled using a top-loading closed cycle refrigerator. Data were recorded at 5, 50, 100, and 150~K using incident neutron energies $E_i=10$, 25, 80, and 120~meV and Gd chopper frequency of 100, 400, 300, and 400~Hz, respectively. The four configurations gave elastic energy resolutions of 0.3, 0.9, 3, and 4~meV, respectively.
The raw data were processed using the MANTID software~\cite{Arnold156}.

The low-temperature $C_{\rm P}(T)$ data are modeled by the full diagonalization (FD) calculation using the \verb|fulldiag| algorithm~\cite{Todo047203} of the \verb|ALPS| package~\cite{Albuquerque1187} for a spin-1/2 XXZ honeycomb system with the lattice size $N = 18$ using Periodic boundary condition.

Muon spin relaxation/rotation ($\mu$SR) experiments were performed on the Flexible Advanced MuSR Environment (FLAME) spectrometer at the Swiss Muon Source (S$\mu$S), Paul Scherrer Institute (PSI), Villigen, Switzerland, under both zero-field (ZF) and longitudinal-field (LF) conditions. The FLAME instrument provides near-ideal ZF conditions, with residual magnetic fields below $\sim 5~\mu$T, achieved through active compensation using three orthogonal vector magnets that allow precise alignment and cancellation of stray fields. The detector system consists of compact scintillator–SiPM modules with a time resolution better than approximately 150\,ps. The initial muon spin polarization was oriented along the incident beam direction. A pellet of polycrystalline BiErGeO$_5$ (10\,mm in diameter) was attached to a copper sample holder using GE varnish and sandwiched between two 25\,$\mu$m thick copper foils to improve thermal contact. The sample was thermally anchored to the mixing chamber of a KelvinoxJT dilution refrigerator installed in an Oxford Variox cryostat, enabling measurements over a temperature range from 30\,mK to 160\,K and in longitudinal magnetic fields up to 1.5\,T. Special attention was paid to achieving thermal equilibrium at each temperature point, particularly below 100\,mK, by allowing sufficient stabilization time. The ZF and LF asymmetry spectra, recorded using forward and backward detector pairs, were analyzed with the MUSRFIT software package~\cite{Suter69}.

\section{Results}
\subsection{Magnetization}
\begin{figure}
	\includegraphics[width=\columnwidth]{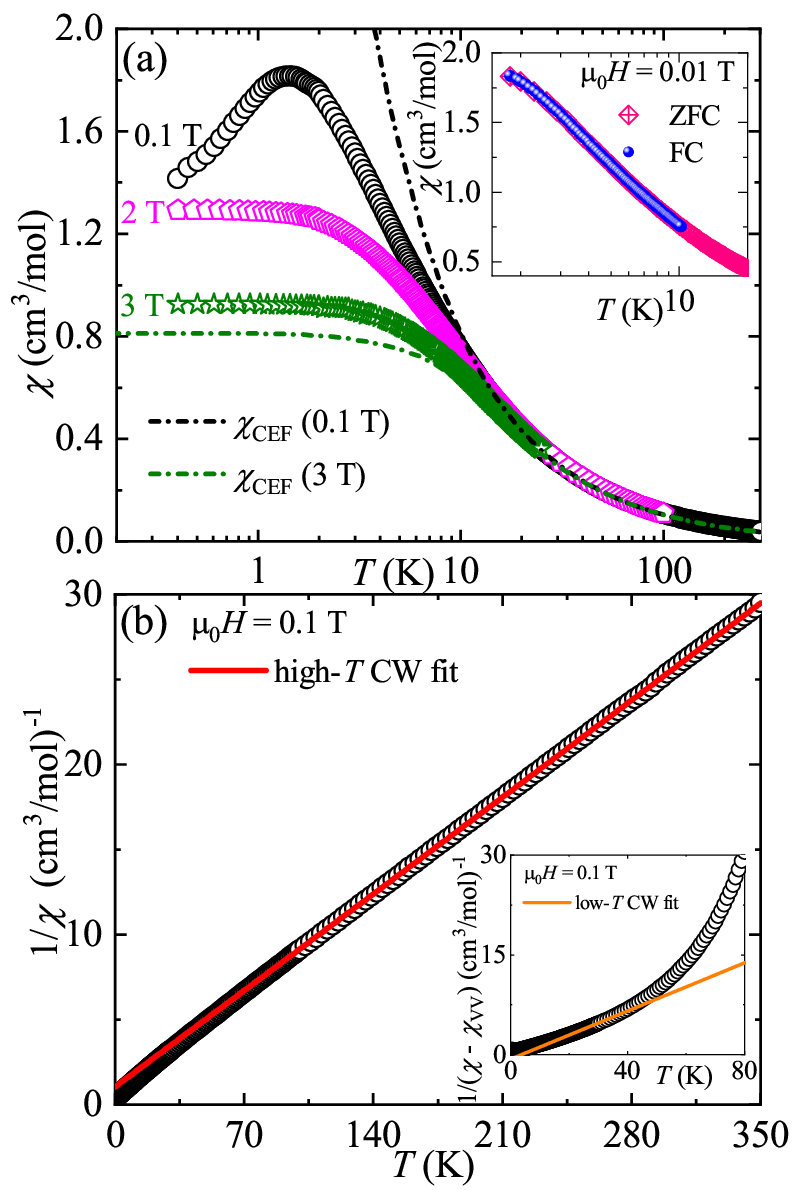}
	\caption{\label{Fig3} (a) $\chi(T)$ of BiErGeO$_5$ measured in different magnetic fields. The dot-dashed lines are the calculated $\chi_{\rm CEF}$ in different magnetic fields. Inset: $\chi(T)$ measured at $\mu_0H=0.01$~T in both ZFC and FC protocols. (b) High-$T$ CW fit to $1/\chi$ data (at $\mu_0H=0.1$~T) extrapolated down to low temperatures. Inset: The low-$T$ CW fit after subtracting the Van-Vleck contribution.}
\end{figure}
The temperature-dependent magnetic susceptibilities $\chi~[\equiv M/H]$ of BiErGeO$_5$ in different magnetic fields are depicted in Fig.~\ref{Fig3}(a). In the lowest field of $\mu_{0}H=0.1$~T, $\chi(T)$ follow a Curie-Weiss (CW) behavior at high temperature and portrays a broad maximum centered around $T^{\rm SR} \simeq 1.4$~K followed by a rapid decrease towards low temperatures. The broad maximum reflects the short-range correlations anticipated for a low-dimensional AFM spin system. With increasing field this maxima is gradually suppressed and is completely disappeared above 2~T. No clear signature of magnetic LRO is evident down to 0.4~K. The inset of Fig.~\ref{Fig3}(a) shows $\chi(T)$ data measured at 100~Oe in zero-field-cooled (ZFC) and field-cooled (FC) conditions. The absence of bifurcation down to 1.8~K excludes the possibility of a spin-glass behavior or spin freezing at low temperatures.

Above 150~K, the inverse susceptibility ($1/\chi$) for $\mu_0H = 0.1$~T was fitted well by the modified CW law
\begin{equation}
	\chi(T)=\chi_0+\frac{C}{T-\theta_{\rm CW}},
	\label{cw}
\end{equation}
where $\chi_0$  is the temperature independent susceptibility and the second term is the CW law. The fit yields $\chi_0 \simeq 1.1 \times 10^{-3}$~cm$^3$/mol, a high-$T$ effective moment $\mu_{\rm eff}^{\rm HT}$ $[=\sqrt{(3k_{\rm B}C/N_{\rm A}\mu_{\rm B}^{2})}$, where $C$, $k_{\rm B}$, $N_{\rm A}$, and $\mu_{\rm B}$ are the Curie constant, Boltzmann constant, Avogadro’s number, and Bohr magneton, respectively]~$\simeq 9.68~\mu_{\rm B}$, and the high-$T$ CW temperature $\theta_{\rm CW}^{\rm HT} \simeq -14.4$~K. This value of $\mu_{\rm eff}^{\rm HT}$ is in good agreement with the expected value, $\mu_{\rm eff} = g\sqrt{J(J+1)}~\simeq 9.58~\mu_{\rm B}$ for Er$^{3+}$ ($J = 15/2$, $g=6/5$) in the $4f^{11}$ configuration~\cite{Xing144413}. Here, the large negative value of $\theta_{\rm CW}^{\rm HT}$ does not signify strong AFM interactions. It rather arises due to CEF excitations at high temperatures where all Kramers doublets become thermally populated and contribute to $\theta_{\rm CW}$. $1/\chi$ is found to deviate from linearity at low temperatures with a clear slope change. After subtraction of the Van-Vleck susceptibility ($\chi_{\rm VV}$) obtained from the $M$ vs $H$ analysis, $1/\chi(T)$ shows a linear regime at low temperatures [inset of Fig.~\ref{Fig3}(b)]. A CW fit in the $T$-range 8-15~K yields $\mu_{\rm eff}^{\rm LT} \simeq 8.53~\mu_{\rm B}$ and $\theta_{\rm CW}^{\rm LT} \simeq -2$~K. The negative value of $\theta_{\rm CW}^{\rm LT}$ reflects a weak AFM exchange coupling between Er$^{3+}$ ions. 

\begin{figure}
	\includegraphics[width=\columnwidth]{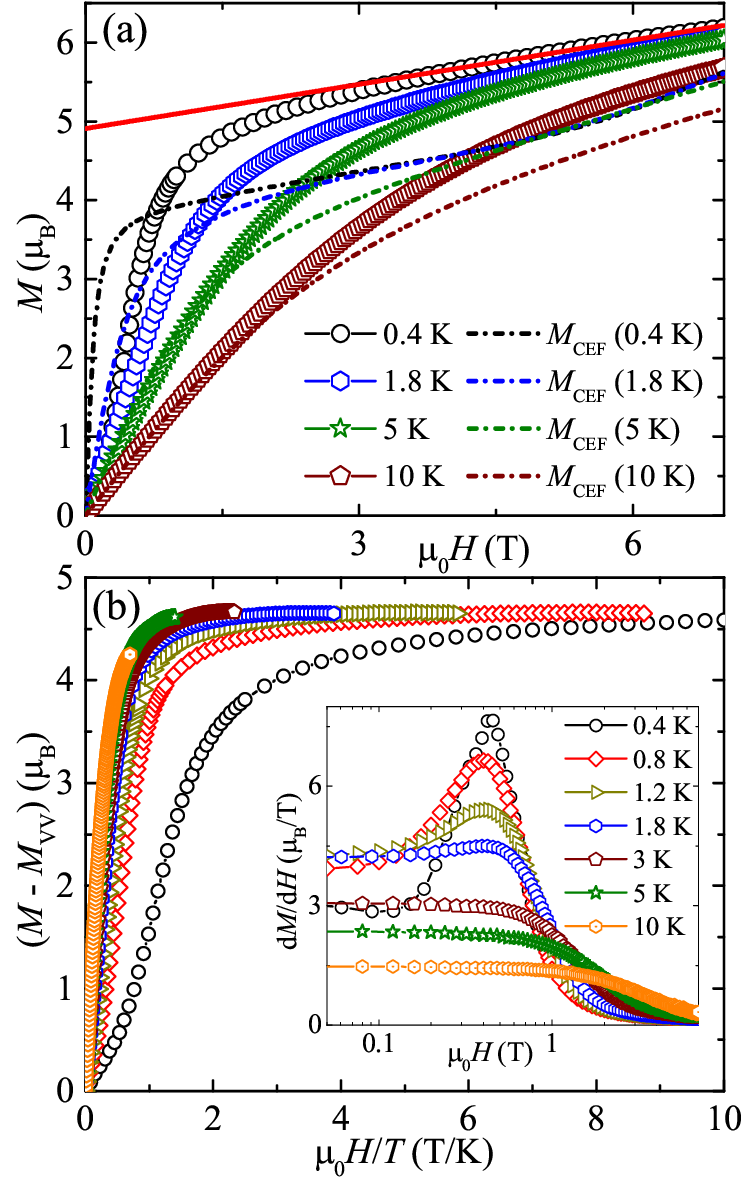}   
	\caption{\label{Fig4} (a) $M$ vs $H$ measured at different temperatures The solid line represents a linear fit to the high-field data for $T=0.4$~K. The dot-dashed lines are the calculated $M_{\rm CEF}$ in different temperatures. (b) ($M-M_{\rm VV}$) vs $\mu_0H/T$ at different temperatures to observe the scaling of magnetization curves in the correlated regime. Inset: d$M$/d$H$ vs $\mu_0H$ plot to highlight the field induced feature.}
\end{figure}
Figure~\ref{Fig4}(a) presents the magnetic isotherm ($M$ vs $\mu_0H$) measured up to 7~T at four different temperatures. The magnetization at $T = 0.4$~K rapidly increases with field along with a distinct curvature in the low field regime. For $\mu_0 H > 1.5$~T, it appears to saturate but a linear increase of magnetisation is still observed in high fields, which can be attributed to Van-Vleck contribution. A linear fit to the high-field ($\geq6$~T) data returns a slope of $\sim 0.087~\mu_{\rm B}/{\rm T}$, which corresponds to $\chi_{\rm VV} = 4.8(1) \times 10^{-3}$~cm$^{3}$/mol. From the $y$-intercept of the linear fit, the saturation magnetization is estimated to be $M_S = 4.93(1)~\mu_{\rm B}$. 



In order to visualize the correlated behavior at low temperatures, the Van-Vleck corrected $M$ (i.e. $M-M_{\rm VV}$) vs $\mu_0 H$ with the $x$-axis scaled with respect to the temperature is shown in Fig.~\ref{Fig4}(b). For $T\geq 1.8$~K, all the $M-M_{\rm VV}$ vs $\mu_0 H/T$ curves collapse onto a single curve, reflecting the paramagnetic nature of the spins. However, for $T< 1.8$~K the curves show clear deviation from this pattern, demonstrating the development of magnetic correlations on a temperature scale comparable to the low $\theta_{\rm CW}^{\rm LT}$ value. Further, the low field curvature or a change of slope in the low temperature $M$ vs $\mu_oH$ curves is well evident from the d$M$/d$H$ vs $\mu_0H$ plot as shown in the inset of Fig.~\ref{Fig4}(b). For $T = 0.4$~K, this feature is manifested as a well defined peak at 0.42~T. As
the temperature rises, the peak intensity gradually decreases and disappears completely for $T> 1.8$~K. This feature may be attributed to a meta-stable field-induced transition, typically expected for anisotropic magnets.

\subsection{Heat capacity}
\begin{figure*}
	\includegraphics[scale=0.7]{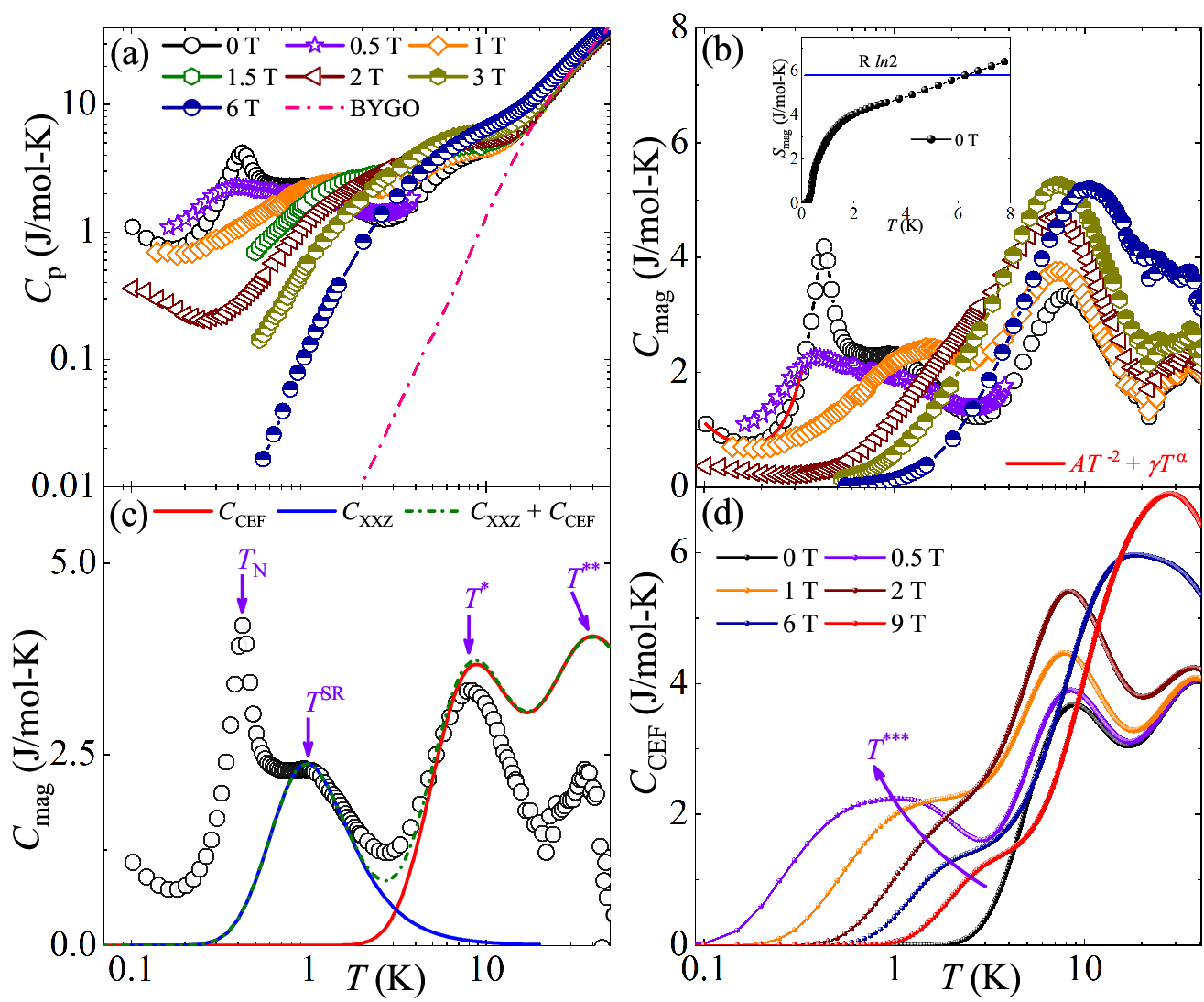}
	\caption{(a) $C_{\rm p}$ vs $T$ for BiErGeO$_5$ measured in different applied fields. The dot-dashed line represents $C_{\rm ph}(T)$ of the nonmagnetic analog BiYGeO$_5$. (b) Magnetic heat capacity $C_{\rm mag}$ vs $T$ in different magnetic fields. The solid line represents a combination of power law and nuclear Schottky fit to the low-$T$ data. Inset: Magnetic entropy $S_{\rm mag}$ vs $T$ in zero field. (c) Zero field $C_{\rm mag}$ vs $T$ highlighting different anomalies. The blue and red solid lines indicate the ED simulation of XXZ model and calculated $C_{\rm CEF}$ in zero field, respectively. The dot-dashed line is the sum of both. (d) Calculated $C_{\rm CEF}$ vs $T$ in different magnetic fields, as discussed in the text.}
	\label{Fig5}
\end{figure*}
To elucidate the effect of low-energy CEF excitations and spin-spin correlations, the temperature-dependent heat capacity [$C_{\rm p}(T)$] of BiErGeO$_5$ measured down to 100~mK under various magnetic fields is presented in Fig.~\ref{Fig5}(a). Zero-field $C_{\rm p}$ exhibits a sharp $\lambda$-like anomaly at $T_{\rm N} \simeq 0.4$~K, indicating the onset of a magnetic LRO. The peak broadens with increasing field and is completely suppressed above $\mu_0H = 0.5$~T. In magnetic insulators, the total $C_{\rm p}(T)$ is the sum of the lattice contribution $C_{\rm ph}(T)$, which dominates at high temperatures and the magnetic part $C_{\rm mag}(T)$, dominating in the low-temperature regime. We estimated $C_{\rm ph}(T)$ by measuring the zero-field heat capacity of the non-magnetic isostructural compound BiYGeO$_5$ [dot-dashed line in Fig.~\ref{Fig5}(a)] down to 2~K. The low-temperature heat capacity data of BiYGeO$_5$ are fitted by $T^{3}$ behaviour which is further extrapolated down to 100~mK. The resulting $C_{\rm ph}(T)$ of BiYGeO$_5$ was scaled with respect to BiErGeO$_5$ by taking the ratio of their atomic masses~\cite{Bouvier13137} and then subtracted from the total heat capacity of BiErGeO$_5$ to obtain $C_{\rm mag}(T)$. $C_{\rm mag}(T)$ at different fields is plotted in Fig.~\ref{Fig5}(b).

Zero-field $C_{\rm mag}(T)$ data shows three broad maxima at around $T^{\rm SR} \simeq 1.1$~K, $T^{\star} \simeq 9$~K, and $T^{\star\star} \simeq 36$~K. The $T^{\rm SR}$ indicates the onset of short-range magnetic correlations between Er$^{3+}$ ions due to the two-dimensionality of the spin lattice which is suppressed with field similar to that observed in the $\chi(T)$ data. 
With increasing field, $T^{\star}$ moves slowly to high temperatures. Surprisingly, another broad maximum ($T^{\star\star\star}$) appears with field below $T^{\star}$ and shifts comparatively faster than $T^{\star}$ towards high temperatures [see Fig.~\ref{Fig5}(d)]. The $T^{\star\star}$ position remains almost unchanged with increase in field. In high magnetic fields ($\mu_0H>$ 2~T), $T^{\star}$ and $T^{\star\star\star}$ are merged into one single broad peak, which shifts towards high-$T$s with further increase in field.

Furthermore, the zero-field $C_{\rm mag}(T)$ exhibits an upturn below $\sim 0.2$~K that can be ascribed to the nuclear Schottky anomaly. Therefore, well below $T_{\rm N}$, $C_{\rm mag}$ was fitted using $C_{\rm mag}(T) = A/T^2 + \gamma T^{\alpha}$ [see the red solid line in Fig.~\ref{Fig5}(b)]. The term $A/T^2$ represents the high-temperature part of the nuclear Schottky anomaly~\cite{Mohanty134408}, while $\gamma T^{\alpha}$ describes the intrinsic power-law behavior. The constant $A$ is associated with the energy gap ($\Delta_{\rm N}$) between the Zeeman-split nuclear levels via the relation $A \propto \Delta_{\rm N}^2$ for $T >> \Delta_{\rm N}$. The obtained fit parameters are $A \simeq 1.1 \times 10^{-3}$~J K/mol, $\gamma \simeq 43.770$~J/mol K$^{3.8}$, and $\alpha \simeq 2.8$. The value of $\alpha$ is close to 3, as expected for a 3D antiferromagnetically ordered state. The obtained $C_{\rm mag}(T)$ is then used to estimate the magnetic entropy [$S_{\rm mag}(T)$] by integrating $C_{\rm mag}(T)/T$ in the measured $T$-range [inset of Fig.~\ref{Fig5}(b)]. The released entropy is close to 1~J/mol-K around $T_{\rm N}$ and increases to the value of $Rln2$ for a Kramer doublet at 6~K.

\subsection{Inelastic neutron scattering}
\begin{figure*}
	\includegraphics[scale=0.9]{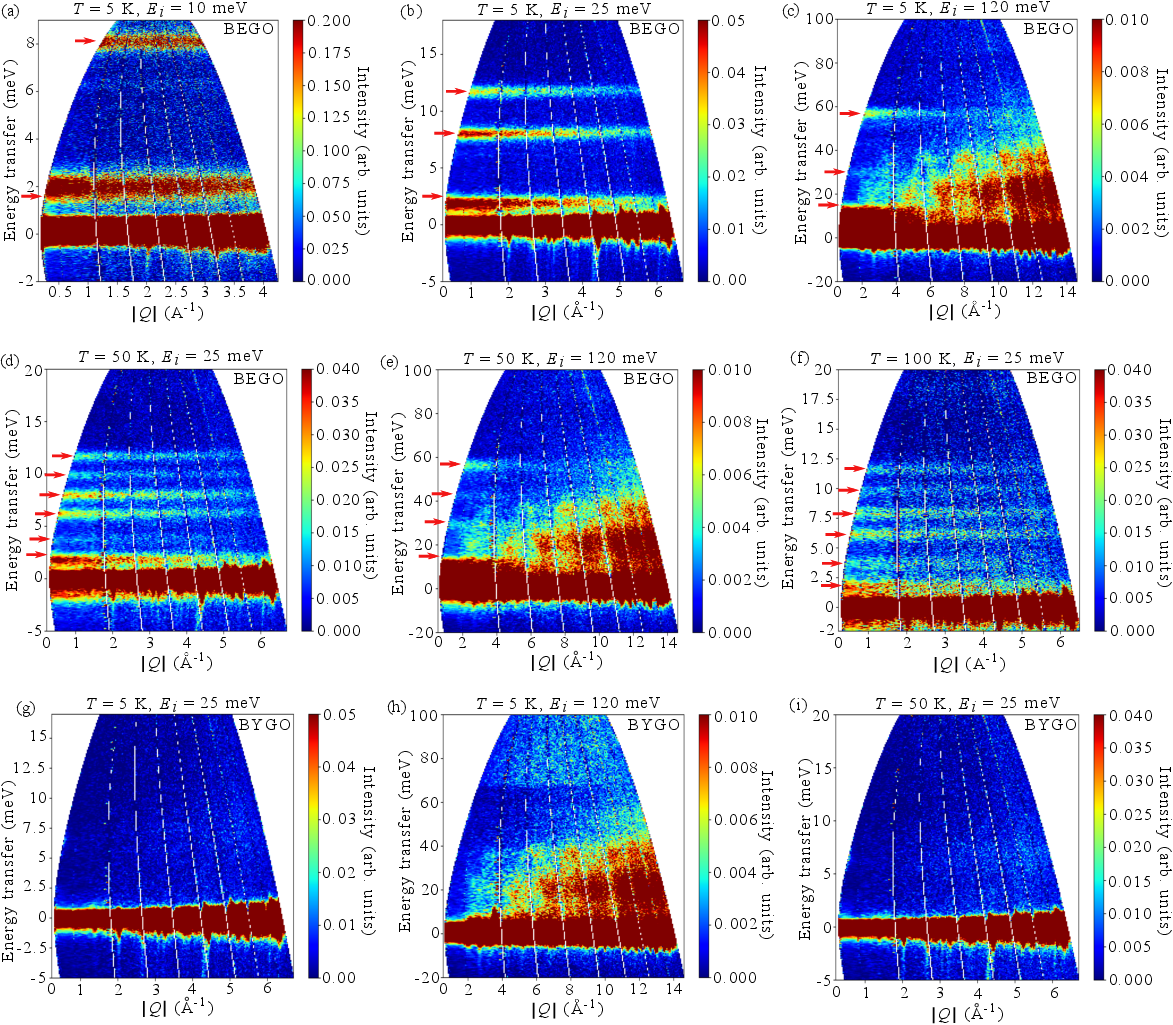}
	\caption{(a-f) Raw INS spectra of BiErGeO$_5$ (BEGO) at different temperatures ($T$ = 5, 50, and 100~K) and incident neutron energies ($E_i$ = 10, 25, and 120~meV). The red arrows indicate the CEF excitations. (g-i) Raw INS spectra of BiYGeO$_5$ (BYGO).}
	\label{Fig6}
\end{figure*}
\begin{figure}
\includegraphics[width=\columnwidth]{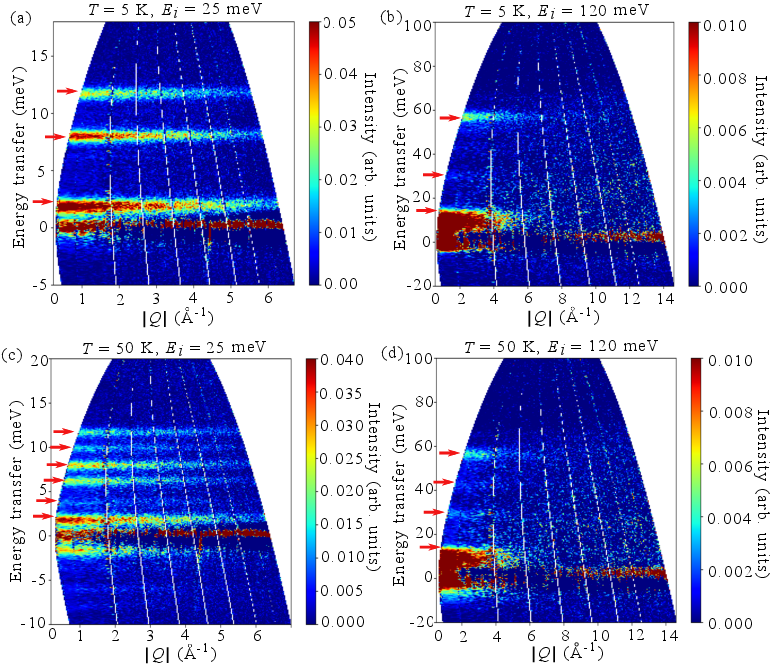} \caption{\label{Fig7} INS spectra of BiErGeO$_5$ after subtracting the phonon scattering contribution at (a) $T = 5$~K, $E_i = 25$~meV, (b) $T = 5$~K, $E_i = 120$~meV, (c) $T = 50$~K, $E_i = 25$~meV, and (d) $T = 50$~K, $E_i = 120$~meV. The red arrows indicate the high energy CEF excitations.}
\end{figure}
Figure~\ref{Fig6} presents the color plots of the INS spectra of BiErGeO$_5$ and BiYGeO$_5$ measured at three different temperatures ($T$ = 5, 50, and 100~K) and three incident neutron energies ($E_{\rm i}$ = 10, 25, and 120~meV). For BiErGeO$_5$, two dispersionless excitations are observed at around 1.8 and 8~meV for $T = 5$~K and $E_{\rm i} = 10$~meV [see Fig.~\ref{Fig6}(a)]. These excitations appear more prominent in $E_{\rm i} = 25$~meV spectrum with an additional excitation near 11.8~meV [see Fig.~\ref{Fig6}(b)]. Furthermore, $E_{\rm i} = 120$~meV spectrum shows a weak excitation at around 29~meV, along with a broad excitation band at $\sim$ 59~meV [see Fig.~\ref{Fig6}(c)]. All these excitations correspond to the transitions between ground state doublet and higher energy CEF levels. Similarly, INS spectra at 50 and 100~K [see Figs.~\ref{Fig6}(d-f)] show several more excitations in the low-$Q$ regime, corresponding to the transitions between intermediate CEF levels. The intensity of these excitations decrease with increasing in $Q$. For the non-magnetic analogue compound BiYGeO$_5$, low-$Q$ excitations are absent and the dispersive excitations in the high-$Q$ regime appear to be phonons [see Figs.~\ref{Fig6}(g-i)].

\begin{figure}
	\includegraphics[width=\columnwidth]{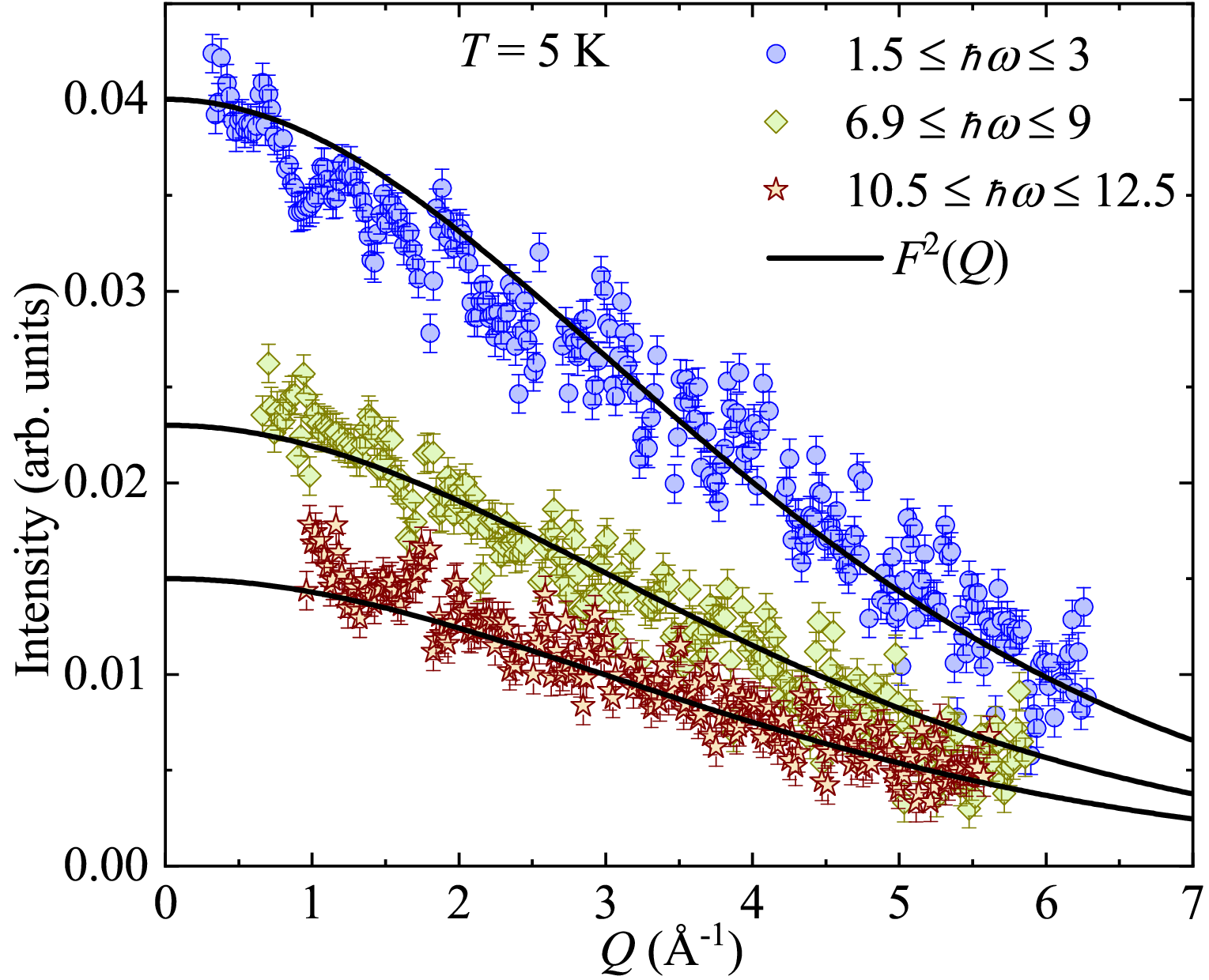}
	\caption{\label{Fig8} $Q$ dependence of INS intensity at $T=$ 5~K obtained by integrating the energy transfer. Solid lines are the square of magnetic form-factor of Er$^{3+}$ ions.}
\end{figure}

In order to separate these high-energy CEF excitations from the phonon scattering contribution, we subtracted the INS spectra of non-magnetic BiYGeO$_5$, in which the excitations are purely phononic, from the spectra of BiErGeO$_5$. Figures~\ref{Fig7}(a-d) depict the color plots of the phonon subtracted INS spectra at $T = 5$ and 50~K for $E_i = 25$ and 120~meV. The intensity of all the CEF excitations decreases with increasing $Q$ ($= |\vec{Q}|$), as the intensity is proportional to square of the magnetic form-factor [$F^{2}(Q)$] in the neutron scattering cross section. We calculated $F^{2}(Q)$ for Er$^{3+}$ ion using the dipole approximation (see Appendix A for details about the magnetic form-factor) and compared with the experimental INS intensity of a few CEF excitations at $T = 5$~K, as shown in Fig.~\ref{Fig8}. The calculated $F^{2}(Q)$ decreases monotonically with increasing $Q$ and agrees very well with the experimental data. The intensity of low-$T$ ($T = 5$~K) [see Fig.~\ref{Fig7}(a-b)] excitations diminish with increase in temperature ($T = 50$~K) [see Fig.~\ref{Fig7}(c-d)] because of thermal broadening (Debye-Waller factor) and depopulation of the ground state Kramers doublet. The intensity of additional excitations due to the transition between intermediate CEF levels also follow the calculated behaviour of $F^{2}(Q)$.

\begin{figure*}
	\includegraphics[scale=0.55]{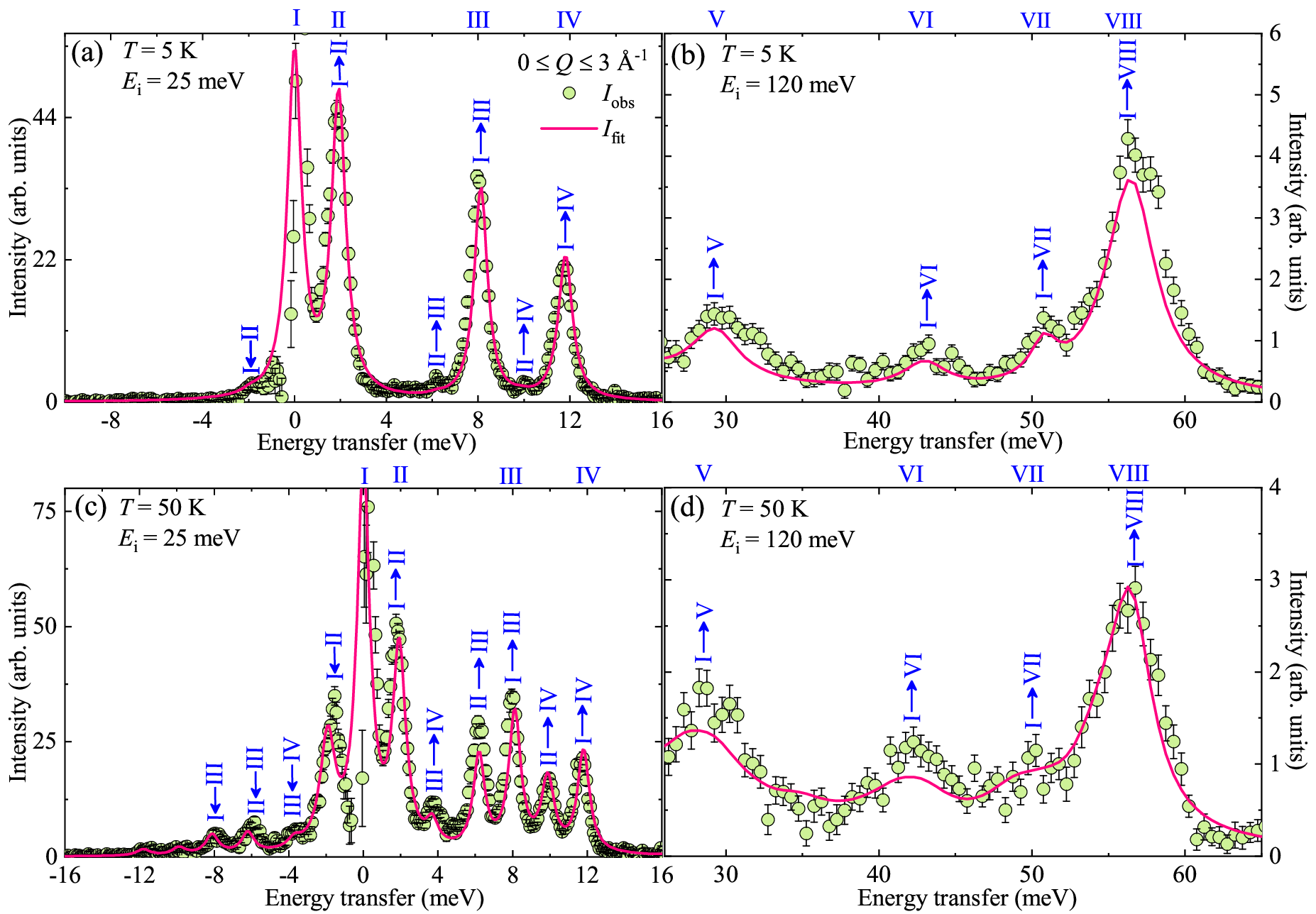}
	\caption{The INS spectral intensity as a function of energy transfer (for $T=$ 5, 50~K and $E_{\rm i}=$ 25, 120~meV) obtained by integrating intensity in the low wave vector regime. The solid line is the corresponding fit using the CEF Hamiltonian. The energies of the eight doublets are shown (I to VIII) on the top. Each peak is labeled with the corresponding transition.}
	\label{Fig9}
\end{figure*}
For a better visualization of all the CEF modes and to fit the INS data, we created a 2D intensity map as a function of energy transfer ($\hbar \omega$) by integrating the phonon-subtracted INS data over the selected wave-vector range (0~$\leq Q \leq$~3~\AA$^{-1}$) as shown in Fig.~\ref{Fig9} for $T = 5$ and 50~K with $E_i = 25$ and 120~meV. The observed strong signal at $\hbar \omega = 0$~meV (I) in Fig.~\ref{Fig9}(a) and (c) correspond to the quasielastic neutron scattering. We observed seven CEF excitations for $T = 5$~K at around 1.8 (II), 8.1 (III), 11.8 (IV), 29.3 (V), 43 (VI), 50.8 (VII), and 56.4~meV (VIII) as expected for Er$^{3+}$-based compounds. All these excitations from the ground state doublet to excited CEF levels (I $\to$ II, III, IV, V, VI, VII, and VIII) occur below 60~meV, which are fully captured by the INS experiments. In addition, two weak excitations at around 6 and 10~meV [see Fig.~\ref{Fig9}(a)] are also detected. The 6~meV peak corresponds to the transition between the first (II) and second (III) excited CEF levels (i.e., II $\to$ III) whereas the 10~meV peak is due to the transition between II and third (IV) excited levels (i.e., II $\to$ IV).

There is a weak peak at around 1.8~meV of the $T = 5$~K spectrum in the negative $\hbar \omega$ region, reflecting transition from the first excited doublet to the ground state doublet (II $\rightarrow$ I). At low temperatures, the excitations arise primarily from the highly populated ground-state doublet with minor contributions from the less populated first excited doublet. On the other hand, at high temperatures the thermal population of the low-lying excited CEF levels (mainly II, III, and IV) increases. As a result, transitions between higher excited states as well as from excited states to the ground state are allowed which give rise to additional peaks in both the positive and negative $\hbar \omega$ regions [see Fig.~\ref{Fig9}(c)].
Two distinct CEF excitation peaks are clearly visible in the negative $\hbar \omega$ regime (e. g. -1.8 and -8~meV) at $T = 50$~K that correspond to the transition from III and II to I, as shown in Fig.~\ref{Fig9}(c). Similarly, we also observed high intensed peaks at 6 and 10~meV as well as weak intensed peaks in the -ve energy transfer region (-6 and -10~meV). Moreover, another new peak appears for $T = 50$~K at 4 and -4~meV due to the transitions from III $\to$ IV and IV $\to$ III, respectively. We used the $T = 5$~K and 50~K INS spectra corresponding to $E_{\rm i} = 25$~ and 120~meV for the CEF analysis as discussed in the next section.

\subsection{CEF analysis}
The INS intensity versus energy transfer data can be analyzed using an appropriate CEF Hamiltonian. According to the Stevens convention, the CEF Hamiltonian can be expressed as~\cite{Stevens209}
\begin{equation}\label{CEF}
	\mathcal{H}_{\rm CEF} = \sum_{l,m}B_l^m\hat{O}_l^m.
\end{equation}
Here, $\hat{O}^m_l$ are the standard Stevens operators~\cite{Huthings227,Stevens209}, which are related to the angular momentum operators~\cite{Newman2000}. $B^m_l$ are the multiplicative factors called CEF parameters, which are related to the electronic structure of the rare-earth materials~\cite{Guchhait144434}. Here, the even integer $l$ varies from 0 to 6 for $f$ electrons and the integer $m$ ranges from $-l$ to $l$. In BiErGeO$_5$, Er$^{3+}$ ion has a low-symmetric $C_{1}$ crystal field environment and the CEF model Hamiltonian for this compound can be expressed as
\begin{align}
	\label{CEF1}
	\begin{split}
	&\mathcal{H}_{\rm CEF} = B_2^0\hat{O}_2^0 + (B_2^1\hat{O}_2^1 + B_2^{-1}\hat{O}_2^{-1}) + (B_2^2\hat{O}_2^2 + B_2^{-2}\hat{O}_2^{-2}) \\ 
	& + B_4^0\hat{O}_4^0 + (B_4^1\hat{O}_4^1 + B_4^{-1}\hat{O}_4^{-1}) + (B_4^2\hat{O}_4^2 + B_4^{-2}\hat{O}_4^{-2})\\ 
	& + (B_4^3\hat{O}_4^3 + B_4^{-3}\hat{O}_4^{-3}) + (B_4^4\hat{O}_4^4 + B_4^{-4}\hat{O}_4^{-4}) + B_6^0\hat{O}_6^0 \\ 
	& + (B_6^1\hat{O}_6^1 + B_6^{-1}\hat{O}_6^{-1}) + (B_6^2\hat{O}_6^2 + B_6^{-2}\hat{O}_6^{-2}) + (B_6^3\hat{O}_6^3 \\ 
	& + B_6^{-3}\hat{O}_6^{-3}) + (B_6^4\hat{O}_6^4 + B_6^{-4}\hat{O}_6^{-4}) + (B_6^5\hat{O}_6^5 + B_6^{-5}\hat{O}_6^{-5}) \\ 
	& + (B_6^6\hat{O}_6^6 + B_6^{-6}\hat{O}_6^{-6}).
   \end{split}
\end{align}

\begin{figure}
	\includegraphics[scale=0.45]{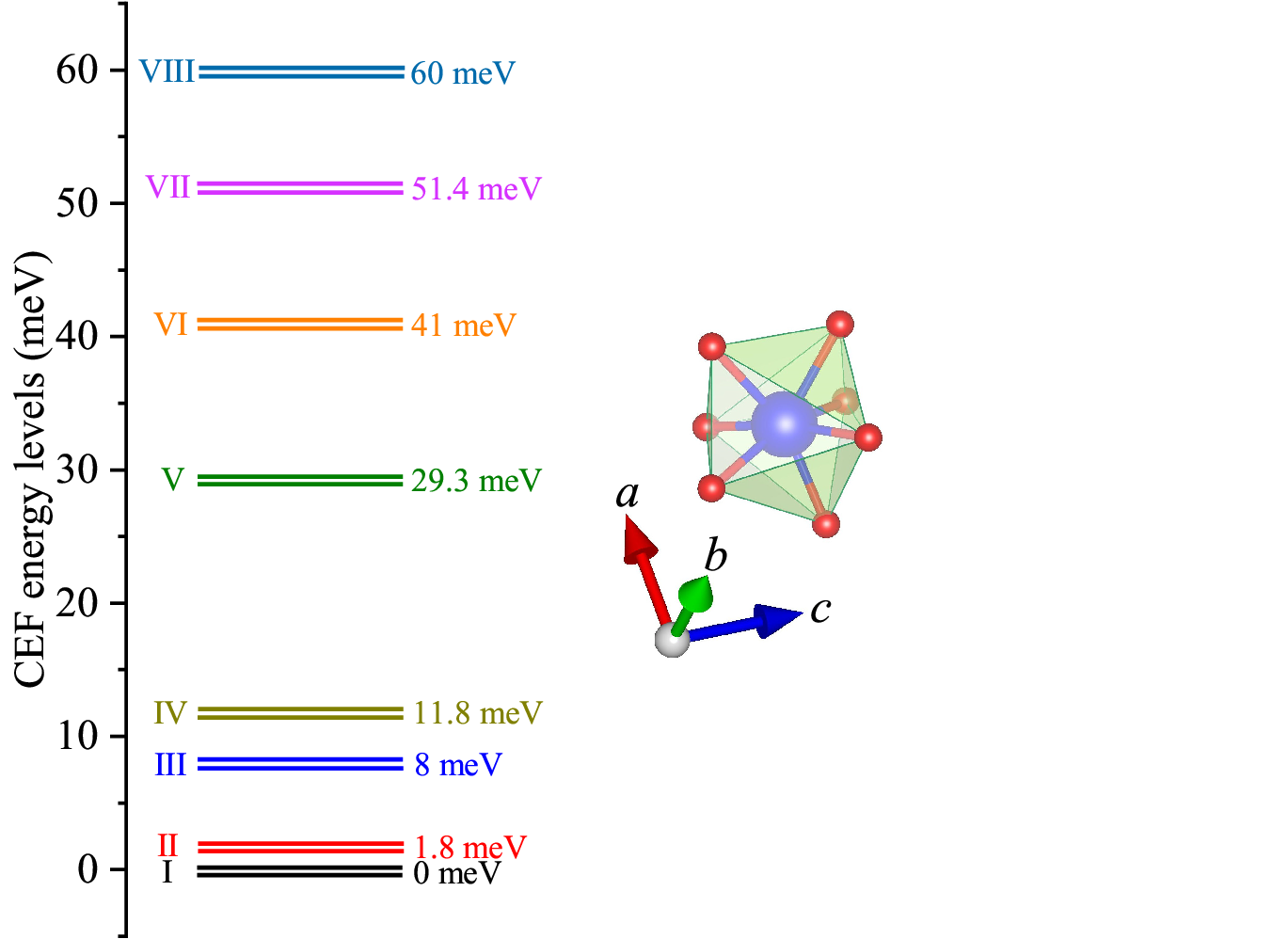}
	\caption{\label{Fig10} Schematic representation of CEF energy levels obtained from the zero field INS data. A distorted ErO$_7$ polyhedron formed by Er$^{3+}$ and O$^{2-}$ ions is shown in the right, that generates the CEF potential.}
\end{figure}

As presented in Figs.~\ref{Fig9}(a-d), we fitted the 5~K and 50~K data simultaneously using the above CEF model with the help of Mantid software~\cite{Arnold156}. The obtained best-fit CEF parameters of this system are tabulated in Table~\ref{CEF_Para}. Next, we diagonalized the CEF Hamiltonian and obtained the CEF energy eigenvalues of the compound. The obtained energy eigenvalues are 0, 1.91, 8.11, 11.79, 29.31, 43, 50.79, and 56.45~meV, corresponding to eight doublets, as depicted in Fig.~\ref{Fig10}. From the CEF Hamiltonian [Eq.~\eqref{CEF1}], the wave functions corresponding to all the Kramers' doublets as linear combinations of the eigenstates $\left|J=15/2,m_J \right\rangle$ can be expressed as
\begin{equation}\label{CEF_wave_vector}
	|\psi_k,\pm\rangle=\sum_{m_J = -15/2}^{m_J = 15/2}C_{m_J}^{k,\pm}\left|J  =15/2, m_J \right\rangle.
\end{equation}
Here, $C_{m_J}^{k,\pm}$ are the weighted coefficients of the eigenstates. The full list of energy eigenvalues and the corresponding coefficients ($C_{m_J}^{k,\pm}$) of different eigenstates for BiErGeO$_5$ are listed in Table~\ref{Eigenvalue_and_Eigervector}. Using the wave-functions of the CEF ground state ($|\psi_0,\pm\rangle $), one can determine the ratio of the anisotropic $g$-factors, $g_{xy}/g_{z} = g_{J}\langle\psi_0,\pm|J_{\pm}|\psi_0,\mp\rangle/2g_{J}\langle\psi_0,\pm|J_z|\psi_0,\pm\rangle = 3.41/2.47 = 1.38$. Here, $z$ and $xy$ directions refer to orientations parallel and perpendicular to the crystallographic $b$-axis, which is normal to the plane of the honeycomb layer.


The XXZ model Hamiltonian of the system can be written as  
\begin{equation}\label{CEF_wave_vector}
	\mathcal{H}^{\rm XXZ} = \sum_{i,j} \left[\mathcal{J}_zS_i^zS_j^z + \mathcal{J}_{xy}\left(S_i^xS_j^x + S_i^yS_j^y \right)\right].
\end{equation}
Here, $\mathcal J_{xy}$ and $\mathcal J_{z}$ represent the in-plane and out-of-plane exchange coupling constants, and $S_i^x$, $S_i^y$, and $S_i^z$ are the spin operators at site $i$ along $x$, $y$, and $z$ directions, respectively. The estimated value of the average exchange interactions at the ground state using the low-temperature powder-averaged CW temperature is $\bar{\mathcal{J}} = (\mathcal{J}_z + 2\mathcal{J}_{xy})/3 = -4\theta_{\rm CW}^{\rm LT}/3 \simeq 2.6$~K. The ratio of the anisotropic exchange interactions is found to be $\mathcal{J}_{xy}/\mathcal{J}_z = g_{xy}^2/g_{z}^2 \simeq 1.9$. Using the above relations we also calculated the anisotropic exchange interactions to be $\mathcal{J}_{xy} \simeq 2.96$~K and $\mathcal{J}_{z} \simeq 1.56$~K.

To examine the effect of CEF excitations on the physical properties, we calculated crystal field magnetic susceptibility [$\chi_{\rm CEF}(T)$], magnetization isotherms [$M_{\rm CEF}(H)$], and heat capacity [$C_{\rm CEF}(T)$] using the CEF energy levels with Zeeman splitting. The details of these calculations are described in Appendix~B. Figures~\ref{Fig3}(a) and \ref{Fig4}(a) illustrate the calculated $\chi_{\rm CEF}(T)$ and $M_{\rm CEF}(H)$ to compare with the experimental data. The low-temperature deviation between the measured and computed results can be attributed to the magnetic exchange coupling between the Er$^{3+}$ ions in BiErGeO$_5$, which is not taken into account for the calculations. Figure~\ref{Fig5}(d) presents $C_{\rm CEF}(T)$ in the low-temperature ($T < 50$~K) regime. In zero-field, the calculations yield two broad maxima in $C_{\rm CEF}$ at around $\sim 8.85$, and 39~K, consistent with the experimental $C_{\rm mag} (T)$ above 3~K. These maxima appear due to the transition from ground state doublet to the first excited and second excited doublets, respectively. The CEF contribution to the heat capacity approaches zero below about 2~K in contrast to the low-temperature rounded maxima at around 1~K observed in the experimental $C_{\rm mag}(T)$ data. This implies that the low-temperature rounded maxima in zero-field $C_{\rm mag}$ is a signature of short-range correlation between the Er$^{3+}$ ions. The FD simulation of $C_{\rm mag}$ for the $S = 1/2$ XXZ model [using Eq.~(\ref{CEF_wave_vector})] with $\mathcal{J}_{xy} = 3$~K and $\mathcal{J}_{z} = 1.6$~K reproduce the low-$T$ broad maxima at zero field. In the presence of magnetic field the degenerate Kramers' doublets split further and the calculated $C_{\rm CEF}$ results in another low-$T$ broad maxima reproducing our experimental $C_{\rm mag}(T)$. At $\mu_0 H = 1$~T, the low-$T$ maximum appears at around $T^{***}\sim 1.4$~K which can be attributed to the transition occurring between Zeeman levels of the ground-state doublet. With increasing field, this maximum shift towards high temperatures, consistent with the experimental $C_{\rm mag}(T)$ curve.

\begin{figure}
	\includegraphics[scale=0.4]{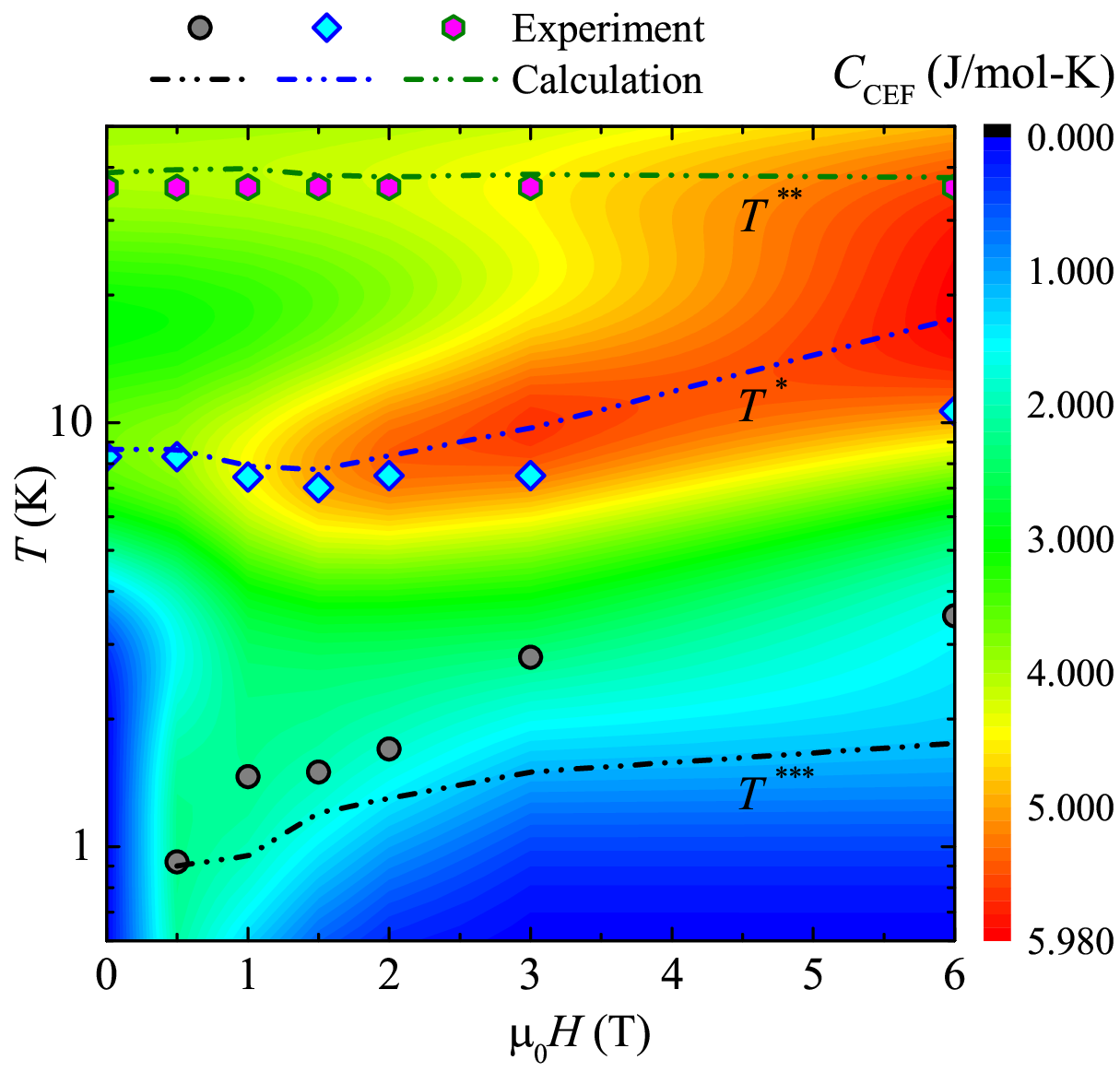}
	\caption{\label{Fig11} 2D contour plot of the $C_{\rm CEF}$ as a function of field and temperature. On the top of this plot, $T^{\star}$, $T^{\star\star}$, and $T^{\star\star\star}$ obtained from the experimental $C_{\rm mag}$ (symbols) [from Fig.~\ref{Fig5}(b)] and calculated $C_{\rm CEF}$ (dot-dashed lines) [from Fig.~\ref{Fig5}(d)] are shown.}
\end{figure}
To facilitate a comparison with the experimental data, we constructed a 2D contour map of $C_{\rm CEF} (T, H)$ along with $T^{\star}$, $T^{\star\star}$, and $T^{\star\star\star}$ obtained from the experimental $C_{\rm mag}(T)$ [from Fig.~\ref{Fig5}(b)], as depicted in Fig.~\ref{Fig11}. The magnetic-field dependence of the three anomalies extracted from the calculations follow the experimental $C_{\rm mag}(T)$ data. The minor deviation between the experimental and theoretical values of $T^{\star}$ and $T^{\star\star}$ may arise from a weak anisotropic exchange interaction among the Er$^{3+}$ ions, which is not incorporated into the present calculations.


\subsection{Muon spin relaxation}
\begin{figure*}
	\includegraphics[width = \linewidth]{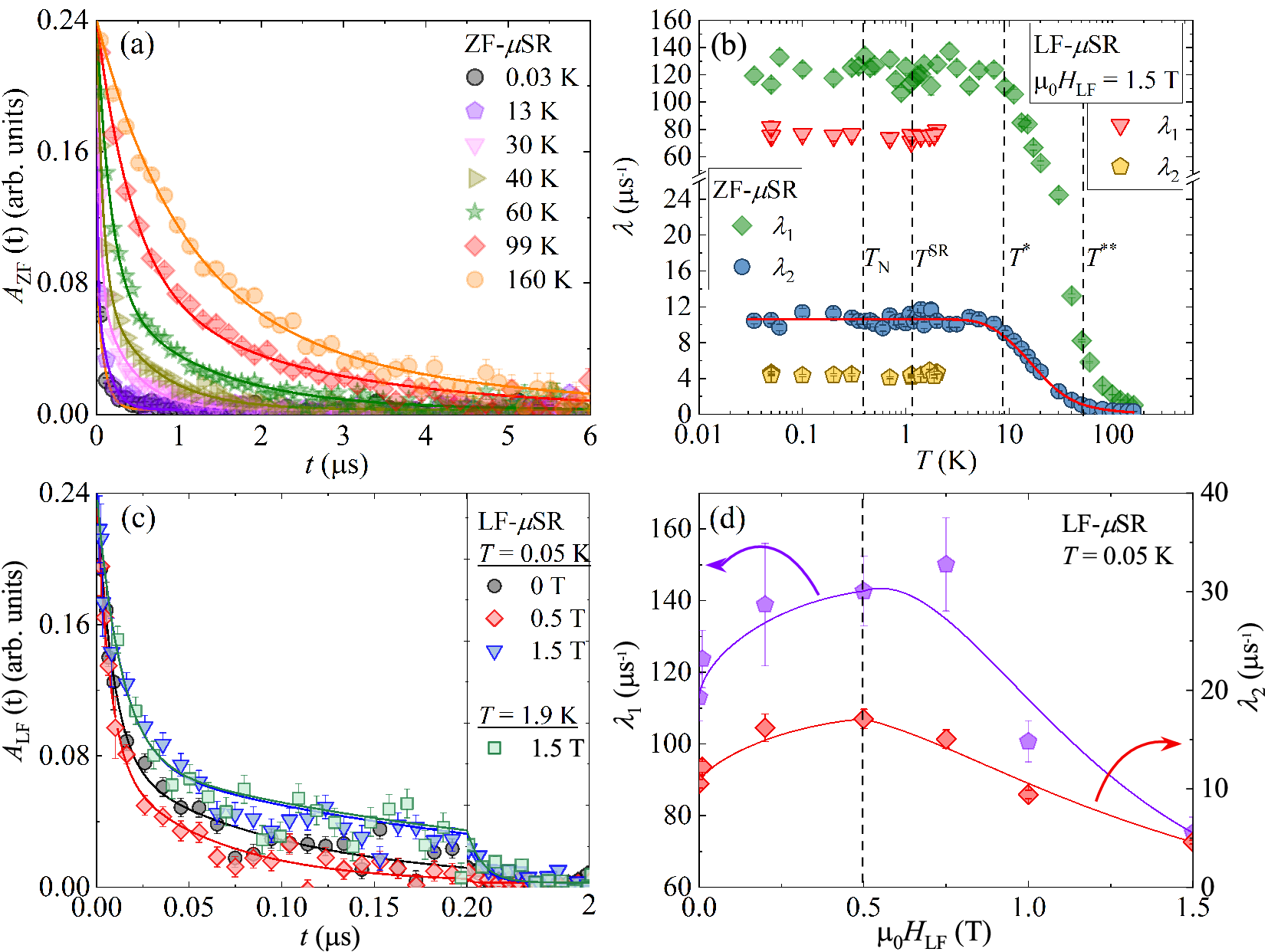}
	\caption{\label{MuSR} (a) Temperature evolution of the zero-field (ZF) $\mu$SR asymmetry spectra of BiErGeO$_5$. (b) Temperature dependence of the muon spin relaxation rates ($\lambda_{1}$ and $\lambda_{2}$) obtained from fits to the ZF data and fixed LF of 1.5~T. The red solid line represent the fit of $\lambda_{2}$ to an Orbach-type relaxation model as described in the text. (c) Representative LF $\mu$SR spectra measured at $T = 0.05$~K under applied fields up to 1.5~T, along with data at $T = 1.9$~K in a field of 1.5~T. Magnification of the short-time behavior ($t \le 0.2\,\mu$s) and long-time tail ($t \ge 0.2\,\mu$s) are shown in left and right panels, respectively. (d) Field dependence of the relaxation rates $\lambda_{1}$ and $\lambda_{2}$ from the LF measurements at $T = 0.05$~K.}
\end{figure*}

To probe the spin dynamics, zero-field (ZF) and longitudinal-field (LF) $\mu$SR measurements were performed on polycrystalline BiErGeO$_5$ sample over a broad temperature and magnetic-field range. The $\mu$SR technique provides a sensitive local probe of internal magnetic fields as low as $\sim 0.1$~G, and accesses the spin dynamics within a unique characteristic time window of $10^{-11}$–$10^{-5}$~s, complementing neutron scattering experiments ($10^{-14}$–$10^{-8}$~s).

Figure~\ref{MuSR}\,(a) presents representative zero-field (ZF) $\mu$SR asymmetry spectra measured between 0.03 and 160~K. Across the entire temperature interval, the asymmetry shows smooth monotonic decay. The ZF data at all temperatures are satisfactorily reproduced using a two-component exponential model of the form~\cite{Sebastian2025}
\begin{equation}
	\label{eq:empirical}
	A_{\rm ZF}(t) = A_{\rm rel}\bigl[f\,e^{-\lambda_1 t} + (1 - f)\,e^{-\lambda_2 t}\bigr] + A_{\rm b},
\end{equation}
where $\lambda_1$ and $\lambda_2$ represent the relaxation rates of the rapidly and slowly decaying components, respectively. The parameter $f$ denotes the fractional contribution of the fast channel, which dominates the short-time response ($t \lesssim 0.03\,\mu$s), while $(1-f)$ accounts for the slower relaxation visible at longer times~\cite{MagarL020409}. The total observable relaxing asymmetry is characterized by $A_{\rm rel} \approx 0.237$, and a small temperature-independent background term $A_{\rm b} \approx 0.003$ arising from muons implanted in the cryostat walls or sample holder. The requirement of two distinct relaxation channels indicates that muons occupy two magnetically inequivalent stopping sites within the sample. Optimal fits were achieved by fixing the fractional amplitude of the fast component to a common value $f \approx 0.685(4)$ for all measured temperatures. The observation $f \approx 2/3$ is consistent with the expected powder-averaged distribution of muon spin components, consisting of transverse (2/3) and longitudinal (1/3) fractions. The transverse component probes both static and dynamic fields, whereas the longitudinal component is sensitive only to dynamical fluctuations.

The temperature dependences of $\lambda_1$ and $\lambda_2$ are presented in Fig.~\ref{MuSR}(b). At high temperatures ($T > 80$~K), both relaxation rates nearly merge and exhibit only a weak temperature dependence, as expected for a paramagnetic state in the motional narrowing regime. Upon cooling, $\lambda$ increases rapidly below about 50~K and becomes nearly temperature independent below $\sim 9$~K. This behavior cannot be attributed solely to exchange-driven spin dynamics or a simple slowing down of spin fluctuations, since both 50~K and 9~K lie well above the energy scale of the powder-averaged exchange coupling, $\bar{\mathcal{J}} = (\mathcal{J}_z + 2\mathcal{J}_{xy})/3 \simeq 2.6$~K. Instead, these temperature scales are close to the characteristic energies associated with excitations from the CEF ground-state doublet to the first and second excited doublets, as determined from our CEF calculations and INS measurements.

The pronounced temperature-dependent decrease of $\lambda$ above $T^{*}$ can therefore be described by an Orbach-type relaxation process, in which the Er$^{3+}$ spins relax via phonon-assisted transitions between the CEF ground-state doublet and excited CEF levels~\cite{Orbach458}. However, as discussed above, the relaxation rate $\lambda_1$ (associated with the $2/3$ transverse component) likely contains both static and dynamic contributions. In order to avoid this static component, we restrict the Orbach analysis to the relaxation rate $\lambda_2(T)$, which predominantly reflects the dynamics. The temperature dependence of $\lambda_2(T)$ was therefore analyzed using a multi-channel Orbach-type relaxation model~\cite{Arh416, Xu064425},
\begin{equation}
\lambda_2(T)=
\left[
\lambda_{2,0}^{-1}
+
\sum_{i=1}^{2}
\frac{\eta_i}{\exp(E_i/T)-1}
\right]^{-1},
\end{equation}
that yields $\eta_1 = 0.192(15)$, $\eta_2 = 2.09(15)$, and $\lambda_{2,0} = 10.6~\mu{\rm s}^{-1}$. Here, $\eta_i$ characterize the spin--lattice coupling strengths and $E_i$ represent the energies of the excited crystal-field levels mediating the relaxation process. Since the excitation energies are independently determined from the CEF analysis and INS measurements, the corresponding energy scales were fixed in the fitting procedure to $E_1\simeq 1.8$~meV and $E_2\simeq 8.0$~meV. Fixing these values reduces the number of free parameters and ensures consistency with the experimentally established CEF scheme. The larger prefactor associated with the $E_2$ channel indicates that the relaxation is predominantly mediated by the CEF excitation near $8$~meV, while the lower-lying level at $1.8$~meV provides a weaker additional relaxation pathway.


Upon further cooling, both $\lambda_1$ and $\lambda_2$ remain nearly temperature independent without showing any clear anomaly at $T_{\rm N}$. As illustrated in Fig.~\ref{MuSR}\,(a - c), the overall time evolution of the asymmetry $A(t)$ remains qualitatively unchanged across $T_{\rm N}$. In particular, the characteristic signatures of a conventional static magnetically ordered state are absent. Neither the powder-averaged long-time $1/3$ tail nor coherent oscillations arising from a well-defined internal magnetic field are observed.

The absence of oscillations in a magnetically ordered phase could, in principle, result from a broad distribution of local magnetic fields at the muon stopping sites or from substantial spatial variations in the magnitude of the ordered moment. The absence of the 1/3 tail however is a clear evidence for dynamics. Another possibility is that the internal field at the muon site is accidentally small due to geometrical cancellation of dipolar contributions. Given that muons in oxides typically localize near oxygen ions, and that in BiErGeO$_5$ the oxygen atoms occupy low-symmetry $8c$ sites within the orthorhombic $Pbca$ structure, such a cancellation appears unlikely. A quantitative determination of the muon stopping site and the corresponding local magnetic fields would be necessary to rigorously assess this scenario.

Similar low-temperature relaxation plateau has been reported in several frustrated magnets that nevertheless exhibit long-range magnetic order~\cite{Xu064425,Cai184415,Bert117203,Lago979,Dalmas127202,Kalvius87}. In some of these systems, the coexistence of magnetic Bragg peaks in neutron diffraction and a temperature-independent relaxation rate in ZF-$\mu$SR has been interpreted as evidence for slow persistent spin dynamics surviving deep inside the ordered state. However, in some cases, the microscopic origin of such plateaus remains debated~\cite{McClarty164216}.

It is important to note that a temperature-independent relaxation at low temperatures does not uniquely imply intrinsic electronic spin dynamics. In systems with non-Kramers doublets, it has been proposed that the implanted muon may locally distort the crystalline electric field, lift the degeneracy of the non-Kramers ground state, and thereby induce a local magnetic moment. This mechanism can give rise to hyperfine-enhanced nuclear magnetism and produce an apparently temperature-independent relaxation rate ~\cite{Foronda017602}. In the present compound, however, the magnetic ion Er$^{3+}$ is a Kramers ion with a Kramers doublet ground state. Because time-reversal symmetry protects a Kramers doublet from splitting by a purely electrostatic perturbation, the muon-induced mechanism discussed in the above work is not directly applicable here. Nevertheless, alternative mechanisms, such as hyperfine-enhanced nuclear contributions or low-energy spin fluctuations, may still influence the observed low-temperature behavior.

To further examine the dynamics of the ground state, we measured the longitudinal-field (LF) dependence of the muon-spin asymmetry $A_{\rm LF}(t)$. If the relaxation at $T = 50$~mK were caused by static internal fields at the muon site, an effective static internal field can be estimated from $H_{\mu}^{s} \simeq \lambda/\gamma_{\mu}$. Using this approximation one obtains $H_{\mu,1}^{s} \simeq 0.14$~T and $H_{\mu,2}^{s} \simeq 0.012$~T for the fast and slow components, respectively. We emphasize that this estimate is meaningful only in the purely static limit; if the relaxation arises from dynamical fluctuations, $\lambda$ depends on both the local-field amplitude and the fluctuation time scale, and thus $\lambda/\gamma_{\mu}$ does not represent a unique internal field. In a static scenario, a complete decoupling is typically expected when the applied longitudinal field exceeds several times the characteristic internal field ($H_{\mu}^{s}$), commonly by about an order of magnitude. Based on the above estimate, fields of order $\sim 1.4$~T would therefore be expected to significantly restore the muon polarization. However, as shown in Fig.~\ref{MuSR}\,(c), only weak decoupling is observed even at the highest applied field of 1.5~T. This absence of strong decoupling provides clear evidence that the muon-spin relaxation is dominated by fluctuating internal fields rather than static local fields. The field-dependent asymmetry spectra were analyzed using the same two-component exponential model employed in the zero-field fits. Notably, the low-temperature saturation of the relaxation rates persists up to at least 1.5~T and to temperatures as high as 1.9~K [see Fig.~\ref{MuSR}(b)], further supporting the presence of persistent slow spin fluctuations in the magnetically ordered state.

The longitudinal-field dependence of the relaxation rates is summarized in Fig.~\ref{MuSR}(d). Within the framework of the Redfield model, the relaxation rate is given by $\lambda = 2\,\Delta^2\,\nu\, /\, [\nu^2 + (\gamma_\mu \mu_{0} H_{\mathrm{LF}})^2]$, where $\Delta$ represents the width of the fluctuating local field distribution and $\nu$ denotes the fluctuation rate. In the fast-fluctuation limit ($\nu \gg \Delta$), this expression predicts a gradual suppression of the relaxation rate with increasing longitudinal field. However, this behavior is not observed here; instead, the relaxation rates (both $\lambda_{1}$ and $\lambda_{2}$) initially increase and exhibit a broad maximum near 0.5~T, rather than undergoing the monotonic suppression. The observed field dependence of $\lambda_{1}$ and $\lambda_{2}$ is consistent with the observations in field-dependent heat-capacity and magnetization measurements and may indicate a field-induced modification of the magnetic ground state.

\section{Discussion and Summary}
The combined magnetization, heat capacity, INS, and $\mu$SR results establish BiErGeO$_5$ as a rare-earth honeycomb antiferromagnet whose ground state is governed by the interplay of strong single-ion anisotropy and competing exchange interactions. The INS spectra reveal eight well-separated CEF doublets of the Er$^{3+}$ ($J=15/2$) ion, reflecting the low-symmetry $C_1$ environment of the ErO$_7$ polyhedra. The extracted CEF wave functions indicate an anisotropic $g$-tensor with $g_{xy}>g_z$, exhibiting predominantly easy-plane character. This anisotropy is further mirrored in the ratio of the exchange parameters, $\mathcal{J}_{xy}/\mathcal{J}_z \simeq 1.9$, obtained from CW analysis and XXZ modeling of the low-$T$ susceptibility.

Although BiErGeO$_5$ undergoes a magnetic LRO at $T_{\rm N} \simeq 0.4$~K, several signatures point towards unconventional low-temperature spin dynamics. Heat capacity data show a broad maximum near $1.1$~K that cannot be accounted for by CEF contributions alone, implying short-range correlations preceding the ordered state. The $\mu$SR spectra lack spontaneous oscillations down to 30~mK, and the relaxation rate remains nearly temperature independent below 7~K. Moreover, the relaxation could not be fully decoupled in a longitudinal field as high as 1.5~T, which is inconsistent with a simple static ordered state. Similar features have also been observed in certain frustrated rare-earth magnets and may reflect slow spin dynamics or moment fragmentation in the ordered phase~\cite{Xu064425,Cai184415,Bert117203,Lago979,Dalmas127202,Kalvius87}. These observations position BiErGeO$_5$ in an intermediate regime where classical LRO coexists with residual quantum fluctuations at low temperatures.

The comparison between BiErGeO$_5$ and its Yb-analogue BiYbGeO$_5$ is illuminating. Substitution of Er for Yb preserves the distorted honeycomb structure but stabilizes magnetic LRO instead of a disordered state previously reported for BiYbGeO$_5$~\cite{Mohanty134408}. This contrasting behaviour highlights the sensitivity of magnetic ground state in distorted honeycomb lattices to the choice of rare-earth ions, particularly to changes in the CEF wave functions and their resulting anisotropic exchange couplings.

In summary, we have investigated the magnetic properties and CEF excitations of the distorted honeycomb magnet BiErGeO$_5$ using magnetization, heat capacity, $\mu$SR, and INS experiments. Eight CEF doublets of Er$^{3+}$ ions are identified from the INS data. The fitted CEF model successfully reproduces the thermodynamic responses above 3~K. At low temperatures, BiErGeO$_5$ undergoes AFM LRO at around $T_{\rm N} \simeq 0.4$~K. However, the presence of a heat-capacity maximum near 1.1~K and persistent spin dynamics observed in $\mu$SR suggest that slow fluctuations survive deep into the ordered phase. The extracted XXZ exchange parameters confirm significant exchange anisotropy ($\mathcal{J}_{xy} > \mathcal{J}_z$) consistent with the shape of the CEF ground state doublet. Overall, BiErGeO$_5$ exemplifies the rich physics of a spin-orbit coupled honeycomb magnet where CEF induced anisotropy and reduced dimensionality conspire to produce a unconventional dynamical behavior even in the presence of a magnetic LRO.

\begin{acknowledgments}
We would like to acknowledge SERB, India (Grant No.~CRG/2022/000997) and DST-FIST (Grant No.~SR/FST/PS-II/2018/54(C)) for the financial support. Experiments at the ISIS Neutron and Muon Source were supported by beamtime allocation RB2510048 from the Science and Technology Facilities Council. Data is available here: https://doi.org/10.5286/ISIS.E.RB2510048.
\end{acknowledgments}


\begin{thebibliography}{59}%
	\makeatletter
	\providecommand \@ifxundefined [1]{%
		\@ifx{#1\undefined}
	}%
	\providecommand \@ifnum [1]{%
		\ifnum #1\expandafter \@firstoftwo
		\else \expandafter \@secondoftwo
		\fi
	}%
	\providecommand \@ifx [1]{%
		\ifx #1\expandafter \@firstoftwo
		\else \expandafter \@secondoftwo
		\fi
	}%
	\providecommand \natexlab [1]{#1}%
	\providecommand \enquote  [1]{``#1''}%
	\providecommand \bibnamefont  [1]{#1}%
	\providecommand \bibfnamefont [1]{#1}%
	\providecommand \citenamefont [1]{#1}%
	\providecommand \href@noop [0]{\@secondoftwo}%
	\providecommand \href [0]{\begingroup \@sanitize@url \@href}%
	\providecommand \@href[1]{\@@startlink{#1}\@@href}%
	\providecommand \@@href[1]{\endgroup#1\@@endlink}%
	\providecommand \@sanitize@url [0]{\catcode `\\12\catcode `\$12\catcode
		`\&12\catcode `\#12\catcode `\^12\catcode `\_12\catcode `\%12\relax}%
	\providecommand \@@startlink[1]{}%
	\providecommand \@@endlink[0]{}%
	\providecommand \url  [0]{\begingroup\@sanitize@url \@url }%
	\providecommand \@url [1]{\endgroup\@href {#1}{\urlprefix }}%
	\providecommand \urlprefix  [0]{URL }%
	\providecommand \Eprint [0]{\href }%
	\providecommand \doibase [0]{https://doi.org/}%
	\providecommand \selectlanguage [0]{\@gobble}%
	\providecommand \bibinfo  [0]{\@secondoftwo}%
	\providecommand \bibfield  [0]{\@secondoftwo}%
	\providecommand \translation [1]{[#1]}%
	\providecommand \BibitemOpen [0]{}%
	\providecommand \bibitemStop [0]{}%
	\providecommand \bibitemNoStop [0]{.\EOS\space}%
	\providecommand \EOS [0]{\spacefactor3000\relax}%
	\providecommand \BibitemShut  [1]{\csname bibitem#1\endcsname}%
	\let\auto@bib@innerbib\@empty
	\bibitem [{\citenamefont {Elliott}(1953)}]{Elliott167}%
	\BibitemOpen
	\bibfield  {author} {\bibinfo {author} {\bibfnamefont {R.~J.}\ \bibnamefont
			{Elliott}},\ }\bibfield  {title} {\bibinfo {title} {Crystal field theory in
			the rare earths},\ }\href {https://doi.org/10.1103/RevModPhys.25.167}
	{\bibfield  {journal} {\bibinfo  {journal} {Rev. Mod. Phys.}\ }\textbf
		{\bibinfo {volume} {25}},\ \bibinfo {pages} {167} (\bibinfo {year}
		{1953})}\BibitemShut {NoStop}%
	\bibitem [{\citenamefont {Fulde}\ and\ \citenamefont
		{Loewenhaupt}(1985)}]{Fulde589}%
	\BibitemOpen
	\bibfield  {author} {\bibinfo {author} {\bibfnamefont {P.}~\bibnamefont
			{Fulde}}\ and\ \bibinfo {author} {\bibfnamefont {M.}~\bibnamefont
			{Loewenhaupt}},\ }\bibfield  {title} {\bibinfo {title} {Magnetic excitations
			in crystal-field split 4$f$ systems},\ }\href
	{https://doi.org/10.1080/00018738500101821} {\bibfield  {journal} {\bibinfo
			{journal} {Adv. Phys.}\ }\textbf {\bibinfo {volume} {34}},\ \bibinfo {pages}
		{589} (\bibinfo {year} {1985})}\BibitemShut {NoStop}%
	\bibitem [{\citenamefont {Li}\ \emph {et~al.}(2016)\citenamefont {Li},
		\citenamefont {Adroja}, \citenamefont {Biswas}, \citenamefont {Baker},
		\citenamefont {Zhang}, \citenamefont {Liu}, \citenamefont {Tsirlin},
		\citenamefont {Gegenwart},\ and\ \citenamefont {Zhang}}]{Li097201}%
	\BibitemOpen
	\bibfield  {author} {\bibinfo {author} {\bibfnamefont {Y.}~\bibnamefont
			{Li}}, \bibinfo {author} {\bibfnamefont {D.}~\bibnamefont {Adroja}}, \bibinfo
		{author} {\bibfnamefont {P.~K.}\ \bibnamefont {Biswas}}, \bibinfo {author}
		{\bibfnamefont {P.~J.}\ \bibnamefont {Baker}}, \bibinfo {author}
		{\bibfnamefont {Q.}~\bibnamefont {Zhang}}, \bibinfo {author} {\bibfnamefont
			{J.}~\bibnamefont {Liu}}, \bibinfo {author} {\bibfnamefont {A.~A.}\
			\bibnamefont {Tsirlin}}, \bibinfo {author} {\bibfnamefont {P.}~\bibnamefont
			{Gegenwart}},\ and\ \bibinfo {author} {\bibfnamefont {Q.}~\bibnamefont
			{Zhang}},\ }\bibfield  {title} {\bibinfo {title} {{Muon Spin Relaxation
				Evidence for the U(1) Quantum Spin-Liquid Ground State in the Triangular
				Antiferromagnet ${\mathrm{YbMgGaO}}_{4}$}},\ }\href
	{https://doi.org/10.1103/PhysRevLett.117.097201} {\bibfield  {journal}
		{\bibinfo  {journal} {Phys. Rev. Lett.}\ }\textbf {\bibinfo {volume} {117}},\
		\bibinfo {pages} {097201} (\bibinfo {year} {2016})}\BibitemShut {NoStop}%
	\bibitem [{\citenamefont {Rau}\ and\ \citenamefont
		{Gingras}(2015)}]{Rau144417}%
	\BibitemOpen
	\bibfield  {author} {\bibinfo {author} {\bibfnamefont {J.~G.}\ \bibnamefont
			{Rau}}\ and\ \bibinfo {author} {\bibfnamefont {M.~J.~P.}\ \bibnamefont
			{Gingras}},\ }\bibfield  {title} {\bibinfo {title} {Magnitude of quantum
			effects in classical spin ices},\ }\href
	{https://doi.org/10.1103/PhysRevB.92.144417} {\bibfield  {journal} {\bibinfo
			{journal} {Phys. Rev. B}\ }\textbf {\bibinfo {volume} {92}},\ \bibinfo
		{pages} {144417} (\bibinfo {year} {2015})}\BibitemShut {NoStop}%
	\bibitem [{\citenamefont {Tomasello}\ \emph {et~al.}(2015)\citenamefont
		{Tomasello}, \citenamefont {Castelnovo}, \citenamefont {Moessner},\ and\
		\citenamefont {Quintanilla}}]{Tomasello155120}%
	\BibitemOpen
	\bibfield  {author} {\bibinfo {author} {\bibfnamefont {B.}~\bibnamefont
			{Tomasello}}, \bibinfo {author} {\bibfnamefont {C.}~\bibnamefont
			{Castelnovo}}, \bibinfo {author} {\bibfnamefont {R.}~\bibnamefont
			{Moessner}},\ and\ \bibinfo {author} {\bibfnamefont {J.}~\bibnamefont
			{Quintanilla}},\ }\bibfield  {title} {\bibinfo {title} {{Single-ion
				anisotropy and magnetic field response in the spin-ice materials
				${\mathrm{Ho}}_{2}{\mathrm{Ti}}_{2}{\mathrm{O}}_{7}$ and
				${\mathrm{Dy}}_{2}{\mathrm{Ti}}_{2}{\mathrm{O}}_{7}$}},\ }\href
	{https://doi.org/10.1103/PhysRevB.92.155120} {\bibfield  {journal} {\bibinfo
			{journal} {Phys. Rev. B}\ }\textbf {\bibinfo {volume} {92}},\ \bibinfo
		{pages} {155120} (\bibinfo {year} {2015})}\BibitemShut {NoStop}%
	\bibitem [{\citenamefont {Petit}\ \emph {et~al.}(2014)\citenamefont {Petit},
		\citenamefont {Robert}, \citenamefont {Guitteny}, \citenamefont {Bonville},
		\citenamefont {Decorse}, \citenamefont {Ollivier}, \citenamefont {Mutka},
		\citenamefont {Gingras},\ and\ \citenamefont {Mirebeau}}]{Petit060410}%
	\BibitemOpen
	\bibfield  {author} {\bibinfo {author} {\bibfnamefont {S.}~\bibnamefont
			{Petit}}, \bibinfo {author} {\bibfnamefont {J.}~\bibnamefont {Robert}},
		\bibinfo {author} {\bibfnamefont {S.}~\bibnamefont {Guitteny}}, \bibinfo
		{author} {\bibfnamefont {P.}~\bibnamefont {Bonville}}, \bibinfo {author}
		{\bibfnamefont {C.}~\bibnamefont {Decorse}}, \bibinfo {author} {\bibfnamefont
			{J.}~\bibnamefont {Ollivier}}, \bibinfo {author} {\bibfnamefont
			{H.}~\bibnamefont {Mutka}}, \bibinfo {author} {\bibfnamefont {M.~J.~P.}\
			\bibnamefont {Gingras}},\ and\ \bibinfo {author} {\bibfnamefont
			{I.}~\bibnamefont {Mirebeau}},\ }\bibfield  {title} {\bibinfo {title} {{Order
				by disorder or energetic selection of the ground state in the $XY$ pyrochlore
				antiferromagnet ${\mathrm{Er}}_{2}{\mathrm{Ti}}_{2}{\mathrm{O}}_{7}$: An
				inelastic neutron scattering study}},\ }\href
	{https://doi.org/10.1103/PhysRevB.90.060410} {\bibfield  {journal} {\bibinfo
			{journal} {Phys. Rev. B}\ }\textbf {\bibinfo {volume} {90}},\ \bibinfo
		{pages} {060410} (\bibinfo {year} {2014})}\BibitemShut {NoStop}%
	\bibitem [{\citenamefont {Gao}\ \emph {et~al.}(2020)\citenamefont {Gao},
		\citenamefont {Xiao}, \citenamefont {Kamazawa}, \citenamefont {Ikeuchi},
		\citenamefont {Biner}, \citenamefont {Kr\"amer}, \citenamefont {R\"uegg},\
		and\ \citenamefont {Arima}}]{Gao024424}%
	\BibitemOpen
	\bibfield  {author} {\bibinfo {author} {\bibfnamefont {S.}~\bibnamefont
			{Gao}}, \bibinfo {author} {\bibfnamefont {F.}~\bibnamefont {Xiao}}, \bibinfo
		{author} {\bibfnamefont {K.}~\bibnamefont {Kamazawa}}, \bibinfo {author}
		{\bibfnamefont {K.}~\bibnamefont {Ikeuchi}}, \bibinfo {author} {\bibfnamefont
			{D.}~\bibnamefont {Biner}}, \bibinfo {author} {\bibfnamefont {K.~W.}\
			\bibnamefont {Kr\"amer}}, \bibinfo {author} {\bibfnamefont {C.}~\bibnamefont
			{R\"uegg}},\ and\ \bibinfo {author} {\bibfnamefont {T.-h.}\ \bibnamefont
			{Arima}},\ }\bibfield  {title} {\bibinfo {title} {{Crystal electric field
				excitations in the quantum spin liquid candidate ${\mathrm{NaErS}}_{2}$}},\
	}\href {https://doi.org/10.1103/PhysRevB.102.024424} {\bibfield  {journal}
		{\bibinfo  {journal} {Phys. Rev. B}\ }\textbf {\bibinfo {volume} {102}},\
		\bibinfo {pages} {024424} (\bibinfo {year} {2020})}\BibitemShut {NoStop}%
	\bibitem [{\citenamefont {Sibille}\ \emph {et~al.}(2018)\citenamefont
		{Sibille}, \citenamefont {Gauthier}, \citenamefont {Yan}, \citenamefont
		{Ciomaga~Hatnean}, \citenamefont {Ollivier}, \citenamefont {Winn},
		\citenamefont {Filges}, \citenamefont {Balakrishnan}, \citenamefont
		{Kenzelmann}, \citenamefont {Shannon},\ and\ \citenamefont
		{Fennell}}]{Sibille711}%
	\BibitemOpen
	\bibfield  {author} {\bibinfo {author} {\bibfnamefont {R.}~\bibnamefont
			{Sibille}}, \bibinfo {author} {\bibfnamefont {N.}~\bibnamefont {Gauthier}},
		\bibinfo {author} {\bibfnamefont {H.}~\bibnamefont {Yan}}, \bibinfo {author}
		{\bibfnamefont {M.}~\bibnamefont {Ciomaga~Hatnean}}, \bibinfo {author}
		{\bibfnamefont {J.}~\bibnamefont {Ollivier}}, \bibinfo {author}
		{\bibfnamefont {B.}~\bibnamefont {Winn}}, \bibinfo {author} {\bibfnamefont
			{U.}~\bibnamefont {Filges}}, \bibinfo {author} {\bibfnamefont
			{G.}~\bibnamefont {Balakrishnan}}, \bibinfo {author} {\bibfnamefont
			{M.}~\bibnamefont {Kenzelmann}}, \bibinfo {author} {\bibfnamefont
			{N.}~\bibnamefont {Shannon}},\ and\ \bibinfo {author} {\bibfnamefont
			{T.}~\bibnamefont {Fennell}},\ }\bibfield  {title} {\bibinfo {title}
		{{Experimental signatures of emergent quantum electrodynamics in
				${\mathrm{Pr}}_{2}{\mathrm{Hf}}_{2}{\mathrm{O}}_{7}$}},\ }\href
	{https://doi.org/10.1038/s41567-018-0116-x} {\bibfield  {journal} {\bibinfo
			{journal} {Nat. Phys.}\ }\textbf {\bibinfo {volume} {14}},\ \bibinfo {pages}
		{711} (\bibinfo {year} {2018})}\BibitemShut {NoStop}%
	\bibitem [{\citenamefont {Li}\ \emph {et~al.}(2017)\citenamefont {Li},
		\citenamefont {Adroja}, \citenamefont {Bewley}, \citenamefont {Voneshen},
		\citenamefont {Tsirlin}, \citenamefont {Gegenwart},\ and\ \citenamefont
		{Zhang}}]{Li107202}%
	\BibitemOpen
	\bibfield  {author} {\bibinfo {author} {\bibfnamefont {Y.}~\bibnamefont
			{Li}}, \bibinfo {author} {\bibfnamefont {D.}~\bibnamefont {Adroja}}, \bibinfo
		{author} {\bibfnamefont {R.~I.}\ \bibnamefont {Bewley}}, \bibinfo {author}
		{\bibfnamefont {D.}~\bibnamefont {Voneshen}}, \bibinfo {author}
		{\bibfnamefont {A.~A.}\ \bibnamefont {Tsirlin}}, \bibinfo {author}
		{\bibfnamefont {P.}~\bibnamefont {Gegenwart}},\ and\ \bibinfo {author}
		{\bibfnamefont {Q.}~\bibnamefont {Zhang}},\ }\bibfield  {title} {\bibinfo
		{title} {{Crystalline Electric-Field Randomness in the Triangular Lattice
				Spin-Liquid ${\mathrm{YbMgGaO}}_{4}$}},\ }\href
	{https://doi.org/10.1103/PhysRevLett.118.107202} {\bibfield  {journal}
		{\bibinfo  {journal} {Phys. Rev. Lett.}\ }\textbf {\bibinfo {volume} {118}},\
		\bibinfo {pages} {107202} (\bibinfo {year} {2017})}\BibitemShut {NoStop}%
	\bibitem [{\citenamefont {Somesh}\ \emph {et~al.}(2023)\citenamefont {Somesh},
		\citenamefont {Islam}, \citenamefont {Mohanty}, \citenamefont {Simutis},
		\citenamefont {Guguchia}, \citenamefont {Wang}, \citenamefont
		{Sichelschmidt}, \citenamefont {Baenitz},\ and\ \citenamefont
		{Nath}}]{Somesh064421}%
	\BibitemOpen
	\bibfield  {author} {\bibinfo {author} {\bibfnamefont {K.}~\bibnamefont
			{Somesh}}, \bibinfo {author} {\bibfnamefont {S.~S.}\ \bibnamefont {Islam}},
		\bibinfo {author} {\bibfnamefont {S.}~\bibnamefont {Mohanty}}, \bibinfo
		{author} {\bibfnamefont {G.}~\bibnamefont {Simutis}}, \bibinfo {author}
		{\bibfnamefont {Z.}~\bibnamefont {Guguchia}}, \bibinfo {author}
		{\bibfnamefont {C.}~\bibnamefont {Wang}}, \bibinfo {author} {\bibfnamefont
			{J.}~\bibnamefont {Sichelschmidt}}, \bibinfo {author} {\bibfnamefont
			{M.}~\bibnamefont {Baenitz}},\ and\ \bibinfo {author} {\bibfnamefont
			{R.}~\bibnamefont {Nath}},\ }\bibfield  {title} {\bibinfo {title} {{Absence
				of magnetic order and emergence of unconventional fluctuations in the
				${J}_{\mathrm{eff}}=\frac{1}{2}$ triangular-lattice antiferromagnet
				${\mathrm{YbBO}}_{3}$}},\ }\href
	{https://doi.org/10.1103/PhysRevB.107.064421} {\bibfield  {journal} {\bibinfo
			{journal} {Phys. Rev. B}\ }\textbf {\bibinfo {volume} {107}},\ \bibinfo
		{pages} {064421} (\bibinfo {year} {2023})}\BibitemShut {NoStop}%
	\bibitem [{\citenamefont {Lhotel}\ \emph {et~al.}(2021)\citenamefont {Lhotel},
		\citenamefont {Mangin-Thro}, \citenamefont {Ressouche}, \citenamefont
		{Steffens}, \citenamefont {Bichaud}, \citenamefont {Knebel}, \citenamefont
		{Brison}, \citenamefont {Marin}, \citenamefont {Raymond},\ and\ \citenamefont
		{Zhitomirsky}}]{Lhotel024427}%
	\BibitemOpen
	\bibfield  {author} {\bibinfo {author} {\bibfnamefont {E.}~\bibnamefont
			{Lhotel}}, \bibinfo {author} {\bibfnamefont {L.}~\bibnamefont {Mangin-Thro}},
		\bibinfo {author} {\bibfnamefont {E.}~\bibnamefont {Ressouche}}, \bibinfo
		{author} {\bibfnamefont {P.}~\bibnamefont {Steffens}}, \bibinfo {author}
		{\bibfnamefont {E.}~\bibnamefont {Bichaud}}, \bibinfo {author} {\bibfnamefont
			{G.}~\bibnamefont {Knebel}}, \bibinfo {author} {\bibfnamefont {J.-P.}\
			\bibnamefont {Brison}}, \bibinfo {author} {\bibfnamefont {C.}~\bibnamefont
			{Marin}}, \bibinfo {author} {\bibfnamefont {S.}~\bibnamefont {Raymond}},\
		and\ \bibinfo {author} {\bibfnamefont {M.~E.}\ \bibnamefont {Zhitomirsky}},\
	}\bibfield  {title} {\bibinfo {title} {{Spin dynamics of the quantum dipolar
				magnet ${\mathrm{Yb}}_{3}{\mathrm{Ga}}_{5}{\mathrm{O}}_{12}$ in an external
				field}},\ }\href {https://doi.org/10.1103/PhysRevB.104.024427} {\bibfield
		{journal} {\bibinfo  {journal} {Phys. Rev. B}\ }\textbf {\bibinfo {volume}
			{104}},\ \bibinfo {pages} {024427} (\bibinfo {year} {2021})}\BibitemShut
	{NoStop}%
	\bibitem [{\citenamefont {Guchhait}\ \emph {et~al.}(2025)\citenamefont
		{Guchhait}, \citenamefont {Kolay}, \citenamefont {Magar},\ and\ \citenamefont
		{Nath}}]{Guchhait214437}%
	\BibitemOpen
	\bibfield  {author} {\bibinfo {author} {\bibfnamefont {S.}~\bibnamefont
			{Guchhait}}, \bibinfo {author} {\bibfnamefont {R.}~\bibnamefont {Kolay}},
		\bibinfo {author} {\bibfnamefont {A.}~\bibnamefont {Magar}},\ and\ \bibinfo
		{author} {\bibfnamefont {R.}~\bibnamefont {Nath}},\ }\bibfield  {title}
	{\bibinfo {title} {{{Magnetic and crystal electric field studies of the
					Yb$^{3+}$-based triangular lattice antiferromagnets NaSrYb(BO$_3$)$_2$ and
					K$_3$YbSi$_2$O$_7$}}},\ }\href {https://doi.org/10.1103/ks6z-6nxj} {\bibfield
		{journal} {\bibinfo  {journal} {Phys. Rev. B}\ }\textbf {\bibinfo {volume}
			{111}},\ \bibinfo {pages} {214437} (\bibinfo {year} {2025})}\BibitemShut
	{NoStop}%
	\bibitem [{\citenamefont {Sebastian}\ \emph
		{et~al.}(2025{\natexlab{a}})\citenamefont {Sebastian}, \citenamefont {Kolay},
		\citenamefont {B}, \citenamefont {Ding}, \citenamefont {Furukawa},\ and\
		\citenamefont {Nath}}]{Sebastian104428}%
	\BibitemOpen
	\bibfield  {author} {\bibinfo {author} {\bibfnamefont {S.~J.}\ \bibnamefont
			{Sebastian}}, \bibinfo {author} {\bibfnamefont {R.}~\bibnamefont {Kolay}},
		\bibinfo {author} {\bibfnamefont {A.}~\bibnamefont {B}}, \bibinfo {author}
		{\bibfnamefont {Q.-P.}\ \bibnamefont {Ding}}, \bibinfo {author}
		{\bibfnamefont {Y.}~\bibnamefont {Furukawa}},\ and\ \bibinfo {author}
		{\bibfnamefont {R.}~\bibnamefont {Nath}},\ }\bibfield  {title} {\bibinfo
		{title} {{Spin fluctuations, absence of magnetic order, and crystal electric
				field studies in the ${\mathrm{Yb}}^{3+}$-based triangular lattice
				antiferromagnet ${\mathrm{Rb}}_{3}\mathrm{Yb}{({\mathrm{VO}}_{4})}_{2}$}},\
	}\href {https://doi.org/10.1103/ydpn-8wgf} {\bibfield  {journal} {\bibinfo
			{journal} {Phys. Rev. B}\ }\textbf {\bibinfo {volume} {112}},\ \bibinfo
		{pages} {104428} (\bibinfo {year} {2025}{\natexlab{a}})}\BibitemShut
	{NoStop}%
	\bibitem [{\citenamefont {Yahne}\ \emph {et~al.}(2020)\citenamefont {Yahne},
		\citenamefont {Sanjeewa}, \citenamefont {Sefat}, \citenamefont {Stadelman},
		\citenamefont {Kolis}, \citenamefont {Calder},\ and\ \citenamefont
		{Ross}}]{Yahne104423}%
	\BibitemOpen
	\bibfield  {author} {\bibinfo {author} {\bibfnamefont {D.~R.}\ \bibnamefont
			{Yahne}}, \bibinfo {author} {\bibfnamefont {L.~D.}\ \bibnamefont {Sanjeewa}},
		\bibinfo {author} {\bibfnamefont {A.~S.}\ \bibnamefont {Sefat}}, \bibinfo
		{author} {\bibfnamefont {B.~S.}\ \bibnamefont {Stadelman}}, \bibinfo {author}
		{\bibfnamefont {J.~W.}\ \bibnamefont {Kolis}}, \bibinfo {author}
		{\bibfnamefont {S.}~\bibnamefont {Calder}},\ and\ \bibinfo {author}
		{\bibfnamefont {K.~A.}\ \bibnamefont {Ross}},\ }\bibfield  {title} {\bibinfo
		{title} {{Pseudospin versus magnetic dipole moment ordering in the isosceles
				triangular lattice material
				${\mathrm{K}}_{3}{\mathrm{Er}(\mathrm{VO}}_{4}{)}_{2}$}},\ }\href
	{https://doi.org/10.1103/PhysRevB.102.104423} {\bibfield  {journal} {\bibinfo
			{journal} {Phys. Rev. B}\ }\textbf {\bibinfo {volume} {102}},\ \bibinfo
		{pages} {104423} (\bibinfo {year} {2020})}\BibitemShut {NoStop}%
	\bibitem [{\citenamefont {Scheie}\ \emph {et~al.}(2020)\citenamefont {Scheie},
		\citenamefont {Garlea}, \citenamefont {Sanjeewa}, \citenamefont {Xing},\ and\
		\citenamefont {Sefat}}]{Scheie144432}%
	\BibitemOpen
	\bibfield  {author} {\bibinfo {author} {\bibfnamefont {A.}~\bibnamefont
			{Scheie}}, \bibinfo {author} {\bibfnamefont {V.~O.}\ \bibnamefont {Garlea}},
		\bibinfo {author} {\bibfnamefont {L.~D.}\ \bibnamefont {Sanjeewa}}, \bibinfo
		{author} {\bibfnamefont {J.}~\bibnamefont {Xing}},\ and\ \bibinfo {author}
		{\bibfnamefont {A.~S.}\ \bibnamefont {Sefat}},\ }\bibfield  {title} {\bibinfo
		{title} {{Crystal-field Hamiltonian and anisotropy in ${\mathrm{KErSe}}_{2}$
				and ${\mathrm{CsErSe}}_{2}$}},\ }\href
	{https://doi.org/10.1103/PhysRevB.101.144432} {\bibfield  {journal} {\bibinfo
			{journal} {Phys. Rev. B}\ }\textbf {\bibinfo {volume} {101}},\ \bibinfo
		{pages} {144432} (\bibinfo {year} {2020})}\BibitemShut {NoStop}%
	\bibitem [{\citenamefont {Gaudet}\ \emph {et~al.}(2018)\citenamefont {Gaudet},
		\citenamefont {Hallas}, \citenamefont {Kolesnikov},\ and\ \citenamefont
		{Gaulin}}]{Gaudet024415}%
	\BibitemOpen
	\bibfield  {author} {\bibinfo {author} {\bibfnamefont {J.}~\bibnamefont
			{Gaudet}}, \bibinfo {author} {\bibfnamefont {A.~M.}\ \bibnamefont {Hallas}},
		\bibinfo {author} {\bibfnamefont {A.~I.}\ \bibnamefont {Kolesnikov}},\ and\
		\bibinfo {author} {\bibfnamefont {B.~D.}\ \bibnamefont {Gaulin}},\ }\bibfield
	{title} {\bibinfo {title} {{Effect of chemical pressure on the crystal
				electric field states of erbium pyrochlore magnets}},\ }\href
	{https://doi.org/10.1103/PhysRevB.97.024415} {\bibfield  {journal} {\bibinfo
			{journal} {Phys. Rev. B}\ }\textbf {\bibinfo {volume} {97}},\ \bibinfo
		{pages} {024415} (\bibinfo {year} {2018})}\BibitemShut {NoStop}%
	\bibitem [{\citenamefont {Champion}\ \emph {et~al.}(2003)\citenamefont
		{Champion}, \citenamefont {Harris}, \citenamefont {Holdsworth}, \citenamefont
		{Wills}, \citenamefont {Balakrishnan}, \citenamefont {Bramwell},
		\citenamefont {\ifmmode \check{C}\else \v{C}\fi{}i\ifmmode~\check{z}\else
			\v{z}\fi{}m\'ar}, \citenamefont {Fennell}, \citenamefont {Gardner},
		\citenamefont {Lago}, \citenamefont {McMorrow}, \citenamefont
		{Orend\'a\ifmmode~\check{c}\else \v{c}\fi{}}, \citenamefont
		{Orend\'a\ifmmode~\check{c}\else \v{c}\fi{}ov\'a}, \citenamefont {Paul},
		\citenamefont {Smith}, \citenamefont {Telling},\ and\ \citenamefont
		{Wildes}}]{Champion020401}%
	\BibitemOpen
	\bibfield  {author} {\bibinfo {author} {\bibfnamefont {J.~D.~M.}\
			\bibnamefont {Champion}}, \bibinfo {author} {\bibfnamefont {M.~J.}\
			\bibnamefont {Harris}}, \bibinfo {author} {\bibfnamefont {P.~C.~W.}\
			\bibnamefont {Holdsworth}}, \bibinfo {author} {\bibfnamefont {A.~S.}\
			\bibnamefont {Wills}}, \bibinfo {author} {\bibfnamefont {G.}~\bibnamefont
			{Balakrishnan}}, \bibinfo {author} {\bibfnamefont {S.~T.}\ \bibnamefont
			{Bramwell}}, \bibinfo {author} {\bibfnamefont {E.}~\bibnamefont {\ifmmode
				\check{C}\else \v{C}\fi{}i\ifmmode~\check{z}\else \v{z}\fi{}m\'ar}}, \bibinfo
		{author} {\bibfnamefont {T.}~\bibnamefont {Fennell}}, \bibinfo {author}
		{\bibfnamefont {J.~S.}\ \bibnamefont {Gardner}}, \bibinfo {author}
		{\bibfnamefont {J.}~\bibnamefont {Lago}}, \bibinfo {author} {\bibfnamefont
			{D.~F.}\ \bibnamefont {McMorrow}}, \bibinfo {author} {\bibfnamefont
			{M.}~\bibnamefont {Orend\'a\ifmmode~\check{c}\else \v{c}\fi{}}}, \bibinfo
		{author} {\bibfnamefont {A.}~\bibnamefont {Orend\'a\ifmmode~\check{c}\else
				\v{c}\fi{}ov\'a}}, \bibinfo {author} {\bibfnamefont {D.~M.}\ \bibnamefont
			{Paul}}, \bibinfo {author} {\bibfnamefont {R.~I.}\ \bibnamefont {Smith}},
		\bibinfo {author} {\bibfnamefont {M.~T.~F.}\ \bibnamefont {Telling}},\ and\
		\bibinfo {author} {\bibfnamefont {A.}~\bibnamefont {Wildes}},\ }\bibfield
	{title} {\bibinfo {title}
		{{${\mathrm{Er}}_{2}{\mathrm{Ti}}_{2}{\mathrm{O}}_{7}:$ Evidence of quantum
				order by disorder in a frustrated antiferromagnet}},\ }\href
	{https://doi.org/10.1103/PhysRevB.68.020401} {\bibfield  {journal} {\bibinfo
			{journal} {Phys. Rev. B}\ }\textbf {\bibinfo {volume} {68}},\ \bibinfo
		{pages} {020401} (\bibinfo {year} {2003})}\BibitemShut {NoStop}%
	\bibitem [{\citenamefont {Cai}\ \emph {et~al.}(2020)\citenamefont {Cai},
		\citenamefont {Lygouras}, \citenamefont {Thomas}, \citenamefont {Wilson},
		\citenamefont {Beare}, \citenamefont {Sharma}, \citenamefont {Marjerrison},
		\citenamefont {Yahne}, \citenamefont {Ross}, \citenamefont {Gong},
		\citenamefont {Uemura}, \citenamefont {Dabkowska},\ and\ \citenamefont
		{Luke}}]{Cai094432}%
	\BibitemOpen
	\bibfield  {author} {\bibinfo {author} {\bibfnamefont {Y.}~\bibnamefont
			{Cai}}, \bibinfo {author} {\bibfnamefont {C.}~\bibnamefont {Lygouras}},
		\bibinfo {author} {\bibfnamefont {G.}~\bibnamefont {Thomas}}, \bibinfo
		{author} {\bibfnamefont {M.~N.}\ \bibnamefont {Wilson}}, \bibinfo {author}
		{\bibfnamefont {J.}~\bibnamefont {Beare}}, \bibinfo {author} {\bibfnamefont
			{S.}~\bibnamefont {Sharma}}, \bibinfo {author} {\bibfnamefont {C.~A.}\
			\bibnamefont {Marjerrison}}, \bibinfo {author} {\bibfnamefont {D.~R.}\
			\bibnamefont {Yahne}}, \bibinfo {author} {\bibfnamefont {K.~A.}\ \bibnamefont
			{Ross}}, \bibinfo {author} {\bibfnamefont {Z.}~\bibnamefont {Gong}}, \bibinfo
		{author} {\bibfnamefont {Y.~J.}\ \bibnamefont {Uemura}}, \bibinfo {author}
		{\bibfnamefont {H.~A.}\ \bibnamefont {Dabkowska}},\ and\ \bibinfo {author}
		{\bibfnamefont {G.~M.}\ \bibnamefont {Luke}},\ }\bibfield  {title} {\bibinfo
		{title} {{$\ensuremath{\mu}\mathrm{SR}$ study of the triangular Ising
				antiferromagnet ${\mathrm{ErMgGaO}}_{4}$}},\ }\href
	{https://doi.org/10.1103/PhysRevB.101.094432} {\bibfield  {journal} {\bibinfo
			{journal} {Phys. Rev. B}\ }\textbf {\bibinfo {volume} {101}},\ \bibinfo
		{pages} {094432} (\bibinfo {year} {2020})}\BibitemShut {NoStop}%
	\bibitem [{\citenamefont {Cai}\ \emph {et~al.}(2019)\citenamefont {Cai},
		\citenamefont {Wilson}, \citenamefont {Beare}, \citenamefont {Lygouras},
		\citenamefont {Thomas}, \citenamefont {Yahne}, \citenamefont {Ross},
		\citenamefont {Taddei}, \citenamefont {Sala}, \citenamefont {Dabkowska},
		\citenamefont {Aczel},\ and\ \citenamefont {Luke}}]{Cai184415}%
	\BibitemOpen
	\bibfield  {author} {\bibinfo {author} {\bibfnamefont {Y.}~\bibnamefont
			{Cai}}, \bibinfo {author} {\bibfnamefont {M.~N.}\ \bibnamefont {Wilson}},
		\bibinfo {author} {\bibfnamefont {J.}~\bibnamefont {Beare}}, \bibinfo
		{author} {\bibfnamefont {C.}~\bibnamefont {Lygouras}}, \bibinfo {author}
		{\bibfnamefont {G.}~\bibnamefont {Thomas}}, \bibinfo {author} {\bibfnamefont
			{D.~R.}\ \bibnamefont {Yahne}}, \bibinfo {author} {\bibfnamefont
			{K.}~\bibnamefont {Ross}}, \bibinfo {author} {\bibfnamefont {K.~M.}\
			\bibnamefont {Taddei}}, \bibinfo {author} {\bibfnamefont {G.}~\bibnamefont
			{Sala}}, \bibinfo {author} {\bibfnamefont {H.~A.}\ \bibnamefont {Dabkowska}},
		\bibinfo {author} {\bibfnamefont {A.~A.}\ \bibnamefont {Aczel}},\ and\
		\bibinfo {author} {\bibfnamefont {G.~M.}\ \bibnamefont {Luke}},\ }\bibfield
	{title} {\bibinfo {title} {{Crystal fields and magnetic structure of the
				Ising antiferromagnet
				${\mathrm{Er}}_{3}{\mathrm{Ga}}_{5}{\mathrm{O}}_{12}$}},\ }\href
	{https://doi.org/10.1103/PhysRevB.100.184415} {\bibfield  {journal} {\bibinfo
			{journal} {Phys. Rev. B}\ }\textbf {\bibinfo {volume} {100}},\ \bibinfo
		{pages} {184415} (\bibinfo {year} {2019})}\BibitemShut {NoStop}%
	\bibitem [{\citenamefont {Xing}\ \emph {et~al.}(2019)\citenamefont {Xing},
		\citenamefont {Sanjeewa}, \citenamefont {Kim}, \citenamefont {Meier},
		\citenamefont {May}, \citenamefont {Zheng}, \citenamefont {Custelcean},
		\citenamefont {Stewart},\ and\ \citenamefont {Sefat}}]{Xing114413}%
	\BibitemOpen
	\bibfield  {author} {\bibinfo {author} {\bibfnamefont {J.}~\bibnamefont
			{Xing}}, \bibinfo {author} {\bibfnamefont {L.~D.}\ \bibnamefont {Sanjeewa}},
		\bibinfo {author} {\bibfnamefont {J.}~\bibnamefont {Kim}}, \bibinfo {author}
		{\bibfnamefont {W.~R.}\ \bibnamefont {Meier}}, \bibinfo {author}
		{\bibfnamefont {A.~F.}\ \bibnamefont {May}}, \bibinfo {author} {\bibfnamefont
			{Q.}~\bibnamefont {Zheng}}, \bibinfo {author} {\bibfnamefont
			{R.}~\bibnamefont {Custelcean}}, \bibinfo {author} {\bibfnamefont {G.~R.}\
			\bibnamefont {Stewart}},\ and\ \bibinfo {author} {\bibfnamefont {A.~S.}\
			\bibnamefont {Sefat}},\ }\bibfield  {title} {\bibinfo {title} {{Synthesis,
				magnetization, and heat capacity of triangular lattice materials
				${\mathrm{NaErSe}}_{2}$ and ${\mathrm{KErSe}}_{2}$}},\ }\href
	{https://doi.org/10.1103/PhysRevMaterials.3.114413} {\bibfield  {journal}
		{\bibinfo  {journal} {Phys. Rev. Mater.}\ }\textbf {\bibinfo {volume} {3}},\
		\bibinfo {pages} {114413} (\bibinfo {year} {2019})}\BibitemShut {NoStop}%
	\bibitem [{\citenamefont {Xing}\ \emph {et~al.}(2021)\citenamefont {Xing},
		\citenamefont {Taddei}, \citenamefont {Sanjeewa}, \citenamefont {Fishman},
		\citenamefont {Daum}, \citenamefont {Mourigal}, \citenamefont {dela Cruz},\
		and\ \citenamefont {Sefat}}]{Xing144413}%
	\BibitemOpen
	\bibfield  {author} {\bibinfo {author} {\bibfnamefont {J.}~\bibnamefont
			{Xing}}, \bibinfo {author} {\bibfnamefont {K.~M.}\ \bibnamefont {Taddei}},
		\bibinfo {author} {\bibfnamefont {L.~D.}\ \bibnamefont {Sanjeewa}}, \bibinfo
		{author} {\bibfnamefont {R.~S.}\ \bibnamefont {Fishman}}, \bibinfo {author}
		{\bibfnamefont {M.}~\bibnamefont {Daum}}, \bibinfo {author} {\bibfnamefont
			{M.}~\bibnamefont {Mourigal}}, \bibinfo {author} {\bibfnamefont
			{C.}~\bibnamefont {dela Cruz}},\ and\ \bibinfo {author} {\bibfnamefont
			{A.~S.}\ \bibnamefont {Sefat}},\ }\bibfield  {title} {\bibinfo {title}
		{{Stripe antiferromagnetic ground state of the ideal triangular lattice
				compound ${\mathrm{KErSe}}_{2}$}},\ }\href
	{https://doi.org/10.1103/PhysRevB.103.144413} {\bibfield  {journal} {\bibinfo
			{journal} {Phys. Rev. B}\ }\textbf {\bibinfo {volume} {103}},\ \bibinfo
		{pages} {144413} (\bibinfo {year} {2021})}\BibitemShut {NoStop}%
	\bibitem [{\citenamefont {Ding}\ \emph {et~al.}(2023)\citenamefont {Ding},
		\citenamefont {Wo}, \citenamefont {Luo}, \citenamefont {Gu}, \citenamefont
		{Gu}, \citenamefont {Bewley}, \citenamefont {Chen},\ and\ \citenamefont
		{Zhao}}]{DingL100411}%
	\BibitemOpen
	\bibfield  {author} {\bibinfo {author} {\bibfnamefont {G.}~\bibnamefont
			{Ding}}, \bibinfo {author} {\bibfnamefont {H.}~\bibnamefont {Wo}}, \bibinfo
		{author} {\bibfnamefont {R.~L.}\ \bibnamefont {Luo}}, \bibinfo {author}
		{\bibfnamefont {Y.}~\bibnamefont {Gu}}, \bibinfo {author} {\bibfnamefont
			{Y.}~\bibnamefont {Gu}}, \bibinfo {author} {\bibfnamefont {R.}~\bibnamefont
			{Bewley}}, \bibinfo {author} {\bibfnamefont {G.}~\bibnamefont {Chen}},\ and\
		\bibinfo {author} {\bibfnamefont {J.}~\bibnamefont {Zhao}},\ }\bibfield
	{title} {\bibinfo {title} {{Stripe order and spin dynamics in the
				triangular-lattice antiferromagnet ${\mathrm{KErSe}}_{2}$: A single-crystal
				study with a theoretical description}},\ }\href
	{https://doi.org/10.1103/PhysRevB.107.L100411} {\bibfield  {journal}
		{\bibinfo  {journal} {Phys. Rev. B}\ }\textbf {\bibinfo {volume} {107}},\
		\bibinfo {pages} {L100411} (\bibinfo {year} {2023})}\BibitemShut {NoStop}%
	\bibitem [{\citenamefont {Xie}\ \emph {et~al.}(2024)\citenamefont {Xie},
		\citenamefont {Zhuo}, \citenamefont {Cai}, \citenamefont {Zhang},\ and\
		\citenamefont {Zhang}}]{Xie117505}%
	\BibitemOpen
	\bibfield  {author} {\bibinfo {author} {\bibfnamefont {M.}~\bibnamefont
			{Xie}}, \bibinfo {author} {\bibfnamefont {W.}~\bibnamefont {Zhuo}}, \bibinfo
		{author} {\bibfnamefont {Y.}~\bibnamefont {Cai}}, \bibinfo {author}
		{\bibfnamefont {Z.}~\bibnamefont {Zhang}},\ and\ \bibinfo {author}
		{\bibfnamefont {Q.}~\bibnamefont {Zhang}},\ }\bibfield  {title} {\bibinfo
		{title} {{Rare-Earth Chalcogenides: An Inspiring Playground for Exploring
				Frustrated Magnetism}},\ }\href
	{https://doi.org/10.1088/0256-307X/41/11/117505} {\bibfield  {journal}
		{\bibinfo  {journal} {Chin. Phys. Lett.}\ }\textbf {\bibinfo {volume} {41}},\
		\bibinfo {pages} {117505} (\bibinfo {year} {2024})}\BibitemShut {NoStop}%
	\bibitem [{\citenamefont {Liu}\ \emph {et~al.}(2024)\citenamefont {Liu},
		\citenamefont {Zhang}, \citenamefont {Yan}, \citenamefont {Li}, \citenamefont
		{Zhang}, \citenamefont {Ji}, \citenamefont {Jin}, \citenamefont {Shi},\ and\
		\citenamefont {Zhang}}]{Liu10}%
	\BibitemOpen
	\bibfield  {author} {\bibinfo {author} {\bibfnamefont {W.}~\bibnamefont
			{Liu}}, \bibinfo {author} {\bibfnamefont {Z.}~\bibnamefont {Zhang}}, \bibinfo
		{author} {\bibfnamefont {D.}~\bibnamefont {Yan}}, \bibinfo {author}
		{\bibfnamefont {J.}~\bibnamefont {Li}}, \bibinfo {author} {\bibfnamefont
			{Z.}~\bibnamefont {Zhang}}, \bibinfo {author} {\bibfnamefont
			{J.}~\bibnamefont {Ji}}, \bibinfo {author} {\bibfnamefont {F.}~\bibnamefont
			{Jin}}, \bibinfo {author} {\bibfnamefont {Y.}~\bibnamefont {Shi}},\ and\
		\bibinfo {author} {\bibfnamefont {Q.}~\bibnamefont {Zhang}},\ }\bibfield
	{title} {\bibinfo {title} {{Finite Temperature Magnetism in the Triangular
				Lattice Antiferromagnet KErTe$_2$}},\ }\href
	{https://doi.org/10.1088/0256-307X/41/9/097503} {\bibfield  {journal}
		{\bibinfo  {journal} {Chin. Phys. Lett.}\ }\textbf {\bibinfo {volume} {41}},\
		\bibinfo {pages} {097503} (\bibinfo {year} {2024})}\BibitemShut {NoStop}%
	\bibitem [{\citenamefont {Liu}\ \emph {et~al.}()\citenamefont {Liu},
		\citenamefont {Zhang}, \citenamefont {Yan}, \citenamefont {Li}, \citenamefont
		{Zhang}, \citenamefont {Ji}, \citenamefont {Jin}, \citenamefont {Shi},\ and\
		\citenamefont {Zhang}}]{Liu2021}%
	\BibitemOpen
	\bibfield  {author} {\bibinfo {author} {\bibfnamefont {W.}~\bibnamefont
			{Liu}}, \bibinfo {author} {\bibfnamefont {Z.}~\bibnamefont {Zhang}}, \bibinfo
		{author} {\bibfnamefont {D.}~\bibnamefont {Yan}}, \bibinfo {author}
		{\bibfnamefont {J.}~\bibnamefont {Li}}, \bibinfo {author} {\bibfnamefont
			{Z.}~\bibnamefont {Zhang}}, \bibinfo {author} {\bibfnamefont
			{J.}~\bibnamefont {Ji}}, \bibinfo {author} {\bibfnamefont {F.}~\bibnamefont
			{Jin}}, \bibinfo {author} {\bibfnamefont {Y.}~\bibnamefont {Shi}},\ and\
		\bibinfo {author} {\bibfnamefont {Q.}~\bibnamefont {Zhang}},\ }\href@noop {}
	{\bibinfo {title} {{{Effects of the Crystalline Electric Field in the
					$KErTe_{2}$ Quantum Spin Liquid Candidate}}}},\ \Eprint
	{https://arxiv.org/abs/2108.09693} {arXiv:2108.09693} \BibitemShut {NoStop}%
	\bibitem [{\citenamefont {Liu}\ \emph {et~al.}(2021)\citenamefont {Liu},
		\citenamefont {Yan}, \citenamefont {Zhang}, \citenamefont {Ji}, \citenamefont
		{Shi}, \citenamefont {Jin},\ and\ \citenamefont {Zhang}}]{Liu107504}%
	\BibitemOpen
	\bibfield  {author} {\bibinfo {author} {\bibfnamefont {W.}~\bibnamefont
			{Liu}}, \bibinfo {author} {\bibfnamefont {D.}~\bibnamefont {Yan}}, \bibinfo
		{author} {\bibfnamefont {Z.}~\bibnamefont {Zhang}}, \bibinfo {author}
		{\bibfnamefont {J.}~\bibnamefont {Ji}}, \bibinfo {author} {\bibfnamefont
			{Y.}~\bibnamefont {Shi}}, \bibinfo {author} {\bibfnamefont {F.}~\bibnamefont
			{Jin}},\ and\ \bibinfo {author} {\bibfnamefont {Q.}~\bibnamefont {Zhang}},\
	}\bibfield  {title} {\bibinfo {title} {{Crystal growth and magnetic
				properties of quantum spin liquid candidate KErTe$_2$}},\ }\href
	{https://doi.org/10.1088/1674-1056/ac1574} {\bibfield  {journal} {\bibinfo
			{journal} {Chin. Phys. B.}\ }\textbf {\bibinfo {volume} {30}},\ \bibinfo
		{pages} {107504} (\bibinfo {year} {2021})}\BibitemShut {NoStop}%
	\bibitem [{\citenamefont {Whitelock}\ \emph {et~al.}()\citenamefont
		{Whitelock}, \citenamefont {Scheie}, \citenamefont {McMaster}, \citenamefont
		{Leahy}, \citenamefont {Xiang}, \citenamefont {Ozerov}, \citenamefont
		{Smirnov}, \citenamefont {Choi}, \citenamefont {dela Cruz}, \citenamefont
		{Ajeesh}, \citenamefont {Krakovsky}, \citenamefont {Rehn}, \citenamefont
		{Xing}, \citenamefont {Sefat},\ and\ \citenamefont {Lee}}]{whitelock2025}%
	\BibitemOpen
	\bibfield  {author} {\bibinfo {author} {\bibfnamefont {H.}~\bibnamefont
			{Whitelock}}, \bibinfo {author} {\bibfnamefont {A.~O.}\ \bibnamefont
			{Scheie}}, \bibinfo {author} {\bibfnamefont {M.}~\bibnamefont {McMaster}},
		\bibinfo {author} {\bibfnamefont {I.~A.}\ \bibnamefont {Leahy}}, \bibinfo
		{author} {\bibfnamefont {L.}~\bibnamefont {Xiang}}, \bibinfo {author}
		{\bibfnamefont {M.}~\bibnamefont {Ozerov}}, \bibinfo {author} {\bibfnamefont
			{D.}~\bibnamefont {Smirnov}}, \bibinfo {author} {\bibfnamefont {E.~S.}\
			\bibnamefont {Choi}}, \bibinfo {author} {\bibfnamefont {C.}~\bibnamefont
			{dela Cruz}}, \bibinfo {author} {\bibfnamefont {M.~O.}\ \bibnamefont
			{Ajeesh}}, \bibinfo {author} {\bibfnamefont {E.~S.}\ \bibnamefont
			{Krakovsky}}, \bibinfo {author} {\bibfnamefont {D.~A.}\ \bibnamefont {Rehn}},
		\bibinfo {author} {\bibfnamefont {J.}~\bibnamefont {Xing}}, \bibinfo {author}
		{\bibfnamefont {A.~S.}\ \bibnamefont {Sefat}},\ and\ \bibinfo {author}
		{\bibfnamefont {M.}~\bibnamefont {Lee}},\ }\href@noop {} {\bibinfo {title}
		{{Uncovering field-induced magnetic phase transition by direct observation of
				the crystal electric-field splitting in a rare-earth magnetic insulator}}},\
	\Eprint {https://arxiv.org/abs/2510.21616} {arXiv:2510.21616} \BibitemShut
	{NoStop}%
	\bibitem [{\citenamefont {Hester}\ \emph
		{et~al.}(2021{\natexlab{a}})\citenamefont {Hester}, \citenamefont {DeLazzer},
		\citenamefont {Lim}, \citenamefont {Brown},\ and\ \citenamefont
		{Ross}}]{Hester125804}%
	\BibitemOpen
	\bibfield  {author} {\bibinfo {author} {\bibfnamefont {G.}~\bibnamefont
			{Hester}}, \bibinfo {author} {\bibfnamefont {T.~N.}\ \bibnamefont
			{DeLazzer}}, \bibinfo {author} {\bibfnamefont {S.~S.}\ \bibnamefont {Lim}},
		\bibinfo {author} {\bibfnamefont {C.~M.}\ \bibnamefont {Brown}},\ and\
		\bibinfo {author} {\bibfnamefont {K.~A.}\ \bibnamefont {Ross}},\ }\bibfield
	{title} {\bibinfo {title} {{Néel ordering in the distorted honeycomb
				pyrosilicate: C–Er$_2$Si$_2$O$_7$}},\ }\href
	{https://doi.org/10.1088/1361-648X/abd5f8} {\bibfield  {journal} {\bibinfo
			{journal} {J. Phys. Condens. Matter}\ }\textbf {\bibinfo {volume} {33}},\
		\bibinfo {pages} {125804} (\bibinfo {year} {2021}{\natexlab{a}})}\BibitemShut
	{NoStop}%
	\bibitem [{\citenamefont {Islam}\ \emph {et~al.}(2024)\citenamefont {Islam},
		\citenamefont {d'Ambrumenil}, \citenamefont {Khalyavin}, \citenamefont
		{Manuel}, \citenamefont {Orlandi}, \citenamefont {Ollivier}, \citenamefont
		{Ciomaga~Hatnean}, \citenamefont {Balakrishnan},\ and\ \citenamefont
		{Petrenko}}]{Islam094420}%
	\BibitemOpen
	\bibfield  {author} {\bibinfo {author} {\bibfnamefont {M.}~\bibnamefont
			{Islam}}, \bibinfo {author} {\bibfnamefont {N.}~\bibnamefont {d'Ambrumenil}},
		\bibinfo {author} {\bibfnamefont {D.~D.}\ \bibnamefont {Khalyavin}}, \bibinfo
		{author} {\bibfnamefont {P.}~\bibnamefont {Manuel}}, \bibinfo {author}
		{\bibfnamefont {F.}~\bibnamefont {Orlandi}}, \bibinfo {author} {\bibfnamefont
			{J.}~\bibnamefont {Ollivier}}, \bibinfo {author} {\bibfnamefont
			{M.}~\bibnamefont {Ciomaga~Hatnean}}, \bibinfo {author} {\bibfnamefont
			{G.}~\bibnamefont {Balakrishnan}},\ and\ \bibinfo {author} {\bibfnamefont
			{O.~A.}\ \bibnamefont {Petrenko}},\ }\bibfield  {title} {\bibinfo {title}
		{{Magnetic structure, excitations, and field-induced transitions in the
				honeycomb lattice compound
				${\mathrm{Er}}_{2}{\mathrm{Si}}_{2}{\mathrm{O}}_{7}$}},\ }\href
	{https://doi.org/10.1103/PhysRevB.109.094420} {\bibfield  {journal} {\bibinfo
			{journal} {Phys. Rev. B}\ }\textbf {\bibinfo {volume} {109}},\ \bibinfo
		{pages} {094420} (\bibinfo {year} {2024})}\BibitemShut {NoStop}%
	\bibitem [{\citenamefont {Hester}\ \emph
		{et~al.}(2021{\natexlab{b}})\citenamefont {Hester}, \citenamefont {DeLazzer},
		\citenamefont {Yahne}, \citenamefont {Sarkis}, \citenamefont {Zhao},
		\citenamefont {Rodriguez~Rivera}, \citenamefont {Calder},\ and\ \citenamefont
		{Ross}}]{Hester405801}%
	\BibitemOpen
	\bibfield  {author} {\bibinfo {author} {\bibfnamefont {G.}~\bibnamefont
			{Hester}}, \bibinfo {author} {\bibfnamefont {T.~N.}\ \bibnamefont
			{DeLazzer}}, \bibinfo {author} {\bibfnamefont {D.~R.}\ \bibnamefont {Yahne}},
		\bibinfo {author} {\bibfnamefont {C.~L.}\ \bibnamefont {Sarkis}}, \bibinfo
		{author} {\bibfnamefont {H.~D.}\ \bibnamefont {Zhao}}, \bibinfo {author}
		{\bibfnamefont {J.~A.}\ \bibnamefont {Rodriguez~Rivera}}, \bibinfo {author}
		{\bibfnamefont {S.}~\bibnamefont {Calder}},\ and\ \bibinfo {author}
		{\bibfnamefont {K.~A.}\ \bibnamefont {Ross}},\ }\bibfield  {title} {\bibinfo
		{title} {{Magnetic properties of the Ising-like rare earth pyrosilicate:
				D-${\mathrm{Er}}_{2}{\mathrm{Si}}_{2}{\mathrm{O}}_{7}$}},\ }\href
	{https://doi.org/10.1088/1361-648X/ac136a} {\bibfield  {journal} {\bibinfo
			{journal} {J. Condens. Matter Phys.}\ }\textbf {\bibinfo {volume} {33}},\
		\bibinfo {pages} {405801} (\bibinfo {year} {2021}{\natexlab{b}})}\BibitemShut
	{NoStop}%
	\bibitem [{\citenamefont {Kitaev}(2006)}]{Kitaev2}%
	\BibitemOpen
	\bibfield  {author} {\bibinfo {author} {\bibfnamefont {A.}~\bibnamefont
			{Kitaev}},\ }\bibfield  {title} {\bibinfo {title} {Anyons in an exactly
			solved model and beyond},\ }\href
	{https://doi.org/https://doi.org/10.1016/j.aop.2005.10.005} {\bibfield
		{journal} {\bibinfo  {journal} {Annals of Physics}\ }\textbf {\bibinfo
			{volume} {321}},\ \bibinfo {pages} {2} (\bibinfo {year} {2006})}\BibitemShut
	{NoStop}%
	\bibitem [{\citenamefont {Trebst}\ and\ \citenamefont
		{Hickey}(2022)}]{Trebst1}%
	\BibitemOpen
	\bibfield  {author} {\bibinfo {author} {\bibfnamefont {S.}~\bibnamefont
			{Trebst}}\ and\ \bibinfo {author} {\bibfnamefont {C.}~\bibnamefont
			{Hickey}},\ }\bibfield  {title} {\bibinfo {title} {Kitaev materials},\ }\href
	{https://doi.org/https://doi.org/10.1016/j.physrep.2021.11.003} {\bibfield
		{journal} {\bibinfo  {journal} {Phys. Rep.}\ }\textbf {\bibinfo {volume}
			{950}},\ \bibinfo {pages} {1} (\bibinfo {year} {2022})}\BibitemShut {NoStop}%
	\bibitem [{\citenamefont {Hermanns}\ \emph {et~al.}(2018)\citenamefont
		{Hermanns}, \citenamefont {Kimchi},\ and\ \citenamefont
		{Knolle}}]{Hermanns17}%
	\BibitemOpen
	\bibfield  {author} {\bibinfo {author} {\bibfnamefont {M.}~\bibnamefont
			{Hermanns}}, \bibinfo {author} {\bibfnamefont {I.}~\bibnamefont {Kimchi}},\
		and\ \bibinfo {author} {\bibfnamefont {J.}~\bibnamefont {Knolle}},\
	}\bibfield  {title} {\bibinfo {title} {Physics of the kitaev model:
			Fractionalization, dynamical correlations, and material connections},\ }\href
	{https://doi.org/10.1146/annurev-conmatphys-033117-053934} {\bibfield
		{journal} {\bibinfo  {journal} {Annu. Rev. Condens. Matter Phys.}\ }\textbf
		{\bibinfo {volume} {9}},\ \bibinfo {pages} {17} (\bibinfo {year}
		{2018})}\BibitemShut {NoStop}%
	\bibitem [{\citenamefont {Mohanty}\ \emph {et~al.}(2023)\citenamefont
		{Mohanty}, \citenamefont {Islam}, \citenamefont {Winterhalter-Stocker},
		\citenamefont {Jesche}, \citenamefont {Simutis}, \citenamefont {Wang},
		\citenamefont {Guguchia}, \citenamefont {Sichelschmidt}, \citenamefont
		{Baenitz}, \citenamefont {Tsirlin}, \citenamefont {Gegenwart},\ and\
		\citenamefont {Nath}}]{Mohanty134408}%
	\BibitemOpen
	\bibfield  {author} {\bibinfo {author} {\bibfnamefont {S.}~\bibnamefont
			{Mohanty}}, \bibinfo {author} {\bibfnamefont {S.~S.}\ \bibnamefont {Islam}},
		\bibinfo {author} {\bibfnamefont {N.}~\bibnamefont {Winterhalter-Stocker}},
		\bibinfo {author} {\bibfnamefont {A.}~\bibnamefont {Jesche}}, \bibinfo
		{author} {\bibfnamefont {G.}~\bibnamefont {Simutis}}, \bibinfo {author}
		{\bibfnamefont {C.}~\bibnamefont {Wang}}, \bibinfo {author} {\bibfnamefont
			{Z.}~\bibnamefont {Guguchia}}, \bibinfo {author} {\bibfnamefont
			{J.}~\bibnamefont {Sichelschmidt}}, \bibinfo {author} {\bibfnamefont
			{M.}~\bibnamefont {Baenitz}}, \bibinfo {author} {\bibfnamefont {A.~A.}\
			\bibnamefont {Tsirlin}}, \bibinfo {author} {\bibfnamefont {P.}~\bibnamefont
			{Gegenwart}},\ and\ \bibinfo {author} {\bibfnamefont {R.}~\bibnamefont
			{Nath}},\ }\bibfield  {title} {\bibinfo {title} {{Disordered ground state in
				the spin-orbit coupled ${J}_{\mathrm{eff}}$ = $\frac{1}{2}$ distorted
				honeycomb magnet ${\mathrm{BiYbGeO}}_{5}$}},\ }\href
	{https://doi.org/10.1103/PhysRevB.108.134408} {\bibfield  {journal} {\bibinfo
			{journal} {Phys. Rev. B}\ }\textbf {\bibinfo {volume} {108}},\ \bibinfo
		{pages} {134408} (\bibinfo {year} {2023})}\BibitemShut {NoStop}%
	\bibitem [{\citenamefont {Carvajal}(1993)}]{Carvajal55}%
	\BibitemOpen
	\bibfield  {author} {\bibinfo {author} {\bibfnamefont {J.~R.}\ \bibnamefont
			{Carvajal}},\ }\bibfield  {title} {\bibinfo {title} {Recent advances in
			magnetic structure determination by neutron powder diffraction},\ }\href
	{https://doi.org/https://doi.org/10.1016/0921-4526(93)90108-I} {\bibfield
		{journal} {\bibinfo  {journal} {Physica B: Condensed Matter}\ }\textbf
		{\bibinfo {volume} {192}},\ \bibinfo {pages} {55} (\bibinfo {year}
		{1993})}\BibitemShut {NoStop}%
	\bibitem [{\citenamefont {Cascales}\ and\ \citenamefont
		{Zaldo}(2003)}]{Cascales262}%
	\BibitemOpen
	\bibfield  {author} {\bibinfo {author} {\bibfnamefont {C.}~\bibnamefont
			{Cascales}}\ and\ \bibinfo {author} {\bibfnamefont {C.}~\bibnamefont
			{Zaldo}},\ }\bibfield  {title} {\bibinfo {title} {{Crystal-field analysis of
				Eu$^{3+}$ energy levels in the new rare-earth RBiY$_{1-x}$R$_x$GeO$_5$
				oxide}},\ }\href {https://doi.org/10.1016/S0022-4596(02)00173-1} {\bibfield
		{journal} {\bibinfo  {journal} {J. Solid State Chem.}\ }\textbf {\bibinfo
			{volume} {171}},\ \bibinfo {pages} {262} (\bibinfo {year}
		{2003})}\BibitemShut {NoStop}%
	\bibitem [{\citenamefont {Le}\ \emph {et~al.}(2023)\citenamefont {Le},
		\citenamefont {Guidi}, \citenamefont {Bewley}, \citenamefont {Stewart},
		\citenamefont {Schooneveld}, \citenamefont {Raspino}, \citenamefont {Pooley},
		\citenamefont {Boxall}, \citenamefont {Gascoyne}, \citenamefont {Rhodes},
		\citenamefont {Moorby}, \citenamefont {Templeman}, \citenamefont {Afford},
		\citenamefont {Waller}, \citenamefont {Zacek},\ and\ \citenamefont
		{Shaw}}]{Le168646}%
	\BibitemOpen
	\bibfield  {author} {\bibinfo {author} {\bibfnamefont {M.}~\bibnamefont
			{Le}}, \bibinfo {author} {\bibfnamefont {T.}~\bibnamefont {Guidi}}, \bibinfo
		{author} {\bibfnamefont {R.}~\bibnamefont {Bewley}}, \bibinfo {author}
		{\bibfnamefont {J.}~\bibnamefont {Stewart}}, \bibinfo {author} {\bibfnamefont
			{E.}~\bibnamefont {Schooneveld}}, \bibinfo {author} {\bibfnamefont
			{D.}~\bibnamefont {Raspino}}, \bibinfo {author} {\bibfnamefont
			{D.}~\bibnamefont {Pooley}}, \bibinfo {author} {\bibfnamefont
			{J.}~\bibnamefont {Boxall}}, \bibinfo {author} {\bibfnamefont
			{K.}~\bibnamefont {Gascoyne}}, \bibinfo {author} {\bibfnamefont
			{N.}~\bibnamefont {Rhodes}}, \bibinfo {author} {\bibfnamefont
			{S.}~\bibnamefont {Moorby}}, \bibinfo {author} {\bibfnamefont
			{D.}~\bibnamefont {Templeman}}, \bibinfo {author} {\bibfnamefont
			{L.}~\bibnamefont {Afford}}, \bibinfo {author} {\bibfnamefont
			{S.}~\bibnamefont {Waller}}, \bibinfo {author} {\bibfnamefont
			{D.}~\bibnamefont {Zacek}},\ and\ \bibinfo {author} {\bibfnamefont
			{R.}~\bibnamefont {Shaw}},\ }\bibfield  {title} {\bibinfo {title} {Upgrade of
			the mari spectrometer at isis},\ }\href
	{https://doi.org/https://doi.org/10.1016/j.nima.2023.168646} {\bibfield
		{journal} {\bibinfo  {journal} {Nuclear Instruments and Methods in Physics
				Research Section A: Accelerators, Spectrometers, Detectors and Associated
				Equipment}\ }\textbf {\bibinfo {volume} {1056}},\ \bibinfo {pages} {168646}
		(\bibinfo {year} {2023})}\BibitemShut {NoStop}%
	\bibitem [{\citenamefont {Arnold}\ \emph {et~al.}(2014)\citenamefont {Arnold},
		\citenamefont {Bilheux}, \citenamefont {Borreguero}, \citenamefont {Buts},
		\citenamefont {Campbell}, \citenamefont {Chapon}, \citenamefont {Doucet},
		\citenamefont {Draper}, \citenamefont {{Ferraz Leal}}, \citenamefont {Gigg},
		\citenamefont {Lynch}, \citenamefont {Markvardsen}, \citenamefont
		{Mikkelson}, \citenamefont {Mikkelson}, \citenamefont {Miller}, \citenamefont
		{Palmen}, \citenamefont {Parker}, \citenamefont {Passos}, \citenamefont
		{Perring}, \citenamefont {Peterson}, \citenamefont {Ren}, \citenamefont
		{Reuter}, \citenamefont {Savici}, \citenamefont {Taylor}, \citenamefont
		{Taylor}, \citenamefont {Tolchenov}, \citenamefont {Zhou},\ and\
		\citenamefont {Zikovsky}}]{Arnold156}%
	\BibitemOpen
	\bibfield  {author} {\bibinfo {author} {\bibfnamefont {O.}~\bibnamefont
			{Arnold}}, \bibinfo {author} {\bibfnamefont {J.}~\bibnamefont {Bilheux}},
		\bibinfo {author} {\bibfnamefont {J.}~\bibnamefont {Borreguero}}, \bibinfo
		{author} {\bibfnamefont {A.}~\bibnamefont {Buts}}, \bibinfo {author}
		{\bibfnamefont {S.}~\bibnamefont {Campbell}}, \bibinfo {author}
		{\bibfnamefont {L.}~\bibnamefont {Chapon}}, \bibinfo {author} {\bibfnamefont
			{M.}~\bibnamefont {Doucet}}, \bibinfo {author} {\bibfnamefont
			{N.}~\bibnamefont {Draper}}, \bibinfo {author} {\bibfnamefont
			{R.}~\bibnamefont {{Ferraz Leal}}}, \bibinfo {author} {\bibfnamefont
			{M.}~\bibnamefont {Gigg}}, \bibinfo {author} {\bibfnamefont {V.}~\bibnamefont
			{Lynch}}, \bibinfo {author} {\bibfnamefont {A.}~\bibnamefont {Markvardsen}},
		\bibinfo {author} {\bibfnamefont {D.}~\bibnamefont {Mikkelson}}, \bibinfo
		{author} {\bibfnamefont {R.}~\bibnamefont {Mikkelson}}, \bibinfo {author}
		{\bibfnamefont {R.}~\bibnamefont {Miller}}, \bibinfo {author} {\bibfnamefont
			{K.}~\bibnamefont {Palmen}}, \bibinfo {author} {\bibfnamefont
			{P.}~\bibnamefont {Parker}}, \bibinfo {author} {\bibfnamefont
			{G.}~\bibnamefont {Passos}}, \bibinfo {author} {\bibfnamefont
			{T.}~\bibnamefont {Perring}}, \bibinfo {author} {\bibfnamefont
			{P.}~\bibnamefont {Peterson}}, \bibinfo {author} {\bibfnamefont
			{S.}~\bibnamefont {Ren}}, \bibinfo {author} {\bibfnamefont {M.}~\bibnamefont
			{Reuter}}, \bibinfo {author} {\bibfnamefont {A.}~\bibnamefont {Savici}},
		\bibinfo {author} {\bibfnamefont {J.}~\bibnamefont {Taylor}}, \bibinfo
		{author} {\bibfnamefont {R.}~\bibnamefont {Taylor}}, \bibinfo {author}
		{\bibfnamefont {R.}~\bibnamefont {Tolchenov}}, \bibinfo {author}
		{\bibfnamefont {W.}~\bibnamefont {Zhou}},\ and\ \bibinfo {author}
		{\bibfnamefont {J.}~\bibnamefont {Zikovsky}},\ }\bibfield  {title} {\bibinfo
		{title} {{Mantid Data analysis and visualization package for neutron
				scattering and $\mu$SR experiments}},\ }\href
	{https://doi.org/https://doi.org/10.1016/j.nima.2014.07.029} {\bibfield
		{journal} {\bibinfo  {journal} {Nucl. Instrum. Methods Phys. Res. Sect. A}\
		}\textbf {\bibinfo {volume} {764}},\ \bibinfo {pages} {156} (\bibinfo {year}
		{2014})}\BibitemShut {NoStop}%
	\bibitem [{\citenamefont {Todo}\ and\ \citenamefont {Kato}(2001)}]{Todo047203}%
	\BibitemOpen
	\bibfield  {author} {\bibinfo {author} {\bibfnamefont {S.}~\bibnamefont
			{Todo}}\ and\ \bibinfo {author} {\bibfnamefont {K.}~\bibnamefont {Kato}},\
	}\bibfield  {title} {\bibinfo {title} {{Cluster Algorithms for General-
				$\mathit{S}$ Quantum Spin Systems}},\ }\href
	{https://doi.org/10.1103/PhysRevLett.87.047203} {\bibfield  {journal}
		{\bibinfo  {journal} {Phys. Rev. Lett.}\ }\textbf {\bibinfo {volume} {87}},\
		\bibinfo {pages} {047203} (\bibinfo {year} {2001})}\BibitemShut {NoStop}%
	\bibitem [{\citenamefont {Albuquerque}\ \emph {et~al.}(2007)\citenamefont
		{Albuquerque}, \citenamefont {Alet}, \citenamefont {Corboz}, \citenamefont
		{Dayal}, \citenamefont {Feiguin}, \citenamefont {Fuchs}, \citenamefont
		{Gamper}, \citenamefont {Gull}, \citenamefont {Gürtler}, \citenamefont
		{Honecker}, \citenamefont {Igarashi}, \citenamefont {Körner}, \citenamefont
		{Kozhevnikov}, \citenamefont {Läuchli}, \citenamefont {Manmana},
		\citenamefont {Matsumoto}, \citenamefont {McCulloch}, \citenamefont {Michel},
		\citenamefont {Noack}, \citenamefont {Pawłowski}, \citenamefont {Pollet},
		\citenamefont {Pruschke}, \citenamefont {Schollwöck}, \citenamefont {Todo},
		\citenamefont {Trebst}, \citenamefont {Troyer}, \citenamefont {Werner},\ and\
		\citenamefont {Wessel}}]{Albuquerque1187}%
	\BibitemOpen
	\bibfield  {author} {\bibinfo {author} {\bibfnamefont {A.}~\bibnamefont
			{Albuquerque}}, \bibinfo {author} {\bibfnamefont {F.}~\bibnamefont {Alet}},
		\bibinfo {author} {\bibfnamefont {P.}~\bibnamefont {Corboz}}, \bibinfo
		{author} {\bibfnamefont {P.}~\bibnamefont {Dayal}}, \bibinfo {author}
		{\bibfnamefont {A.}~\bibnamefont {Feiguin}}, \bibinfo {author} {\bibfnamefont
			{S.}~\bibnamefont {Fuchs}}, \bibinfo {author} {\bibfnamefont
			{L.}~\bibnamefont {Gamper}}, \bibinfo {author} {\bibfnamefont
			{E.}~\bibnamefont {Gull}}, \bibinfo {author} {\bibfnamefont {S.}~\bibnamefont
			{Gürtler}}, \bibinfo {author} {\bibfnamefont {A.}~\bibnamefont {Honecker}},
		\bibinfo {author} {\bibfnamefont {R.}~\bibnamefont {Igarashi}}, \bibinfo
		{author} {\bibfnamefont {M.}~\bibnamefont {Körner}}, \bibinfo {author}
		{\bibfnamefont {A.}~\bibnamefont {Kozhevnikov}}, \bibinfo {author}
		{\bibfnamefont {A.}~\bibnamefont {Läuchli}}, \bibinfo {author}
		{\bibfnamefont {S.}~\bibnamefont {Manmana}}, \bibinfo {author} {\bibfnamefont
			{M.}~\bibnamefont {Matsumoto}}, \bibinfo {author} {\bibfnamefont
			{I.}~\bibnamefont {McCulloch}}, \bibinfo {author} {\bibfnamefont
			{F.}~\bibnamefont {Michel}}, \bibinfo {author} {\bibfnamefont
			{R.}~\bibnamefont {Noack}}, \bibinfo {author} {\bibfnamefont
			{G.}~\bibnamefont {Pawłowski}}, \bibinfo {author} {\bibfnamefont
			{L.}~\bibnamefont {Pollet}}, \bibinfo {author} {\bibfnamefont
			{T.}~\bibnamefont {Pruschke}}, \bibinfo {author} {\bibfnamefont
			{U.}~\bibnamefont {Schollwöck}}, \bibinfo {author} {\bibfnamefont
			{S.}~\bibnamefont {Todo}}, \bibinfo {author} {\bibfnamefont {S.}~\bibnamefont
			{Trebst}}, \bibinfo {author} {\bibfnamefont {M.}~\bibnamefont {Troyer}},
		\bibinfo {author} {\bibfnamefont {P.}~\bibnamefont {Werner}},\ and\ \bibinfo
		{author} {\bibfnamefont {S.}~\bibnamefont {Wessel}},\ }\bibfield  {title}
	{\bibinfo {title} {{The ALPS project release 1.3: Open-source software for
				strongly correlated systems}},\ }\href
	{https://doi.org/https://doi.org/10.1016/j.jmmm.2006.10.304} {\bibfield
		{journal} {\bibinfo  {journal} {J. Magn. Magn. Mater.}\ }\textbf {\bibinfo
			{volume} {310}},\ \bibinfo {pages} {1187} (\bibinfo {year}
		{2007})}\BibitemShut {NoStop}%
	\bibitem [{\citenamefont {Suter}\ and\ \citenamefont {Wojek}(2012)}]{Suter69}%
	\BibitemOpen
	\bibfield  {author} {\bibinfo {author} {\bibfnamefont {A.}~\bibnamefont
			{Suter}}\ and\ \bibinfo {author} {\bibfnamefont {B.~M.}\ \bibnamefont
			{Wojek}},\ }\bibfield  {title} {\bibinfo {title} {{Musrfit: A Free
				Platform-Independent Framework for $\mu$SR Data Analysis}},\ }\href
	{https://doi.org/https://doi.org/10.1016/j.phpro.2012.04.042} {\bibfield
		{journal} {\bibinfo  {journal} {Phys. Procedia}\ }\textbf {\bibinfo {volume}
			{30}},\ \bibinfo {pages} {69} (\bibinfo {year} {2012})}\BibitemShut {NoStop}%
	\bibitem [{\citenamefont {Bouvier}\ \emph {et~al.}(1991)\citenamefont
		{Bouvier}, \citenamefont {Lethuillier},\ and\ \citenamefont
		{Schmitt}}]{Bouvier13137}%
	\BibitemOpen
	\bibfield  {author} {\bibinfo {author} {\bibfnamefont {M.}~\bibnamefont
			{Bouvier}}, \bibinfo {author} {\bibfnamefont {P.}~\bibnamefont
			{Lethuillier}},\ and\ \bibinfo {author} {\bibfnamefont {D.}~\bibnamefont
			{Schmitt}},\ }\bibfield  {title} {\bibinfo {title} {Specific heat in some
			gadolinium compounds. i. experimental},\ }\href
	{https://doi.org/10.1103/PhysRevB.43.13137} {\bibfield  {journal} {\bibinfo
			{journal} {Phys. Rev. B}\ }\textbf {\bibinfo {volume} {43}},\ \bibinfo
		{pages} {13137} (\bibinfo {year} {1991})}\BibitemShut {NoStop}%
	\bibitem [{\citenamefont {Stevens}(1952)}]{Stevens209}%
	\BibitemOpen
	\bibfield  {author} {\bibinfo {author} {\bibfnamefont {K.~W.~H.}\
			\bibnamefont {Stevens}},\ }\bibfield  {title} {\bibinfo {title} {Matrix
			elements and operator equivalents connected with the magnetic properties of
			rare earth ions},\ }\href {https://doi.org/10.1088/0370-1298/65/3/308}
	{\bibfield  {journal} {\bibinfo  {journal} {Proc. Phys. Soc. Section A}\
		}\textbf {\bibinfo {volume} {65}},\ \bibinfo {pages} {209} (\bibinfo {year}
		{1952})}\BibitemShut {NoStop}%
	\bibitem [{\citenamefont {Hutchings}(1964)}]{Huthings227}%
	\BibitemOpen
	\bibfield  {author} {\bibinfo {author} {\bibfnamefont {M.}~\bibnamefont
			{Hutchings}},\ }\href
	{https://doi.org/https://doi.org/10.1016/S0081-1947(08)60517-2} {\emph
		{\bibinfo {title} {Point-Charge Calculations of Energy Levels of Magnetic
				Ions in Crystalline Electric Fields}}},\ \bibinfo {series} {Solid State
		Physics}, Vol.~\bibinfo {volume} {16}\ (\bibinfo  {publisher} {Academic
		Press},\ \bibinfo {year} {1964})\ p.\ \bibinfo {pages} {227}\BibitemShut
	{NoStop}%
	\bibitem [{\citenamefont {Newman}\ and\ \citenamefont {Ng}(2000)}]{Newman2000}%
	\BibitemOpen
	\bibfield  {author} {\bibinfo {author} {\bibfnamefont {D.~J.}\ \bibnamefont
			{Newman}}\ and\ \bibinfo {author} {\bibfnamefont {B.}~\bibnamefont {Ng}},\
	}\href {https://doi.org/https://doi.org/10.1017/CBO9780511524295} {\emph
		{\bibinfo {title} {Crystal Field Handbook}}}\ (\bibinfo  {publisher}
	{Cambridge University Press},\ \bibinfo {year} {2000})\BibitemShut {NoStop}%
	\bibitem [{\citenamefont {Guchhait}\ \emph {et~al.}(2024)\citenamefont
		{Guchhait}, \citenamefont {Painganoor}, \citenamefont {Islam}, \citenamefont
		{Sichelschmidt}, \citenamefont {Le}, \citenamefont {Aouane}, \citenamefont
		{Christensen},\ and\ \citenamefont {Nath}}]{Guchhait144434}%
	\BibitemOpen
	\bibfield  {author} {\bibinfo {author} {\bibfnamefont {S.}~\bibnamefont
			{Guchhait}}, \bibinfo {author} {\bibfnamefont {A.}~\bibnamefont
			{Painganoor}}, \bibinfo {author} {\bibfnamefont {S.~S.}\ \bibnamefont
			{Islam}}, \bibinfo {author} {\bibfnamefont {J.}~\bibnamefont
			{Sichelschmidt}}, \bibinfo {author} {\bibfnamefont {M.~D.}\ \bibnamefont
			{Le}}, \bibinfo {author} {\bibfnamefont {M.}~\bibnamefont {Aouane}}, \bibinfo
		{author} {\bibfnamefont {N.~B.}\ \bibnamefont {Christensen}},\ and\ \bibinfo
		{author} {\bibfnamefont {R.}~\bibnamefont {Nath}},\ }\bibfield  {title}
	{\bibinfo {title} {{Magnetic and crystal electric field studies of the rare
				earth based square lattice antiferromagnet NdKNaNbO$_5$}},\ }\href
	{https://doi.org/10.1103/PhysRevB.110.144434} {\bibfield  {journal} {\bibinfo
			{journal} {Phys. Rev. B}\ }\textbf {\bibinfo {volume} {110}},\ \bibinfo
		{pages} {144434} (\bibinfo {year} {2024})}\BibitemShut {NoStop}%
	\bibitem [{\citenamefont {Sebastian}\ \emph
		{et~al.}(2025{\natexlab{b}})\citenamefont {Sebastian}, \citenamefont {Islam},
		\citenamefont {Kolay}, \citenamefont {Mohanty}, \citenamefont {Ding},
		\citenamefont {Skourski}, \citenamefont {Sichelschmidt}, \citenamefont
		{Baenitz}, \citenamefont {Krieger}, \citenamefont {Hicken}, \citenamefont
		{Luetkens}, \citenamefont {Tsirlin}, \citenamefont {Furukawa},\ and\
		\citenamefont {Nath}}]{Sebastian2025}%
	\BibitemOpen
	\bibfield  {author} {\bibinfo {author} {\bibfnamefont {S.~J.}\ \bibnamefont
			{Sebastian}}, \bibinfo {author} {\bibfnamefont {S.~S.}\ \bibnamefont
			{Islam}}, \bibinfo {author} {\bibfnamefont {R.}~\bibnamefont {Kolay}},
		\bibinfo {author} {\bibfnamefont {S.}~\bibnamefont {Mohanty}}, \bibinfo
		{author} {\bibfnamefont {Q.~P.}\ \bibnamefont {Ding}}, \bibinfo {author}
		{\bibfnamefont {Y.}~\bibnamefont {Skourski}}, \bibinfo {author}
		{\bibfnamefont {J.}~\bibnamefont {Sichelschmidt}}, \bibinfo {author}
		{\bibfnamefont {M.}~\bibnamefont {Baenitz}}, \bibinfo {author} {\bibfnamefont
			{J.~A.}\ \bibnamefont {Krieger}}, \bibinfo {author} {\bibfnamefont {T.~J.}\
			\bibnamefont {Hicken}}, \bibinfo {author} {\bibfnamefont {H.}~\bibnamefont
			{Luetkens}}, \bibinfo {author} {\bibfnamefont {A.~A.}\ \bibnamefont
			{Tsirlin}}, \bibinfo {author} {\bibfnamefont {Y.}~\bibnamefont {Furukawa}},\
		and\ \bibinfo {author} {\bibfnamefont {R.}~\bibnamefont {Nath}},\ }\href
	{https://arxiv.org/abs/2511.07844} {\bibinfo {title} {{Inhomogeneous dynamic
				state in the double trillium lattice antiferromagnet KBaFe$_2$(PO$_4$)$_3$}}}
	(\bibinfo {year} {2025}{\natexlab{b}}),\ \Eprint
	{https://arxiv.org/abs/2511.07844} {arXiv:2511.07844 [cond-mat.mtrl-sci]}
	\BibitemShut {NoStop}%
	\bibitem [{\citenamefont {Magar}\ \emph {et~al.}(2026)\citenamefont {Magar},
		\citenamefont {Somesh}, \citenamefont {Saravanan}, \citenamefont
		{Sichelschmidt}, \citenamefont {Skourski}, \citenamefont {Telling},
		\citenamefont {Ginga}, \citenamefont {Tsirlin},\ and\ \citenamefont
		{Nath}}]{MagarL020409}%
	\BibitemOpen
	\bibfield  {author} {\bibinfo {author} {\bibfnamefont {A.}~\bibnamefont
			{Magar}}, \bibinfo {author} {\bibfnamefont {K.}~\bibnamefont {Somesh}},
		\bibinfo {author} {\bibfnamefont {M.~P.}\ \bibnamefont {Saravanan}}, \bibinfo
		{author} {\bibfnamefont {J.}~\bibnamefont {Sichelschmidt}}, \bibinfo {author}
		{\bibfnamefont {Y.}~\bibnamefont {Skourski}}, \bibinfo {author}
		{\bibfnamefont {M.~T.~F.}\ \bibnamefont {Telling}}, \bibinfo {author}
		{\bibfnamefont {V.~A.}\ \bibnamefont {Ginga}}, \bibinfo {author}
		{\bibfnamefont {A.~A.}\ \bibnamefont {Tsirlin}},\ and\ \bibinfo {author}
		{\bibfnamefont {R.}~\bibnamefont {Nath}},\ }\bibfield  {title} {\bibinfo
		{title} {{Proximate spin-liquid behavior in the double trillium lattice
				antiferromagnet
				${\mathrm{K}}_{2}{\mathrm{Co}}_{2}{({\mathrm{SO}}_{4})}_{3}$}},\ }\href
	{https://doi.org/10.1103/m8mj-slvl} {\bibfield  {journal} {\bibinfo
			{journal} {Phys. Rev. B}\ }\textbf {\bibinfo {volume} {113}},\ \bibinfo
		{pages} {L020409} (\bibinfo {year} {2026})}\BibitemShut {NoStop}%
	\bibitem [{\citenamefont {Orbach}(1961)}]{Orbach458}%
	\BibitemOpen
	\bibfield  {author} {\bibinfo {author} {\bibfnamefont {R.}~\bibnamefont
			{Orbach}},\ }\bibfield  {title} {\bibinfo {title} {{Spin-lattice relaxation
				in rare-earth salts}},\ }\href
	{https://doi.org/https://doi.org/10.1098/rspa.1961.0211} {\bibfield
		{journal} {\bibinfo  {journal} {Proc. Phys. Soc. A}\ }\textbf {\bibinfo
			{volume} {264}},\ \bibinfo {pages} {458} (\bibinfo {year}
		{1961})}\BibitemShut {NoStop}%
	\bibitem [{\citenamefont {Arh}\ \emph {et~al.}(2022)\citenamefont {Arh},
		\citenamefont {Sana}, \citenamefont {Pregelj}, \citenamefont {Khuntia},
		\citenamefont {Jagli{\v{c}}i{\'{c}}}, \citenamefont {Le}, \citenamefont
		{Biswas}, \citenamefont {Manuel}, \citenamefont {Mangin-Thro}, \citenamefont
		{Ozarowski},\ and\ \citenamefont {Zorko}}]{Arh416}%
	\BibitemOpen
	\bibfield  {author} {\bibinfo {author} {\bibfnamefont {T.}~\bibnamefont
			{Arh}}, \bibinfo {author} {\bibfnamefont {B.}~\bibnamefont {Sana}}, \bibinfo
		{author} {\bibfnamefont {M.}~\bibnamefont {Pregelj}}, \bibinfo {author}
		{\bibfnamefont {P.}~\bibnamefont {Khuntia}}, \bibinfo {author} {\bibfnamefont
			{Z.}~\bibnamefont {Jagli{\v{c}}i{\'{c}}}}, \bibinfo {author} {\bibfnamefont
			{M.~D.}\ \bibnamefont {Le}}, \bibinfo {author} {\bibfnamefont {P.~K.}\
			\bibnamefont {Biswas}}, \bibinfo {author} {\bibfnamefont {P.}~\bibnamefont
			{Manuel}}, \bibinfo {author} {\bibfnamefont {L.}~\bibnamefont {Mangin-Thro}},
		\bibinfo {author} {\bibfnamefont {A.}~\bibnamefont {Ozarowski}},\ and\
		\bibinfo {author} {\bibfnamefont {A.}~\bibnamefont {Zorko}},\ }\bibfield
	{title} {\bibinfo {title} {The ising triangular-lattice antiferromagnet
			neodymium heptatantalate as a quantum spin liquid candidate},\ }\href
	{https://doi.org/10.1038/s41563-021-01169-y} {\bibfield  {journal} {\bibinfo
			{journal} {Nat. Mater.}\ }\textbf {\bibinfo {volume} {21}},\ \bibinfo {pages}
		{416} (\bibinfo {year} {2022})}\BibitemShut {NoStop}%
	\bibitem [{\citenamefont {Xu}\ \emph {et~al.}(2016)\citenamefont {Xu},
		\citenamefont {Balz}, \citenamefont {Baines}, \citenamefont {Luetkens},\ and\
		\citenamefont {Lake}}]{Xu064425}%
	\BibitemOpen
	\bibfield  {author} {\bibinfo {author} {\bibfnamefont {J.}~\bibnamefont
			{Xu}}, \bibinfo {author} {\bibfnamefont {C.}~\bibnamefont {Balz}}, \bibinfo
		{author} {\bibfnamefont {C.}~\bibnamefont {Baines}}, \bibinfo {author}
		{\bibfnamefont {H.}~\bibnamefont {Luetkens}},\ and\ \bibinfo {author}
		{\bibfnamefont {B.}~\bibnamefont {Lake}},\ }\bibfield  {title} {\bibinfo
		{title} {Spin dynamics of the ordered dipolar-octupolar
			pseudospin-$\frac{1}{2}$ pyrochlore
			${\text{nd}}_{2}{\text{zr}}_{2}{\text{o}}_{7}$ probed by muon spin
			relaxation},\ }\href {https://doi.org/10.1103/PhysRevB.94.064425} {\bibfield
		{journal} {\bibinfo  {journal} {Phys. Rev. B}\ }\textbf {\bibinfo {volume}
			{94}},\ \bibinfo {pages} {064425} (\bibinfo {year} {2016})}\BibitemShut
	{NoStop}%
	\bibitem [{\citenamefont {Bert}\ \emph {et~al.}(2006)\citenamefont {Bert},
		\citenamefont {Mendels}, \citenamefont {Olariu}, \citenamefont {Blanchard},
		\citenamefont {Collin}, \citenamefont {Amato}, \citenamefont {Baines},\ and\
		\citenamefont {Hillier}}]{Bert117203}%
	\BibitemOpen
	\bibfield  {author} {\bibinfo {author} {\bibfnamefont {F.}~\bibnamefont
			{Bert}}, \bibinfo {author} {\bibfnamefont {P.}~\bibnamefont {Mendels}},
		\bibinfo {author} {\bibfnamefont {A.}~\bibnamefont {Olariu}}, \bibinfo
		{author} {\bibfnamefont {N.}~\bibnamefont {Blanchard}}, \bibinfo {author}
		{\bibfnamefont {G.}~\bibnamefont {Collin}}, \bibinfo {author} {\bibfnamefont
			{A.}~\bibnamefont {Amato}}, \bibinfo {author} {\bibfnamefont
			{C.}~\bibnamefont {Baines}},\ and\ \bibinfo {author} {\bibfnamefont {A.~D.}\
			\bibnamefont {Hillier}},\ }\bibfield  {title} {\bibinfo {title} {Direct
			evidence for a dynamical ground state in the highly frustrated
			${\mathrm{tb}}_{2}{\mathrm{sn}}_{2}{\mathrm{o}}_{7}$ pyrochlore},\ }\href
	{https://doi.org/10.1103/PhysRevLett.97.117203} {\bibfield  {journal}
		{\bibinfo  {journal} {Phys. Rev. Lett.}\ }\textbf {\bibinfo {volume} {97}},\
		\bibinfo {pages} {117203} (\bibinfo {year} {2006})}\BibitemShut {NoStop}%
	\bibitem [{\citenamefont {Lago}\ \emph {et~al.}(2005)\citenamefont {Lago},
		\citenamefont {Lancaster}, \citenamefont {Blundell}, \citenamefont
		{Bramwell}, \citenamefont {Pratt}, \citenamefont {Shirai},\ and\
		\citenamefont {Baines}}]{Lago979}%
	\BibitemOpen
	\bibfield  {author} {\bibinfo {author} {\bibfnamefont {J.}~\bibnamefont
			{Lago}}, \bibinfo {author} {\bibfnamefont {T.}~\bibnamefont {Lancaster}},
		\bibinfo {author} {\bibfnamefont {S.~J.}\ \bibnamefont {Blundell}}, \bibinfo
		{author} {\bibfnamefont {S.~T.}\ \bibnamefont {Bramwell}}, \bibinfo {author}
		{\bibfnamefont {F.~L.}\ \bibnamefont {Pratt}}, \bibinfo {author}
		{\bibfnamefont {M.}~\bibnamefont {Shirai}},\ and\ \bibinfo {author}
		{\bibfnamefont {C.}~\bibnamefont {Baines}},\ }\bibfield  {title} {\bibinfo
		{title} {{Magnetic ordering and dynamics in the XY pyrochlore
				antiferromagnet: a muon-spin relaxation study of
				${\mathrm{Er}}_{2}{\mathrm{Ti}}_{2}{\mathrm{O}}_{7}$ and
				${\mathrm{Er}}_{2}{\mathrm{Sn}}_{2}{\mathrm{O}}_{7}$}},\ }\href
	{https://doi.org/10.1088/1361-648X/ac136a} {\bibfield  {journal} {\bibinfo
			{journal} {J. Condens. Matter Phys.}\ }\textbf {\bibinfo {volume} {17}},\
		\bibinfo {pages} {979} (\bibinfo {year} {2005})}\BibitemShut {NoStop}%
	\bibitem [{\citenamefont {Dalmas~de R\'eotier}\ \emph
		{et~al.}(2006)\citenamefont {Dalmas~de R\'eotier}, \citenamefont {Yaouanc},
		\citenamefont {Keller}, \citenamefont {Cervellino}, \citenamefont {Roessli},
		\citenamefont {Baines}, \citenamefont {Forget}, \citenamefont {Vaju},
		\citenamefont {Gubbens}, \citenamefont {Amato},\ and\ \citenamefont
		{King}}]{Dalmas127202}%
	\BibitemOpen
	\bibfield  {author} {\bibinfo {author} {\bibfnamefont {P.}~\bibnamefont
			{Dalmas~de R\'eotier}}, \bibinfo {author} {\bibfnamefont {A.}~\bibnamefont
			{Yaouanc}}, \bibinfo {author} {\bibfnamefont {L.}~\bibnamefont {Keller}},
		\bibinfo {author} {\bibfnamefont {A.}~\bibnamefont {Cervellino}}, \bibinfo
		{author} {\bibfnamefont {B.}~\bibnamefont {Roessli}}, \bibinfo {author}
		{\bibfnamefont {C.}~\bibnamefont {Baines}}, \bibinfo {author} {\bibfnamefont
			{A.}~\bibnamefont {Forget}}, \bibinfo {author} {\bibfnamefont
			{C.}~\bibnamefont {Vaju}}, \bibinfo {author} {\bibfnamefont {P.~C.~M.}\
			\bibnamefont {Gubbens}}, \bibinfo {author} {\bibfnamefont {A.}~\bibnamefont
			{Amato}},\ and\ \bibinfo {author} {\bibfnamefont {P.~J.~C.}\ \bibnamefont
			{King}},\ }\bibfield  {title} {\bibinfo {title} {{Spin Dynamics and Magnetic
				Order in Magnetically Frustrated
				${\mathrm{Tb}}_{2}{\mathrm{Sn}}_{2}{\mathrm{O}}_{7}$}},\ }\href
	{https://doi.org/10.1103/PhysRevLett.96.127202} {\bibfield  {journal}
		{\bibinfo  {journal} {Phys. Rev. Lett.}\ }\textbf {\bibinfo {volume} {96}},\
		\bibinfo {pages} {127202} (\bibinfo {year} {2006})}\BibitemShut {NoStop}%
	\bibitem [{\citenamefont {Kalvius}\ \emph {et~al.}(2010)\citenamefont
		{Kalvius}, \citenamefont {Krimmel}, \citenamefont {Hartmann}, \citenamefont
		{Litterst}, \citenamefont {Wäppling}, \citenamefont {Wagner}, \citenamefont
		{Tsurkan},\ and\ \citenamefont {Loidl}}]{Kalvius87}%
	\BibitemOpen
	\bibfield  {author} {\bibinfo {author} {\bibfnamefont {G.~M.}\ \bibnamefont
			{Kalvius}}, \bibinfo {author} {\bibfnamefont {A.}~\bibnamefont {Krimmel}},
		\bibinfo {author} {\bibfnamefont {O.}~\bibnamefont {Hartmann}}, \bibinfo
		{author} {\bibfnamefont {F.~J.}\ \bibnamefont {Litterst}}, \bibinfo {author}
		{\bibfnamefont {R.}~\bibnamefont {Wäppling}}, \bibinfo {author}
		{\bibfnamefont {F.~E.}\ \bibnamefont {Wagner}}, \bibinfo {author}
		{\bibfnamefont {V.}~\bibnamefont {Tsurkan}},\ and\ \bibinfo {author}
		{\bibfnamefont {A.}~\bibnamefont {Loidl}},\ }\bibfield  {title} {\bibinfo
		{title} {{Frustration driven magnetic states of A-site spinels probed by
				$\mu$SR}},\ }\href {https://doi.org/10.1140/epjb/e2010-00262-7} {\bibfield
		{journal} {\bibinfo  {journal} {Eur. Phys. J. B.}\ }\textbf {\bibinfo
			{volume} {77}},\ \bibinfo {pages} {87} (\bibinfo {year} {2010})}\BibitemShut
	{NoStop}%
	\bibitem [{\citenamefont {McClarty}\ \emph {et~al.}(2011)\citenamefont
		{McClarty}, \citenamefont {Cosman}, \citenamefont {Del~Maestro},\ and\
		\citenamefont {Gingras}}]{McClarty164216}%
	\BibitemOpen
	\bibfield  {author} {\bibinfo {author} {\bibfnamefont {P.~A.}\ \bibnamefont
			{McClarty}}, \bibinfo {author} {\bibfnamefont {J.~N.}\ \bibnamefont
			{Cosman}}, \bibinfo {author} {\bibfnamefont {A.~G.}\ \bibnamefont
			{Del~Maestro}},\ and\ \bibinfo {author} {\bibfnamefont {M.~J.~P.}\
			\bibnamefont {Gingras}},\ }\bibfield  {title} {\bibinfo {title} {{Calculation
				of the expected zero-field muon relaxation rate in the geometrically
				frustrated rare earth pyrochlore Gd$_2$Sn$_2$O$_7$ antiferromagnet}},\ }\href
	{https://doi.org/10.1088/0953-8984/23/16/164216} {\bibfield  {journal}
		{\bibinfo  {journal} {J. Phys. Condens. Matter}\ }\textbf {\bibinfo {volume}
			{23}},\ \bibinfo {pages} {164216} (\bibinfo {year} {2011})}\BibitemShut
	{NoStop}%
	\bibitem [{\citenamefont {Foronda}\ \emph {et~al.}(2015)\citenamefont
		{Foronda}, \citenamefont {Lang}, \citenamefont {M\"oller}, \citenamefont
		{Lancaster}, \citenamefont {Boothroyd}, \citenamefont {Pratt}, \citenamefont
		{Giblin}, \citenamefont {Prabhakaran},\ and\ \citenamefont
		{Blundell}}]{Foronda017602}%
	\BibitemOpen
	\bibfield  {author} {\bibinfo {author} {\bibfnamefont {F.~R.}\ \bibnamefont
			{Foronda}}, \bibinfo {author} {\bibfnamefont {F.}~\bibnamefont {Lang}},
		\bibinfo {author} {\bibfnamefont {J.~S.}\ \bibnamefont {M\"oller}}, \bibinfo
		{author} {\bibfnamefont {T.}~\bibnamefont {Lancaster}}, \bibinfo {author}
		{\bibfnamefont {A.~T.}\ \bibnamefont {Boothroyd}}, \bibinfo {author}
		{\bibfnamefont {F.~L.}\ \bibnamefont {Pratt}}, \bibinfo {author}
		{\bibfnamefont {S.~R.}\ \bibnamefont {Giblin}}, \bibinfo {author}
		{\bibfnamefont {D.}~\bibnamefont {Prabhakaran}},\ and\ \bibinfo {author}
		{\bibfnamefont {S.~J.}\ \bibnamefont {Blundell}},\ }\bibfield  {title}
	{\bibinfo {title} {Anisotropic local modification of crystal field levels in
			pr-based pyrochlores: A muon-induced effect modeled using density functional
			theory},\ }\href {https://doi.org/10.1103/PhysRevLett.114.017602} {\bibfield
		{journal} {\bibinfo  {journal} {Phys. Rev. Lett.}\ }\textbf {\bibinfo
			{volume} {114}},\ \bibinfo {pages} {017602} (\bibinfo {year}
		{2015})}\BibitemShut {NoStop}%
	\bibitem [{\citenamefont {Boothroyd}(2020)}]{Boothroyd2020}%
	\BibitemOpen
	\bibfield  {author} {\bibinfo {author} {\bibfnamefont {A.}~\bibnamefont
			{Boothroyd}},\ }\href {https://books.google.co.in/books?id=FvTuDwAAQBAJ}
	{\emph {\bibinfo {title} {Principles of Neutron Scattering from Condensed
				Matter}}}\ (\bibinfo  {publisher} {OUP Oxford},\ \bibinfo {year}
	{2020})\BibitemShut {NoStop}%
	\bibitem [{\citenamefont {Prince}(2004)}]{Prince2004}%
	\BibitemOpen
	\bibfield  {author} {\bibinfo {author} {\bibfnamefont {E.}~\bibnamefont
			{Prince}},\ }\href {https://books.google.co.in/books?id=60FoFEGyShIC} {\emph
		{\bibinfo {title} {International Tables for Crystallography}}}\ (\bibinfo
	{publisher} {Springer Netherlands},\ \bibinfo {year} {2004})\BibitemShut
	{NoStop}%
\end{thebibliography}
%

\appendix
\onecolumngrid 
\section{INS spectra}

The intensity of CEF excitation in phonon subtracted data is proportional to $F^2(Q)$ as shown in Fig.~\ref{Fig8}. one can calculate $F^2(Q)$ using the following expression~\cite{Boothroyd2020}
\begin{align}
\begin{split}
F(Q)=  \left[\sum_{i = 1}^3A_ie^{-a_i|\vec{Q}|^2}+D\right]+ 
  \frac{2-g_J}{g_J}\left[\sum_{i = 1}^3A_i^{\prime}|\vec{Q}|^2e^{-a_i^{\prime}|\vec{Q}|^2}+D^{\prime}|\vec{Q}|^2\right]
\end{split}
\end{align}
where, $A_1 = 0.0586$, $A_2 = 0.3540$, $A_3 = 0.6126$, $a_1 = 17.9802$, $a_2 = 7.0964$, $a_3 = 2.7482$, $A_1^{\prime} = 0.1710$, $A_2^{\prime} = 0.9879$, $A_3^{\prime} = 0.9044$, $a_1^{\prime} = 18.5340$, $a_2^{\prime} = 6.6250$, $a_3^{\prime} = 2.1000$, $D = -0.0251$, and $D^{\prime} = 0.0278$ are the magnetic form factor coefficients of Er$^{3+}$~\cite{Prince2004}. Here, $g_J$ is the Land\'e g-factor of Er$^{3+}$.

\section{Physical Properties}
The magnetization is estimated by calculating the expectation value of total angular momentum ($\hat{J}$) with components $J_x$, $J_y$, and $J_z$ as~\cite{Guchhait214437}
\begin{equation}
   M_{\rm CEF}(T,H)
= \frac{N_{\rm A}\,g\,\mu_{\rm B}}{Z}
\sum_k e^{-\frac{E_k(H)}{k_{\rm B}T}}
\langle \psi_k(H) \lvert \hat{J}_{\alpha~=~x,~y,~z} \rvert \psi_k(H)\rangle.
	\label{M_CEF} 
\end{equation}
Here, $Z = \sum_ke^{-E_k(H)/k_{\rm B}T}$ is the partition function, where the summation is taken over all the energy states. $|\psi_k(H)\rangle$ and $E_k(H)$ are the $k^{\rm th}$ eigenstate and eigenvalue of the effective Hamiltonian $\mathcal{H}_{\rm eff} = \mathcal{H}_{\rm CEF} + g\mu_{\rm B}\vec{B}.\vec{J}$, with an applied field $\vec{B}$. $\chi_{\rm CEF}$ can be calculated by taking the first derivative of $M_{\rm CEF}(T,H)$ with respect to $H$. Heat capacity ($C_{\rm CEF}$) for a $N$ level system can be written as 
\begin{equation}
  C_{\rm CEF}(T, H) = \frac{R}{(Zk_{\rm B}T)^2} \sum_{n>m}^{N}[E_n(H) - E_m(H)]^2e^{-\frac{[E_n(H) + E_m(H)]}{k_{\rm B}T}},
 \label{C_CEF}  
\end{equation}
where, $R$ is the universal gas constant. Here, $E_n$ and $E_m$ are the energy of the $n^{th}$ and $m^{th}$ CEF levels, respectively~\cite{Guchhait144434,Guchhait214437}.

\begin{table*}[ptb]
	\caption{Fitted CEF parameters for BiErGeO$_5$.}
	\label{CEF_Para}
	\begin{ruledtabular}
		\begin{tabular}{cccccc}
			$B_l^m$ (meV) & Values & $B_l^m$ (meV) & Values & $B_l^m$ (meV) & Values \\\hline
			$B_2^0$ & -9.73031$\times10^{-2}$ & $B_6^1$ & 3.54671$\times10^{-5}$ & $B_4^{-2}$ & -2.79004$\times10^{-3}$ \\
			$B_2^1$ & -2.12865$\times10^{-1}$ & $B_6^2$ & -3.53401$\times10^{-5}$ & $B_4^{-3}$ & 4.27442$\times10^{-3}$  \\
			$B_2^2$ & -8.56659$\times10^{-2}$ & $B_6^3$ & 3.82072$\times10^{-5}$ & $B_4^{-4}$ & 7.65228$\times10^{-4}$ \\
			$B_4^0$ & -2.78673$\times10^{-4}$ & $B_6^4$ & -2.39736$\times10^{-6}$ & $B_6^{-1}$ & 2.62382$\times10^{-5}$ \\
			$B_4^1$ & -2.16909$\times10^{-4}$ & $B_6^5$ & -3.14546$\times10^{-5}$ & $B_6^{-2}$ & 2.82909$\times10^{-5}$ \\
			$B_4^2$ & 5.00072$\times10^{-4}$ & $B_6^6$ & 1.35961$\times10^{-4}$ & $B_6^{-3}$ & -3.87787$\times10^{-5}$ \\
			$B_4^3$ & -6.39910$\times10^{-3}$ & $B_2^{-1}$ & 3.3494$\times10^{-2}$ & $B_6^{-4}$ & 3.81230$\times10^{-5}$ \\
			$B_4^4$ & -1.40734$\times10^{-3}$ & $B_2^{-2}$ & 3.10207$\times10^{-2}$ & $B_6^{-5}$ & -1.93094$\times10^{-4}$  \\
			$B_6^0$ & -4.21501$\times10^{-6}$ & $B_4^{-1}$ & -1.62383$\times10^{-3}$ & $B_6^{-6}$ & 5.92283$\times10^{-5}$ \\
		\end{tabular}
	\end{ruledtabular}
\end{table*}

\begin{table*}
	\caption{Energy eigenvalues and the coefficients ($C_{m_J}^{k,\pm}$) corresponding to different eigenstates of the CEF Hamiltonian for BiErGeO$_5$.}
	\label{Eigenvalue_and_Eigervector}
	\begin{ruledtabular}
		\begin{tabular}{c|cccccccc}
			$E$ (meV) & $|\frac{15}{2}\rangle$ & $|\frac{13}{2}\rangle$ & $|\frac{11}{2}\rangle$ & $|\frac{9}{2}\rangle$ & $|\frac{7}{2}\rangle$ & $|\frac{5}{2}\rangle$ & $|\frac{3}{2}\rangle$ & $|\frac{1}{2}\rangle$ \tabularnewline
			\hline 
			0.00 & 0.03 & -0.07-0.02i & 0.25-0.04i & -0.14+0.36i & -0.02-0.34i & 0.03-0.02i & 0.02+0.08i & -0.19+0.02i \tabularnewline
			0.00 & -0.44 & -0.11+0.05i & -0.24-0.016i & 0.08+0.14i & 0.11+0.007i & 0.17-0.27i & 0.065-0.42i & -0.11-0.13i  \tabularnewline
			1.8 & -0.6 & 0.03-0.05i & -0.06+0.078i & -0.044-0.16i & -0.07-0.4i & -0.044+0.033i & 0.07+0.09i & 0.17+0.07i \tabularnewline
			1.8 & 0.07 & -0.12-0.18i & 0.22+0.083i & -0.11+0.13i & -0.016-0.06i & 0.29-0.3i & -0.1+0.1i & 0.12+0.14i \tabularnewline
			8 & -0.5 & 0.11-0.12i & 0.04+0.2i & -0.15-0.04i & 0.12+0.2i & -0.054-0.08i & 0.13+0.31i & -0.2-0.05i \tabularnewline
			8 & 0.079 & 0.14-0.02i & 0.23+0.04i & -0.01-0.29i & 0.07-0.05i & -0.19-0.33i & 0.001+0.2i & 0.2-0.13i \tabularnewline
			11.8 & 0.045 & -0.04+0.3i & -0.025+0.13i & 0.22+0.17i & 0.28-0.16i & -0.19-0.07i & -0.11+0.009i & -0.22-0.32i \tabularnewline
			11.8 & 0.34 & 0.2-0.18i & -0.11+0.26i & -0.22-0.11i & 0.21+0.012i & 0.02-0.07i & 0.18+0.04i & -0.21-0.04i \tabularnewline
			29.3 & 0.186 & 0.05-0.06i & -0.26+0.26i & 0.26+0.09i & 0.014-0.29i & 0.11-0.1i & 0.23-0.063i & 0.13-0.17i \tabularnewline
			29.3 & 0.0017 & 0.3-0.54i & -0.13+0.074i & 0.19+0.13i & -0.04+0.13i & -0.11-0.11i & 0.07+0.07i & 0.06-0.03i \tabularnewline
			41 & 0.003 & 0.26+0.28i & 0.37+0.064i & 0.016-0.26i & 0.11+0.05i & 0.24-0.27i & 0.3-0.18i & 0.09-0.1i \tabularnewline
			41 & -0.0497 & 0.072+0.25i & 0.12+0.17i & -0.14+0.24i & 0.24+0.12i & -0.013+0.11i & 0.12+0.15i & 0.09+0.14i \tabularnewline
			51.4 & 0.145 & 0.13-0.11i & 0.43-0.05i & 0.31+0.092i & 0.18+0.42i & -0.05+0.28i & -0.12-0.16i & 0.02-0.12i \tabularnewline
			51.4 & -0.0054 & 0.11+0.07i & -0.054-0.12i & -0.19+0.09i & 0.011+0.18i & 0.03-0.38i & -0.17+0.03i & 0.06+0.16i \tabularnewline
			56.45 & 0.057 & -0.03+0.05i & -0.23-0.09i & -0.051-0.21i & 0.05+0.13i & -0.21-0.02i & 0.43-0.29i & -0.05+0.22i \tabularnewline
			56.45 & 0 & -0.09-0.03i & -0.045+0.011i & 0.074+0.18i & -0.06i+0.15i & 0.14-.17i & 0.03+0.005i & -0.6-0.005i \tabularnewline
	\end{tabular}\end{ruledtabular}
    
\vspace{2ex} 

    \begin{ruledtabular}
    	\begin{tabular}{c|cccccccc}
    		$E$ (meV) & $|-\frac{1}{2}\rangle$ & $|-\frac{3}{2}\rangle$ & $|-\frac{5}{2}\rangle$ & $|-\frac{7}{2}\rangle$ & $|-\frac{9}{2}\rangle$ & $|-\frac{11}{2}\rangle$ & $|-\frac{13}{2}\rangle$ & $|-\frac{15}{2}\rangle$ \tabularnewline
    		\hline 
    		0.00 & -0.12+0.12i & -0.03-0.42i & 0.14+0.28i & -0.11-0.002i & 0.09-0.13i & 0.24+0.005i & -0.11-0.06i & 0.43+0.04i  \tabularnewline
    		0.00  & 0.19+0.039i & 0.03-0.08i & -0.03-0.02i & -0.05+0.34i & 0.11+0.37i & 0.25+0.06i & 0.07-0.01i & 0.03+0.003i \tabularnewline
    		1.8 & 0.143 - 0.12i & 0.087+0.12i & 0.24+0.34i & 0.026-0.058i & -0.09-0.15i & -0.23+0.05i & -0.15+0.16i & -0.07-0.011i \tabularnewline
    		1.8 & -0.18+0.045i & 0.081-0.081i & 0.039+0.039i & -0.14+0.38i & 0.07-0.15i & -0.05-0.08i & -0.024-0.052i & -0.59-0.09i \tabularnewline
    		8 & -0.23+0.073i & -0.16-0.13i & -0.13-0.36i & 0.086-0.021i & -0.22-0.19i & 0.12-0.2i & 0.07-0.12i & 0.051-0.06i \tabularnewline
    		8 & -0.096+0.19i & 0.15+0.29i & 0.028+0.094j & 0.083+0.22j & -0.065+0.14i & 0.14 + 0.16i & 0.16-0.009i & 0.32-0.38i \tabularnewline
    		11.8 & -0.19-0.088i & -0.17-0.076i & -0.025+0.068i & -0.18-0.12i & -0.24-0.03i & -0.06+0.27i & 0.04+0.25i & -0.27-0.2i \tabularnewline
    		11.8 & 0.36-0.13i & -0.08-0.07i & 0.2+0.057i & 0.14+0.3i & -0.27+0.008i & 0.06-0.12i & -0.17+0.3i & 0.04+0.03i \tabularnewline
    		29.3 & -0.022-0.067i & 0.1+0.012i & 0.15+0.009i & 0.07-0.12i & -0.23-0.06i & -0.03-0.14i & 0.18-0.06i & 0.0012+0.001i \tabularnewline
    		29.3 & -0.04+0.21i & -0.11-0.22i & -0.0053+0.15i & 0.21-0.21i & 0.24+0.13i & -0.019+0.37i & -0.013+0.08i & -0.12-0.14i \tabularnewline
    		41 & 0.11+0.13i & -0.11-0.15i & 0.11+0.016i & -0.056-0.26i & 0.27-0.076i & -0.14-0.16i & 0.22+0.13i & -0.013+0.048i \tabularnewline
    		41 & 0.12-0.066i & -0.25+0.25i & 0.33-0.16i & 0.022+0.12i & 0.25+0.05i & -0.03+0.37i & -0.21-0.33i & -0.001+0.003i \tabularnewline
    		51.4 & -0.02-0.17i & 0.165-0.004i & 0.19+0.32i & 0.07+0.16i & -0.21+0.006i & -0.003-0.13i & 0.065-0.12i & 0.005-0.002i \tabularnewline
    		51.4 & -0.072-0.095i & -0.035+0.19i & 0.17+0.23i & -0.02-0.45i & -0.24+0.22i & 0.4-0.14i & -0.17-0.04i & -0.13+0.06i \tabularnewline
    		56.45 & -0.58-0.15i & -0.035-0.004i & 0.09+0.19i & 0.018+0.16i & 0.12-0.16i & 0.04+0.023i & -0.1+0.008i & 0.0+0.0i \tabularnewline
    		56.45 & -0.008+0.23i & 0.34+0.4i & 0.2+0.036i & 0.09-0.11i & 0.11-0.19i & -0.25+0.024i & 0.018+0.063i & 0.05+0.015i \tabularnewline
    	\end{tabular}\end{ruledtabular}
\end{table*}

\end{document}